\documentclass[format=acmsmall, review=false, screen=true]{acmart}

\AtBeginDocument{%
  }

\usepackage{makecell}
\usepackage[colorinlistoftodos]{todonotes}
\usepackage{listings}
\usepackage{xcolor} 
\usepackage{tabularx}
\usepackage{booktabs}
\usepackage{subfigure}
\usepackage{mfirstuc}
\usepackage{url}
\usepackage{hyperref}
\usepackage[most]{tcolorbox}
\usepackage{enumitem}
\usepackage{xltabular}
\usepackage{array}
\usepackage{float}
\usepackage{multirow}
\usepackage{cleveref}
\usepackage{subcaption}
\usepackage{cleveref}

\Crefname{table}{Table}{Tables}

\definecolor{codegreen}{rgb}{0,0.6,0}
\definecolor{codegray}{rgb}{0.5,0.5,0.5}
\definecolor{codepurple}{rgb}{0.58,0,0.82}
\definecolor{backcolour}{rgb}{0.95,0.95,0.92}

\begin{document}

\newcommand{\aigc}{AI-generated code}
\newcommand{\bp}{bug patterns}
\newcommand{\cgm}{code generation methods}
\newcommand{\NoS}{126}
\newcommand{\AItools}{code generation tools}

\newcommand{\R}[1]{{\color{black}#1}} 

\newcommand{\update}[1]{{\color{black}#1}}

\title{A Survey of Bugs in AI-Generated Code}

\author{Ruofan Gao}
\orcid{0009-0009-0093-5790}
\email{r.gao@massey.ac.nz}
\affiliation{%
\
  \institution{School of Mathematical and Computational Sciences, Massey University}
  \city{Palmerston North}
  \country{New Zealand}
}

\author{Amjed Tahir}
\orcid{0000-0001-9454-1366}
\email{a.tahir@massey.ac.nz}
\affiliation{%
\
  \institution{School of Mathematical and Computational Sciences, Massey University}
  \country{New Zealand}
}

\author{Peng Liang}
\orcid{0000-0002-2056-5346}
\email{liangp@whu.edu.cn}
\affiliation{%
  \institution{School of Computer Science, Wuhan University}
  \country{China}
}

\author{Teo Susnjak}
\orcid{0000-0001-9416-1435}
\email{t.susnjak@massey.ac.nz}
\affiliation{%
\
  \institution{School of Mathematical and Computational Sciences, Massey University}
  \country{New Zealand}
}

\author{Ali Soltanian Fard Jahromi}
\orcid{0009-0003-6162-4570}
\email{21011981@massey.ac.nz}
\affiliation{%
\
  \institution{School of Mathematical and Computational Sciences, Massey University}
  \country{New Zealand}
}
\author{Foutse Khomh}
\orcid{0000-0002-5704-4173}
\email{foutse.khomh@polymtl.ca}
\affiliation{%
  \institution{SWAT Laboratory, Polytechnique Montr\'eal}
  \country{Canada}
}


\acmJournal{TOSEM}

\newcommand{\Foutse}[1]{\textcolor{blue}{{\it [Foutse: #1]}}}

\newcommand{\Amjed}[1]{\textcolor{violet}{{\it [Amjed: #1]}}}

\newcommand{\boxformat}{
 colback=lightgray!20,
    colframe=darkgray,
    boxrule=0.5mm,
    arc=2mm
}
\
\renewcommand{\shortauthors}{Gao et al.}

\begin{abstract}
\update{
AI coding systems 
are increasingly used to help developers write code at scale, improving software development efficiency and boosting productivity. While these models and agents have shown strong performance, concerns persist regarding the quality of the \aigc{}. 
Although prior work has reported a range of quality issues associated with \aigc{}, these findings remain scattered and lack a systematic synthesis. A comprehensive review is lacking that demonstrates the bug types, patterns, distribution, their correlation with specific AI coding systems, and the remediation strategies proposed to address them.
To fill this gap, we conduct a systematic literature review of existing studies on \aigc{} to develop a comprehensive understanding of bugs in the generated code and provide a foundation for future improvements in AI-assisted coding.
Our results show that bugs are widespread in \aigc{}, with \textit{functional} and \textit{stability} bugs 
emerging as the most frequently reported categories across the reviewed studies, a substantial proportion of which is attributed to model hallucination. 
Model size, instruction-following capability, and the quality and diversity of training data are key factors influencing bug occurrence. To mitigate bugs, prompt engineering techniques are widely used; however, certain domain-specific or semantically complex issues remain difficult to resolve automatically and still require human intervention. 
}

\end{abstract}





\maketitle

\section{Introduction}
\update{AI-powered code generation systems are revolutionizing software development. Rapid advances in Large Language Models (LLMs), along with the growing adoption of LLM-based coding tools and agents, have substantially expanded the capabilities and applications of AI-assisted software development. The underlying models, built using transfer learning to learn from existing code examples, can then generate code from natural-language descriptions or other programming contexts such as function signatures and test cases.} These tools and models have become invaluable resources for developers, boosting productivity by automating coding tasks, suggesting code snippets for code generation \cite{yan2023closer}, code completion \cite{yu2024fight}, code translation \cite{weisz2021perfection}, and assisting in program debugging and code repair tasks \cite{ma2024teach,li2023hitchhikersguideprogramanalysis,ribeiro2023large, sobania2023analysisautomaticbugfixing}. The most recent Stack Overflow Developer Survey (2025) \cite{stackoverflow2025ai} reported that 84\% of developers use or plan to use AI tools, yet more developers distrust AI-generated outputs (46\%) than trust them (33\%), a significant increase from the 2024 survey, indicating a substantial adoption–trust gap.
Several code generation models have emerged, each offering distinct features and capabilities tailored to different stages of software development (such as the GPT family \cite{brown2020language}, Claude \cite{anthropic2024claude}, Gemini \cite{team2023gemini}, Llama \cite{rozière2024codellamaopenfoundation}, 
DeepSeek-Coder \cite{guo2024deepseekcoderlargelanguagemodel}, among others). 

While AI code generation systems have demonstrated remarkable performance in various code-related tasks, there are still some key challenges  \cite{chong2024artificialintelligencegeneratedcodeconsidered, khan2024assessing}. The accuracy and correctness of the \aigc{} are still significant concerns \cite{COTRONEO2024112113}, as such code often contains bugs \cite{tambon2025bugs,yeticstiren2023evaluating} and security vulnerabilities \cite{majdinasab2024assessing,fu2025security,wang2024your}. These bugs often arise because the underlying models are trained on publicly available code from public code repositories such as GitHub, Stack Overflow, and other coding-related platforms, which are known to contain buggy code \cite{ray2014large} and security issues \cite{rokon2020sourcefinder}.
Such bugs can lead to runtime crashes or unexpected behaviors \cite{patel2024comparative}, thereby increasing the debugging and repair workload for developers, reducing productivity, and potentially delaying project timelines and incurring additional costs \cite{vaithilingam2022expectation}. The \aigc{} often lacks readability and consistency \cite{dantas2023assessing}, as it may employ non-standard naming conventions or fail to conform to the team's coding style \cite{liu2024refining}, thereby negatively impacting software maintainability.

To address the growing concerns about the reliability of \aigc{}, prior research has examined key aspects of generated code, including correctness~\cite{COTRONEO2024112113} and code quality~\cite{liu2024refining}. These studies have examined different models, programming languages, and specific bug categories to evaluate the quality of generated code across various contexts. However, this body of work remains fragmented, with each study focusing on isolated aspects, such as a specific model type, bug type, or domain-specific use case. As a result, we still lack a holistic and consolidated understanding of the types, patterns, and root causes of bugs that emerge in \aigc{}. A comprehensive synthesis of existing findings is crucial at this stage, especially given the accelerating adoption of \cgm{} in real-world development pipelines. Without a unified view of recurring failure modes and their implications, it is difficult to build effective evaluation frameworks, develop targeted mitigation strategies, or design more robust and reliable code-generation systems. Furthermore, identifying cross-cutting bug patterns and failure modes may reveal model-agnostic weaknesses or systematic issues that siloed studies have so far missed.

To fill this critical gap, this paper presents the first systematic literature review (SLR) on bugs in \aigc{}. Our goal is to aggregate and analyze prior empirical findings to uncover common bug types, recurring patterns, bug-prone AI coding systems, and their potential causes. We also examine how studies currently address these bugs, including fixes and mitigation strategies, and assess their applicability across settings. This review provides a much-needed foundation for understanding the failure landscape of \aigc{}, offering actionable insights for researchers and practitioners working to improve the reliability of current and future \cgm{}.


The main \textbf{contributions} of this survey can be summarised as follows:
\update{
\begin{itemize}
 \item  A comprehensive review of \NoS{} studies on bugs in \aigc{}, systematically examining bugs with respect to programming languages, datasets, detection methods, bug types and AI coding systems.

\item A systematic classification of bugs in \aigc{}, providing a structured basis for further analysis and comparison.

\item  An analysis of the tendencies of different AI coding systems in generating buggy code and the potential factors underlying these tendencies.

\item An analysis of existing bug mitigation approaches in the surveyed studies, highlighting potential directions for future research on bug mitigation for \aigc{}.
\end{itemize}
}

The remainder of this paper is organized as follows: Section \ref{sec:background} provides an introduction to the background. Section \ref{sec:methodology} outlines the methodology used in this systematic literature review. Section \ref{sec:results} presents the relevant research results on bug types, AI coding systems and bug mitigation methods. Section \ref{sec:discussion} summarizes the findings, limitations, and future research. Section \ref{sec:threats} discusses potential threats to validity of this study. Section \ref{sec:relatedwork} discusses the related work. Section \ref{sec:conclusions} provides the final conclusions.
\section{Background}
\label{sec:background}
This section provides the background information needed to understand the current status of bugs in AI-generated code. Section~\ref{sec:codingtasks} reviews the widespread application of various AI models and tools across different coding tasks. Section~\ref{sec:challenges} outlines key challenges in AI-assisted code generation, including concerns about the correctness and security of AI-generated code. Section~\ref{sec:programrepair} presents AI-driven program repair and distinguishes it from program repair targeting AI-generated code.

\subsection{Using AI for Coding Tasks}
\label{sec:codingtasks}

AI code generation models and tools are widely used to generate code due to their ability to generate source code from natural language prompts. These models and tools have shown promising results in assisting software development tasks, including code generation and completion. The GPT family comprises general-purpose models that can assist with various coding tasks. GPT-3 \cite{brown2020language} introduced basic code generation and completion capabilities. The subsequent GPT-3.5 model significantly improved both natural language understanding and code generation. GPT-4 \cite{openai2024gpt4technicalreport} continued to increase in sophistication, handling more complex coding scenarios with higher accuracy and reliability.

Meanwhile, Codex \cite{chen2021evaluatinglargelanguagemodels} further expanded its capabilities and domain expertise by using GitHub open-source projects to generate code from natural language descriptions, perform code completion, and facilitate simple refactoring.  
On the other hand, the Claude family \cite{anthropic2024claude} series of models focused more on safety and reducing hallucination, making them suitable for tasks such as code interpretation, unit test generation, and understanding multi-file projects. Later iterations, like Claude 3, enabled support for large-scale project development collaboration and cross-file and module collaborative code generation.
The Gemini series has evolved considerably since its initial release \cite{team2023gemini}. Gemini 1.5 introduced extended context window support, enabling processing of substantially longer code sequences. Gemini 2.0 further advanced the model's reasoning capabilities, demonstrating progressive improvements in code understanding and complex problem-solving across iterations. Unlike proprietary commercial models like Gemini and Claude, Code Llama \cite{rozière2024codellamaopenfoundation} is an open-source model series specifically optimized for code-related tasks, allowing for local deployment and customization. Likewise, DeepSeek-Coder \cite{guo2024deepseekcoderlargelanguagemodel}, one of the strongest open-source coding models to date, even surpasses traditional closed-source leaders on certain benchmarks, offering a clear cost advantage. Distinct from general-purpose or multimodal models, OpenAI o-series models (o1 \cite{openai2024openaio1card}, o3, and o4-mini \cite{OpenAI2025o3miniSystemCardOnline}) are designed with a focus on reasoning. With reasoning and problem-solving capabilities, they provide more intelligent and accurate support, especially in complex problem-solving, scientific research, data analysis, and other fields. O-series models are not limited to natural language processing (NLP), but also extend to multimodal processing capabilities such as image understanding.

In the field of AI-assisted coding tools, GitHub Copilot \cite{githubcopilot2022}, developed by GitHub and OpenAI, is a code completion and automated programming assistant that integrates with popular IDEs, including VS Code and IntelliJ IDE. It enables developers to leverage various LLMs from leading providers. Unlike GitHub Copilot, Cursor \cite{cursor2025} is an AI-driven code editor designed to improve development efficiency. 
It supports intelligent completion and code chat, understands the entire project structure, and provides more accurate bug fixes and refactoring suggestions. 
For organizations with stricter privacy or compliance requirements, tools like Tabnine \cite{tabnine} offer local deployment options that ensure model inference can run entirely on-premises.
For individual developers and small teams with limited budgets, Codeium \cite{codeium} provides free AI-assisted code completion, generation, and chat functionalities. Additionally, Amazon CodeWhisperer \cite{aws_codewhisperer} is an AI tool integrated into Amazon Web Services (AWS), particularly suited for developers working within the AWS ecosystem. It aims to simplify cloud application development by recommending cloud solutions, API integrations, and security practices.

These models and tools have promoted significant progress in the field of code generation. Developers can use them to generate code snippets, automatically complete functions, and implement specific algorithms, significantly improving programming efficiency.

\subsection{Challenges of AI-Assisted Code Generation}
\label{sec:challenges}

A growing body of research has highlighted significant challenges associated with \aigc{}. When writing code manually, developers draw on their own expertise, domain knowledge, and contextual understanding. In contrast, AI-assisted programming requires developers not only to write code but also to critically assess and validate AI-generated outputs. Despite the rapid adoption and widespread use of AI code generation tools, they offer no guarantees regarding the correctness or reliability of the produced code~ \cite{AutomaticProgramming}.

Empirical studies have shown that \aigc{} may contain bugs, including functional bugs \cite{tambon2025bugs} and security vulnerabilities~\cite{pearce2025asleep}. Some of these issues stem from flaws in the human-written training data itself, while others emerge from inherent limitations of current models, such as hallucinations, lack of deep semantic reasoning, and incomplete understanding of program intent ~\cite{fu2025security}. These challenges highlight the need for systematic quality assurance practices when integrating \aigc{} into software development workflows.

Quality issues mainly involve inadequate code reliability and correctness, which directly affect software maintenance, development, and use \cite{liu2024refining}. Common problems include logical bugs, code duplication, inconsistent coding styles, and performance issues. Among them, logical bugs can allow code to run but produce incorrect results, usually due to errors in the algorithm or data-processing logic \cite{ettles2018common}. In addition, code duplication means writing the same logic in multiple places, which increases modification complexity and the risk of bugs \cite{allamanis2019adverse}. Meanwhile, inconsistent coding styles reduce readability, making the code harder to understand and maintain \cite{wang2024functionalcorrectnessinvestigatingcoding}. Moreover, performance issues often stem from a lack of optimization, causing the program to become slow or consume excessive resources when processing large amounts of data or handling high concurrency. Solving these quality issues is crucial to maintaining the software's health and sustainability.

Security issues, in particular, have been a focus of \aigc{} studies, especially when sensitive data or critical services are involved \cite{ramirez2024state,fu2025security}. Vulnerabilities and security weaknesses in the generated code can lead to data leaks, service interruptions, or unauthorized access to the system. Previous studies have shown that tools such as Copilot have suffered from security weaknesses that can lead to significant issues. 
In a study of GitHub Copilot, Pearce et al. \cite{pearce2025asleep} found that about 40\% of the generated code contained vulnerabilities. Copilot has attempted to implement some improvements by adding a security layer to filter out some of the vulnerable code (e.g., \cite{natella2024ai}). However, recent work has shown that even with these improvements, the model still often produces insecure code \cite{majdinasab2024assessing}. 

\aigc{} may not fully meet developers' needs, resulting in additional time spent adjusting or modifying the generated code to meet requirements \cite{SERGEYUK2025107610}. 
Developers may spend time debugging and fixing potential logical and implementation bugs in the generated code, which reduces efficiency and increases development time. 
Even if the code is correct, a lack of clear comments or explanations may lead developers to use it without fully understanding it, which, in turn, affects code quality and system maintainability. In addition, developers may over-rely on AI models to generate code and overlook the need for an in-depth understanding of the generated code, resulting in a degradation of developer skills \cite{tang2024developer}. When AI models fail to generate suitable code, developers struggle to solve problems effectively. Traceability issues also arise, as many developers may not explain which parts of the code are auto-generated \cite{kashif2025developers}, making it harder to track and maintain in the future.

Most AI models are trained on public datasets, such as GitHub, Stack Overflow, and other coding-related platforms, which are known to contain both buggy code \cite{ray2014large} and security flaws \cite{verdi2020empirical}. Thus, these models may also produce erroneous code when deployed after training. Different models tend to produce different types and patterns of bugs, and the presence of bugs in the generated code can significantly reduce productivity and increase developers’ debugging and repair workloads, affecting
project schedules and costs.

\subsection{AI-Driven Program Repair}
\label{sec:programrepair}

Automatic program repair (APR) refers to the process of identifying and repairing bugs or errors in software code through automated means. The process aims to improve software reliability and performance, minimize human error, and reduce maintenance costs. Program repair typically uses techniques such as static analysis, dynamic analysis, search algorithms, and machine learning to generate effective repair solutions that automatically replace or modify defective code fragments. The application of AI in software engineering is also reflected in AI-driven program repair. Generative AI-driven tools can automatically detect bugs, vulnerabilities, or inefficiencies in the code and provide solutions.
Existing APR techniques are categorized into four main types: search-based, constraint-based, pattern-based, and learning-based \cite{huang2024evolving}.
The advances in LLMs have also led to a rapid increase in LLM-driven APR techniques, significantly facilitating software development and maintenance \cite{xia2023automated}. The application of LLMs in APR involves multiple strategies. Zero-shot learning enables the model to perform tasks without explicit examples \cite{xia2022less}, while fine-tuning can improve performance by adapting the pre-trained model to specific program repair scenarios \cite{huang2023empirical}. Additionally, a small amount of supervised learning utilizes limited examples to enhance the model's generalization ability \cite{allamanis2021self}. In terms of input form, LLMs can handle multiple types of input, including directly providing bug information and code snippets, guiding the generation of repair code through explicit prompts, indicating the code parts that the model needs to fill in, interacting with the model, and designing the input format by considering program structure characteristics \cite{li2025context}. These techniques enable LLMs to effectively cope with a variety of repair scenarios, covering 18 different bug types, such as semantic bugs, security vulnerabilities, static warnings, syntax bugs, type bugs, performance bugs, and hardware bugs, thereby significantly improving the efficiency and accuracy of program repair \cite{zhang2024systematic}. 

Unlike reviews of AI for program repair, this review explores methods for repairing AI-generated code. An online survey of 34 software practitioners and researchers using LLMs showed that most respondents considered fixing errors and bugs in LLM-generated code moderately difficult \cite{tambon2025bugs}. Some participants manually corrected bugs by reading stack traces and searching error messages online, while others combined manual inspection with LLM prompts to correct the code. Regarding code diagnosis, respondents indicated that most bug patterns were easy to identify, especially when using IDEs with real-time code checking features. Syntax bugs and low-level bugs were the simplest and easiest to fix. Hard-to-diagnose bug patterns include misunderstandings, unconfirmed assumptions, and omitted edge cases, which are more similar to bugs and errors that developers might make. Respondents also reported that bugs in the misunderstanding category were harder to fix than other types, as these bugs indicated a significant discrepancy between the generated code and the code specified in the prompt. In such cases, correcting the code required re-prompting or extensive manual intervention. For certain bug types, such as hallucinated objects, participants noted that providing more information in the prompt often helped the LLM correct the bug.
\section{Study Design}
\label{sec:methodology}
\begin{figure}[htbp]
\centering
\includegraphics[width=\linewidth]{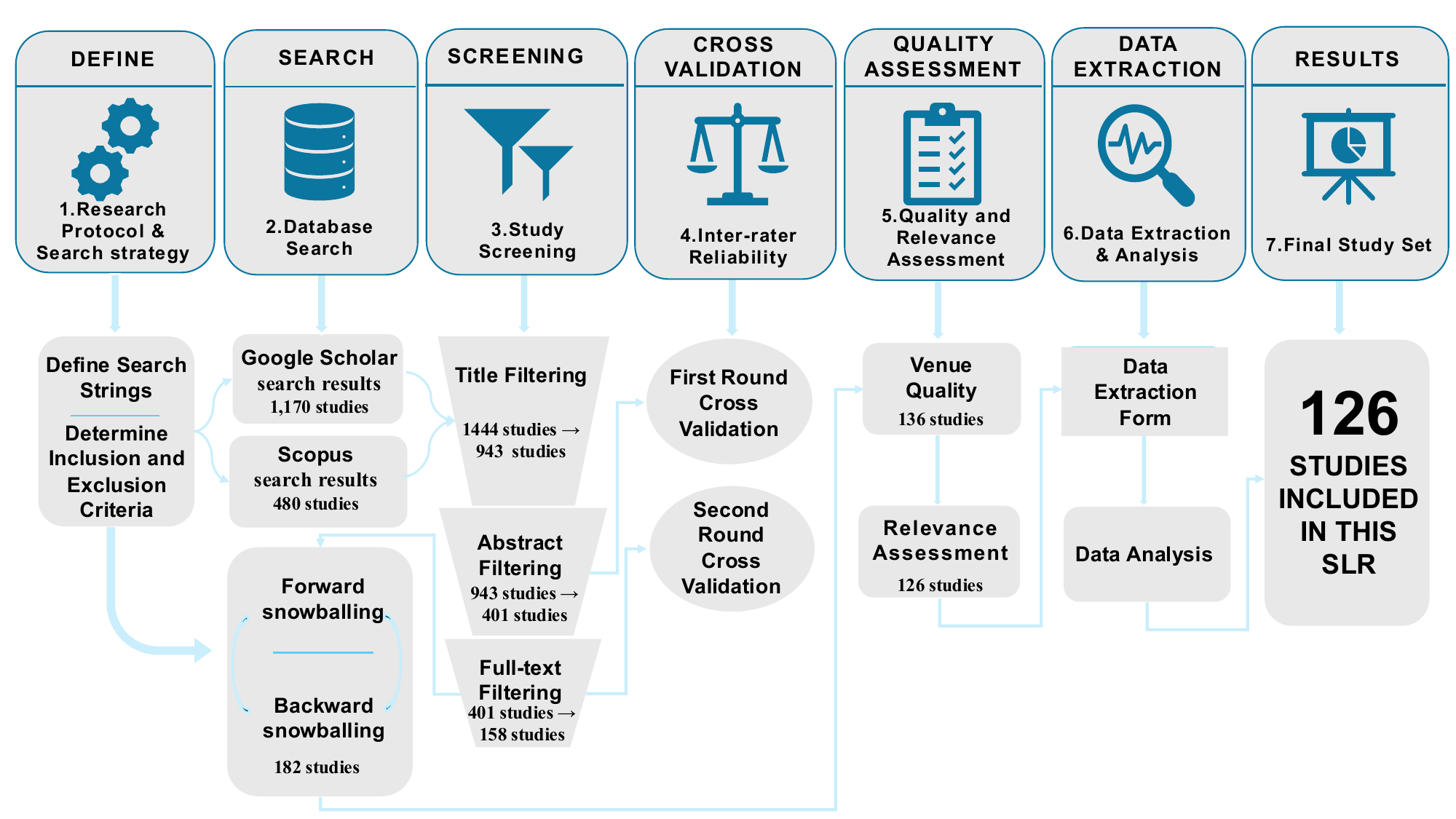}
\caption{The workflow of this systematic literature review}
\label{fig:process}
\end{figure}

The \textit{goal} of this study is to develop a comprehensive understanding of bugs in \aigc{}, including the types that commonly occur, the coding systems in which they appear, and proposed mitigation strategies. 
To this end, we conducted a Systematic Literature Review (SLR) - a well-established method in software engineering research for rigorously identifying, analyzing, and synthesizing relevant empirical evidence across a body of literature. Our SLR follows the methodological guidelines established by Kitchenham and Charters~\cite{KitchenhamGuidelines}, as well as Wohlin’s snowballing strategy~\cite{wohlin2014guidelines}, to ensure both rigor and coverage in identifying and synthesizing relevant studies. Following previous SLRs, such as Harzevili et al. \cite{shiri2024systematic}, Tahir et al. \cite{TAHIR2023111837}, and Fawzy et al. \cite{fawzy2025exploring}, we designed our search process based on the practices of these SLRs in keyword combinations, database selection, filtration strategy, and study selection. \update{Our review covers the period between 2020 and June 2026. The review process consists of two phases. The initial review covers studies published from January 2020 to April 2025. To keep the review up to date, we then conduct a follow-up search covering the period between April 2025 to June 2026.} Figure~\ref{fig:process} provides an overview of our review process, which is detailed in the following subsections.


\subsection{Research Questions}

By applying an SLR, we aim to address three important aspects of bugs in \aigc: 1) understanding bug types and patterns, 2) identifying the AI coding systems used, and 3) any fix or mitigation strategies that address those bugs. We list the corresponding three RQs below, together with the rationale for each RQ. These RQs guide the scope of our review and inform the design of our search, selection, and data extraction processes.

\noindent\textbf{RQ1: What are the bug types and patterns found in \aigc{}?}

\update{This RQ aims to provide a comprehensive overview of the types of bugs present in AI-generated code. To understand the nature of bugs across different AI coding systems, we summarize bug types and patterns from the reviewed studies and classify them into several categories based on their characteristics. We synthesize and organize the bug types reported in prior software engineering studies \cite{CATOLINO2019165, gyimesi2021bugsjs, tan2014bug}. We also utilize the classification of Tambon et al. \cite{tambon2025bugs}, which is specific to bugs found in \aigc{} in GitHub projects. }

\update{\noindent\textbf{RQ2: Which AI coding systems are more prone to produce bugs and why?}

Understanding the bug proneness of different AI coding systems is crucial for improving code-generation capability. Although AI models have made significant progress in code generation, these systems still vary in correctness, accuracy, reliability, and security. Given the diversity of existing models, their outputs may exhibit different types of bugs and defects depending on contextual factors such as prompt design, decoding temperature, or model-specific parameters. Identifying their bug-proneness can guide future selection, improvement, and application of AI \cgm{} and systems.}


\noindent\textbf{RQ3: What methods have been used to mitigate or fix bugs in \aigc{}?}

This RQ aims to collect strategies for mitigating or fixing bugs in \aigc~ (e.g., through program analysis and automatic program repair, prompt engineering, etc.). We collected and analyzed all the methods discussed in the selected studies that aim to improve the AI code generation capabilities. Investigating these methods is crucial to improving the robustness of \aigc{}, ensuring safer adoption in production environments, and guiding future research on trustworthy AI code generation.


%

\subsection{Search Strategy}
We selected two popular search engines as our search data sources: Google Scholar and Scopus. Both search engines have been used in similar reviews \cite{negri2024systematic, hirsch2022systematic}.
Google Scholar indexes a wide range of literature relevant to computer science and software engineering \cite{neuhaus2006depth}, including key journals and conference proceedings. It allows for constructing complex keyword searches using Boolean operators, supporting searching titles and full text. It also provides advanced search options that can be refined by author, publication, and time. Its search results are typically sorted by relevance, making it easier for researchers to find the most pertinent literature. 
Scopus 
allows access to high-quality peer-reviewed journals and conference proceedings. Scopus's search logic is more precise, supporting boolean operators and field searches, allowing for accurate control over the search scope \cite{baas2020scopus}. 
Combining Google Scholar and Scopus complements the breadth and depth of literature retrieval. Google Scholar and Scopus cover all major publishers that publish software engineering research \cite{neuhaus2006depth}, such as IEEE Xplore, ACM digital library, SpringerLink, and ScienceDirect. 

We divided our topic into three key dimensions, including ``\textit{AI tools}'', ``\textit{generation task}'', and ``\textit{issue focus}'', to construct the search strings in both Google Scholar and Scopus. We also included variations of keywords to represent ``\textit{bug}'', including ``\textit{fault}'', ``\textit{flaw}'', and ``\textit{defects}''. The keywords we used are listed in Table~\ref{tab:keywords}.
We constructed complex search strings containing specific keywords using Boolean operators to retrieve literature from Google Scholar and Scopus. We experimented with various keyword combinations and conducted several pilot runs to refine the search strings. 

Initially, we connected the three keyword groups using the ``\textit{OR}'' operator. Then we combined the groups as a whole using the ``\textit{AND}'' operator. Scopus offers many search scopes, such as Authors, Article title, Abstract, Keywords, and All fields. We tried several of the search scopes mentioned above and evaluated the relevance of the returned articles. Ultimately, we found that searching within Article title, Abstract, Keywords yielded the best results, which we then employed in our search. 
When using Google Scholar, since the search scope was anywhere in the article, which resulted in many irrelevant search results, we adjusted our search string by combining the ``\textit{AI tools}'' and ``\textit{generation task}'' keyword groups into merged terms like ``\textit{\aigc}''. Then, we linked them with the issue-focused keyword group using the ``\textit{AND}'' operator to create a new search string.

\begin{table}[]
\caption{Keywords and synonyms, abbreviations, and other alternatives}
  \label{tab:keywords}
\resizebox{\columnwidth}{!}{%
\begin{tabular}{@{}ll@{}}
\toprule
\textbf{Component} & \textbf{Keywords and corresponding synonyms, abbreviations, and other alternatives} \\ \midrule
AI Tools & LLM, AI, Artificial Intelligence, automatic, LLMs, large language models, ChatGPT, Copilot \\ \midrule
Generation Tasks & generated code, generated programs, code generation, code generated \\
\midrule
Issue Focus & bug, buggy code, bug patterns, incorrect code, code quality issues, defects, fault, flaw \\ \bottomrule
\end{tabular}%
}
\end{table}

We set our search to start from January 2020, as most key developments with \aigc{} emerged with the development of LLM models for coding tasks, which started with some milestone work since 2020 that promoted LLM research for code tasks, such as the release of GPT-3 \cite{brown2020language} and CodeBERT \cite{feng2020codebertpretrainedmodelprogramming}. Table~\ref {tab:strings} shows the search strings used to identify relevant studies in both Scopus and Google Scholar. Due to the differences in search scope and rules of the two search engines, slight variations between the two search strings were required (for example, using the keyword ``\textit{automatic}'' in Google Scholar yielded too many irrelevant studies, while it can be used in Scopus because its search scope is limited and does not perform a full-text search). 

\begin{table}[]
\caption{Search strings across data sources}
 \label{tab:strings}
 \resizebox{\columnwidth}{!}{%
\begin{tabular}{ l  p{12cm} }
\toprule
\textbf{Search Engine} & \textbf{Search String} \\ \midrule
Scopus & TITLE-ABS-KEY ( ( ``AI'' OR ``Artificial Intelligence'' OR ``automatic'' OR ``LLMs'' OR ``large language models'' OR ``ChatGPT'' OR ``Copilot'' ) AND ( ``generated programs'' OR ``code generation'' OR ``generated code'' OR ``code generated'' ) AND ( ``bug*'' OR ``incorrect code'' OR ``code quality'' OR ``defects'' OR ``fault'' OR ``flaw'' ) ) AND PUBYEAR > 2019 AND PUBYEAR < 2027 AND ( LIMIT-TO ( DOCTYPE , ``cp'' ) OR LIMIT-TO ( DOCTYPE , ``ar'' ) ) \\ \midrule
Google Scholar & ``AI-generated code'' OR ``LLM-generated code'' OR ``large language models generated code'' OR ``ChatGPT-generated code'' OR ``Copilot-generated code'' OR ``buggy code'' OR ``bug patterns'' OR ``incorrect code'' OR ``code quality issues'' OR  ``defects'' OR ``fault'' OR ``flaw'' after:2019 \\ \bottomrule
\end{tabular}%
 }
\end{table}

\update{We performed the first-phase search on April 2, 2025, followed by the second-phase search on June 25, 2026, retrieving a total of 1,170 studies from Google Scholar (526 in the first phase and 643 in the second phase) and 480 from Scopus (200 in the first phase and 280 in the second phase).
Because the same study may appear in multiple search engines, we manually removed duplicate results from the two search engines (33 in Phase 1 and 70 in Phase 2). After removing duplicates, we identified 1,444 unique studies: 660 from Phase 1 (493 from Google Scholar and 167 from Scopus) and 783 from Phase 2 (573 from Google Scholar and 210 from Scopus).}

\subsection{Selection Criteria}
To ensure the quality and relevance of our review, we developed inclusion and exclusion criteria and applied them when selecting studies. 
The purpose is to systematically identify the studies most relevant to our RQs and exclude those that do not meet the selection criteria.

We included studies that met the following criteria:
\begin{enumerate}[label=I\arabic*]
  \item \update{Studies investigating bugs in \aigc{}, including studies on code generation by LLMs, coding tools, or coding agents, as well as studies on buggy code generation and test generation, provided that they contain concrete evidence or analysis of bugs in \aigc{}.}
  \item Published in peer-reviewed journals, conference proceedings, or workshops. 
  \item Published in English.
\end{enumerate}

We excluded studies that met any of the following exclusion criteria:

\begin{enumerate}[label=E\arabic*]
  \item Studies that utilize AI and LLMs for bug localization, reduction, and program repair (where the subject code is not necessarily AI-generated).
  \item Studies that are not accessible electronically, or the full text is not available for download.
\item Studies that focus mainly on security issues in \aigc{}. Several reviews have extensively discussed the security of \aigc{}, such as \cite{negri2024systematic, basic2025vulnerabilitiesremediationsystematicliterature, articlecybersecurity}. Therefore, we consider this topic outside our scope.
\item \update{Studies available only as preprints (including those on arXiv), as they are not peer-reviewed. If a study initially appeared on arXiv but was later published in a peer-reviewed venue, we include the published version.}
\end{enumerate}





\subsection{Filtration Process}
We manually filtered the studies we obtained from both search engines. Two of the authors carried out this filtration process. We first reviewed the titles and abstracts to exclude studies that did not meet the inclusion criteria. The remaining studies were then reviewed in full text to further assess their relevance. The third co-author cross-validated the results at each stage and used Cohen’s Kappa (k) to assess consistency between the two authors. When inter-rater consistency was low (Cohen's Kappa < 0.8), the two authors discussed and resolved any disagreements about study inclusion, and refined the inclusion criteria accordingly.


We note that filtering studies based solely on titles can omit relevant studies, as titles may not contain enough information to determine relevance. Thus, we took a more inclusive approach and included studies related to bugs in \aigc~ (as long as the title contains coding tasks and AI-related keywords). The title filtration process resulted in the selection of \update{943 studies (382 in Phase 1 and 561 in Phase 2).}  

We then filtered studies based on their abstracts. 
The first author conducted the first round of abstract filtration after thoroughly reading the abstracts of all 943 selected studies. We performed another cross-validation (with another co-author) to ensure the reliability and consistency of the literature filtration process. We randomly selected ten studies from each search engine (20 in total), including five marked as relevant by the first author and another five marked as irrelevant. 
The two authors discussed all cases of disagreement to reach consensus. After filtering studies by abstracts, we identified \update{401 studies (178 in Phase 1 and 223 in Phase 2)} that were retained for full-text review. 


Our last filtration stage involved a full-text review to determine whether the studies met our inclusion criteria. 
Similar to the previous stages, after the first author conducted the full-text screening of all 402 studies, we cross-validated with another co-author. We initially disagreed about some studies that mainly focus on the code generation task. For example, the terms ``buggy code generation'' and ``test generation'' may refer to different types of source code generation, yet they could still be relevant to our RQs. Therefore, after discussion, we decided to include studies that address these topics. After conducting the full-text filtration and cross-validation processes, we identified \update{158 studies (72 in Phase 1 and 86 in Phase 2)} that were relevant to our review.

\subsection{Snowballing}
To enhance the comprehensiveness of the studies included in this review, we employed snowball sampling using the guidelines of Wohlin et al. \cite{wohlin2014guidelines}. We performed \textit{backward} and \textit{forward} snowballing on all 158 selected studies. In the backward snowballing process, we reviewed the list of references in the selected studies. In the forward snowballing, we checked which studies cited the studies we included in both Google Scholar and Scopus. We then reviewed the titles, abstracts, and full texts of all new studies identified through the snowballing process and applied our inclusion and exclusion criteria to each study. We added the relevant studies to our final list. 

We encountered many duplicates and had already included studies during the snowballing process. After excluding these, we identified \update{24 new studies (21 in Phase 1 and 3 in Phase 2), bringing the total to 182 studies.}

\subsection{Data Extraction and Analysis}
Our data extraction was conducted by two of the co-authors. This was because of the large amount of data we needed to extract from each study and to minimize bias and errors. Each of the two authors reviewed half of the studies and independently extracted data from the selected studies to answer our three RQs, using a predefined data extraction form with the following key attributes:

\begin{itemize}
    \item \textbf{Basic information}: authors, year of publication, title, and publication venue.

    \item \textbf{Research objectives}: stating the study's main objectives and research questions.

    \item \textbf{Methodology}: describing the code generation tasks, the datasets and programming languages used, and the employed bug detection methods and techniques.

    \item \textbf{Bug types and patterns}: to address RQ1, we recorded the types of bugs detected and relevant descriptions. 

    \item \textbf{AI coding systems}: to address RQ2, we extracted information on the specific LLMs/coding tools/coding agents involved and their performance with respect to bug generation.

    \item \textbf{Bug mitigation/fixing methods}: to address RQ3, we examined whether the studies proposed methods to mitigate or fix bugs, and recorded the methods and the types of bugs addressed. 

    \item \textbf{Additional information}: summarizing the main findings, comparisons with other studies, and any identified limitations or shortcomings.
\end{itemize}

After completing data extraction for all studies, the two co-authors cross-validated the results by randomly selecting 20 studies from each other's assigned portions and independently re-extracting the data. Discrepancies were identified in the extracted data for five studies. These discrepancies were resolved through discussion and rechecking the original papers, and the final agreed-upon data were used in the subsequent analysis. To assess the quality of the publication venues of the selected studies, we considered their journal and conference rankings. We used the SJR ranking system\footnote{\url{https://www.scimagojr.com/journalrank.php}} to evaluate journal quality and the CORE ranking\footnote{\url{https://portal.core.edu.au/conf-ranks/}} for conference proceedings. \update{Specifically, we used Q1/Q2 journals and CORE A*/A conferences as minimum venue-quality criteria to ensure that the included studies were published in well-established venues. After evaluating all publication venues, 46 studies (14 in Phase 1 and 32 in Phase 2) were excluded because they did not satisfy the venue-selection requirements, leaving 136 studies (79 in Phase 1 and 57 in Phase 2) published in reputable venues.}



In addition, we conducted a qualitative assessment based on each study's topic and its relevance to bugs in \aigc{}. If a study focuses explicitly on bugs in \aigc{}, indicating strong alignment with our research topic, we labeled it as “Focus”. If a study’s main topic is not directly about bugs in \aigc{} but it discusses bug-related issues in some sections, we labeled it as “Discuss”. If a study only briefly mentions some bugs in a few sentences and has limited relevance to our study, we labeled it as “Mention” and considered it for exclusion. \update{As a result, 66 studies (39 from Phase 1 and 27 from Phase 2) were labeled as Focus, 60 (32 from Phase 1 and 28 from Phase 2) as Discuss, and 11 (8 from Phase 1 and 3 from Phase 2) as Mention. After excluding the 10 `Mention' studies, we retained a total of \NoS{} studies (71 from Phase 1 and 55 from Phase 2). We then used a bottom-up approach to identify each bug for data analysis. We synthesized and organized the identified bug types based on existing bug classifications in software engineering \cite{CATOLINO2019165, gyimesi2021bugsjs, tan2014bug}. We also incorporated insights from the classification \cite{tambon2025bugs} specific to bugs found in \aigc{} in GitHub projects.}

\section{Results}
\label{sec:results}
\subsection{Overview}
\subsubsection{Timeline}
\begin{figure}[htbp]
    \centering
    \resizebox{0.99\textwidth}{!}{ 
    \subfigure[The distribution over time]{
        \includegraphics[width=0.45\textwidth]{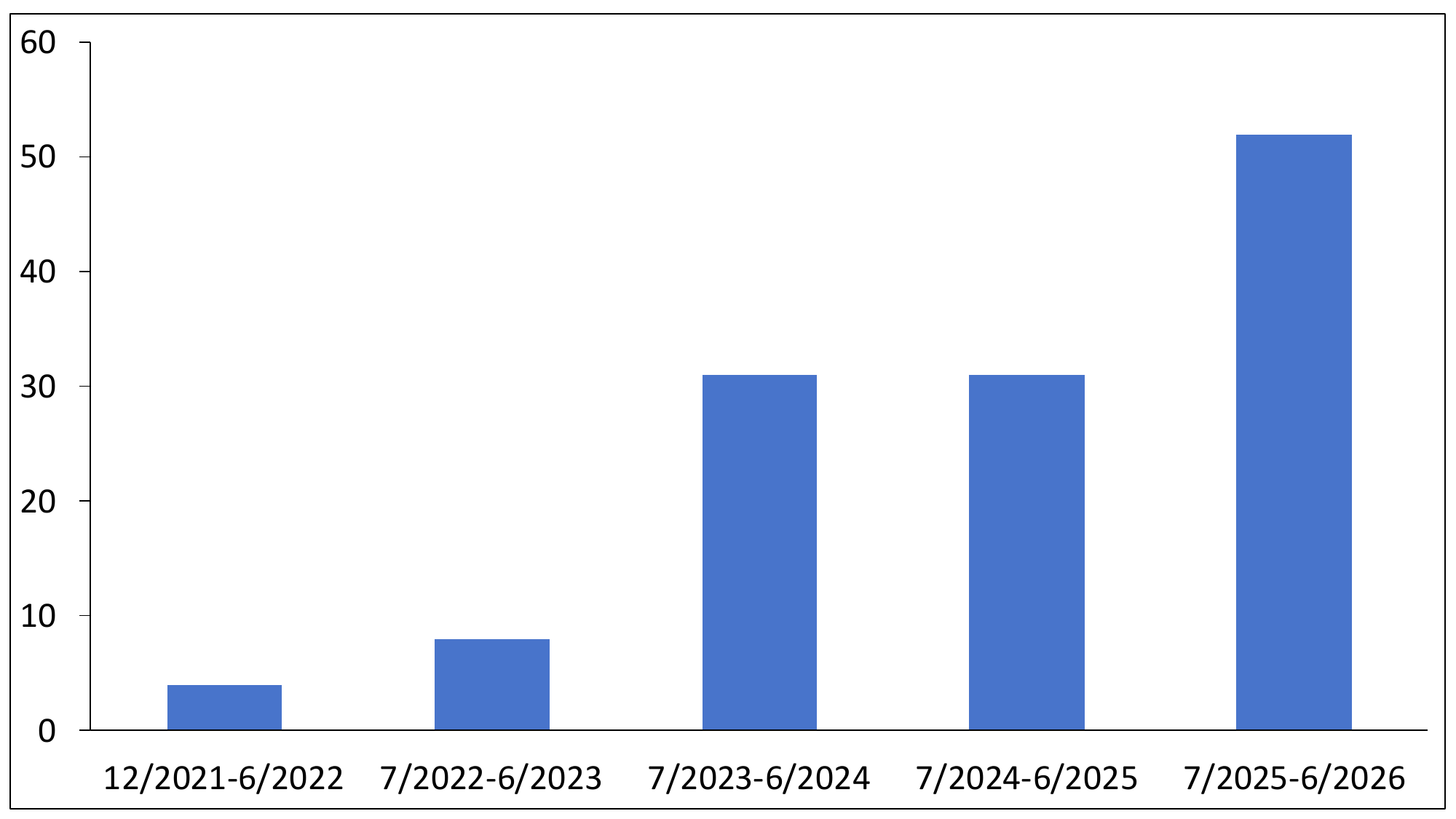}
        \label{fig:year}
    }
    \subfigure[The distribution across venues]{
        \includegraphics[width=0.45\textwidth]{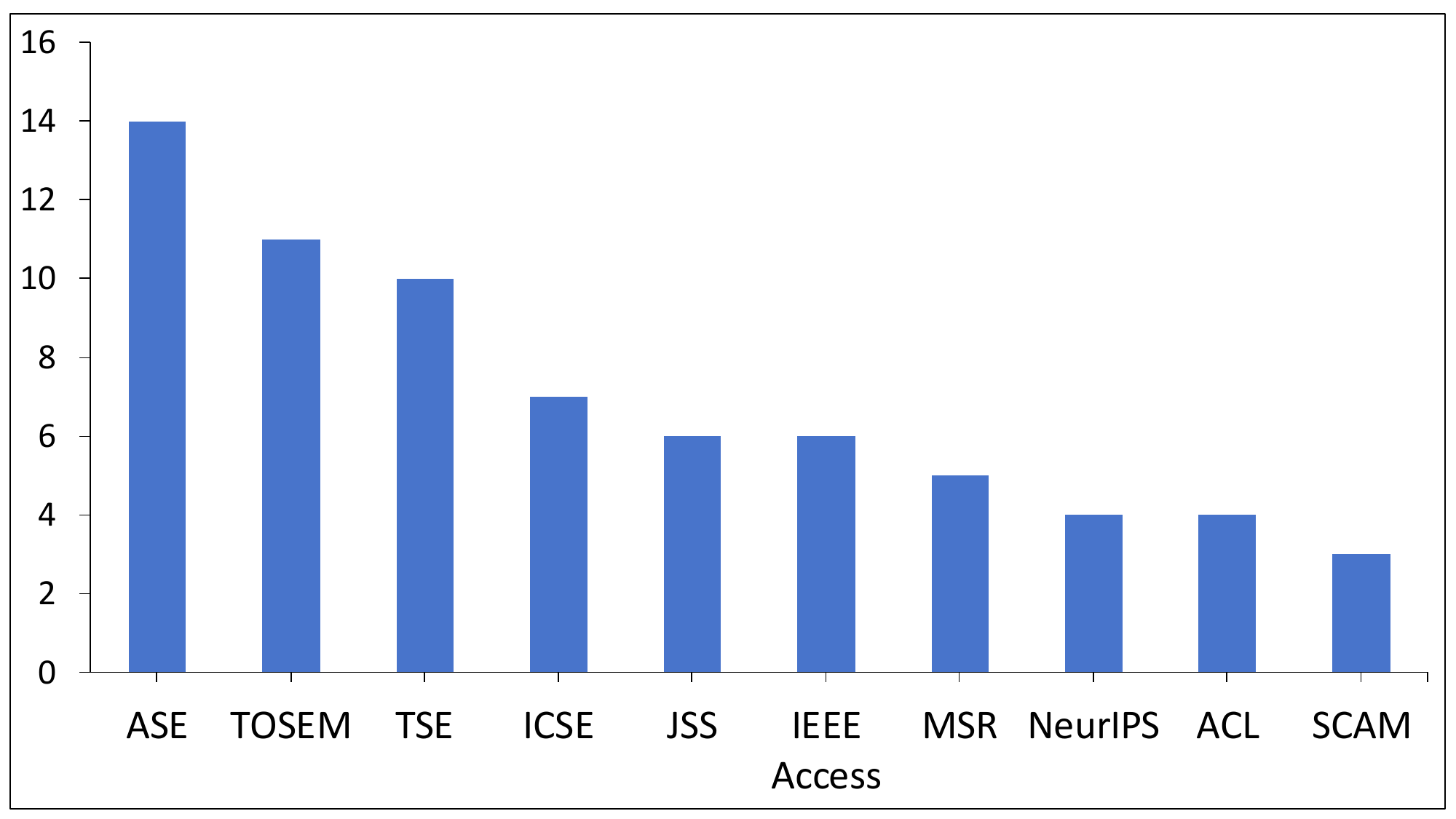}
        \label{fig:venue}
    }
    }
    \caption{Distribution of the selected studies across venues and years}
    \label{fig:venueandyear}
\end{figure}

\update{As shown in Figure~\ref{fig:year}, the studies included in this review range from December 2021 to June 2026. To better capture the development trend, we group the studies into five consecutive one-year periods. We identified only 4 studies during the earliest period (December 2021 to June 2022), which indicates that research on bugs in \aigc{} was still in its infancy. This is followed by a modest increase from July 2022 to June 2023, with 8 studies. This period coincides with the broader adoption of tools such as ChatGPT and GitHub Copilot, which brought increased attention to the reliability and correctness of AI-generated code. Growth accelerates between July 2023 and June 2024, when the number of studies rises to 31. This marks the beginning of a period of rapid expansion in this research field. Between July 2024 and June 2025, the number of relevant studies remained stable at 31. As expected with the recent development of AI models and coding tools, the most substantial growth occurred in the most recent period, from July 2025 to June 2026 (52, around 41\% of studies). This sharp increase indicates that the topic is increasingly gaining significant traction, driven by the rapid advancement and widespread use of LLM-based coding tools.}

\subsubsection{Venues}
We conducted a high-level analysis of the publication venues in which the \NoS{} selected studies appeared, to demonstrate the breadth and quality of our selection. We grouped the publication venues based on their Scimago Journal ranking\footnote{\url{https://www.scimagojr.com/journalrank.php}}, Google Scholar Metrics\footnote{\url{https://scholar.google.com/citations?view_op=top_venues}}, and the ICORE/CORE's conference ranking\footnote{\url{https://portal.core.edu.au/conf-ranks/}}. 
Figure~\ref{fig:venue} shows the publication venues (note: we present only venues that appear at least three times). 
We found that most of the studies we included appear in reputable software engineering conference venues, 
including the International Conference on Software Engineering (ICSE), the International Conference on Automated Software Engineering (ASE) and the International Working Conference on Mining Software Repositories (MSR). 
Publications also appear in renowned software engineering journals, 
including ACM Transactions on Software Engineering and Methodology (TOSEM), IEEE Transactions on Software Engineering (TSE), and Journal of Systems and Software (JSS). 
In addition, several publications also appear in top-ranked AI and NLP conferences such as Conference on Neural Information Processing Systems
(NeurIPS) and the Annual Meeting of the Association for Computational Linguistics (ACL). 
\update{The studies published in higher-ranked venues are generally subject to rigorous review processes and tend to provide more detailed methodology, stronger evaluation protocols, and more comprehensive validation, which can enhance the reliability of the extracted findings.}

\subsubsection{Programming Language}
We further analyzed the studies' context in terms of the programming languages used, as some bugs are more language-related and may affect programs written in certain languages more than others.  We noticed that several studies have discussed bugs in the context of multiple languages. We analyzed the frequency of programming languages reported in studies that provide this information, noting that some studies may reference multiple languages. 
\update{Figure~\ref{fig:PL} shows the top languages used in these studies. As shown in the figure, Python is the predominant programming language, with almost half of the studies (48\%) exploring bug aspects in Python. This is followed by Java (27\%), JavaScript (8\%), C\texttt{++} (8\%), C (4\%), and TypeScript (3\%).} 
Python is a language used for machine learning and data science, characterized by its dynamic nature and relatively concise syntax. Its dominance reflects the preferences of both the research community and LLM developers. Since most research on LLMs and code generation originates from the AI community, Python has naturally become the primary focus of experiments. Many existing code generation and repair benchmarks are also built around Python, encouraging new studies to adopt these established benchmarks, creating a Python-centric research cycle.

\subsubsection{Datasets}
To better understand the scope of bugs in \aigc{} across studies, we analyzed the datasets used in studies that provide this information. Datasets serve as benchmarks that enable systematic and quantitative analysis of bug distributions and characteristics in \aigc{}. We observed that some standard coding datasets have been used more frequently than others. 
Figure~\ref{fig:datasets} illustrates the frequency of datasets used across studies. Most studies (50\%) rely on existing, publicly available, widely used datasets, such as HumanEval \cite{chen2021evaluatinglargelanguagemodels}, APPS \cite{coooper2021code}, DevGPT \cite{xiao2024devgpt}, and MBPP \cite{austin2021program}. 
The three datasets focus exclusively on Python programming tasks. However, Koohestani et al. \cite{koohestani2025benchmarking} reported that HumanEval suffers from several inherent flaws and inconsistencies, including incorrect test cases, insufficient test coverage, erroneous reference solutions, and imprecise problem definitions. They also noted that MBPP exhibits notable shortcomings, such as poor syntax (e.g., excessive whitespace and Python method names starting with uppercase letters—an uncommon convention outside of classes), as well as uncaught bugs and edge cases that break implementations. Moreover, we observed that HumanEval, MBPP, and APPS rely heavily on execution-based metrics such as Pass@k while overlooking code quality and semantic nuances \cite{hu2025assessing}. \update{Beyond these, other studies (e.g., \cite{tian-etal-2024-debugbench, swaraj2026trust}) constructed new datasets by sourcing programming problems from programming competition platforms (7\%) or community platforms (12\%). Education-oriented studies (11\%) often drew upon course assignments or specific programming tasks as datasets. In addition, some studies (e.g.,~\cite{hendrycks2021measuring, honarvar2025question}) developed self-constructed datasets (17\%), while others (such as \cite{baralla2024assessing}) did not specify the datasets used in their experiments (7\%).}

\subsubsection{Bug Detection Methods}
\begin{figure}[htbp]
    \centering
    \subfigure[The frequency of programming languages]{
        \includegraphics[width=0.45\textwidth]{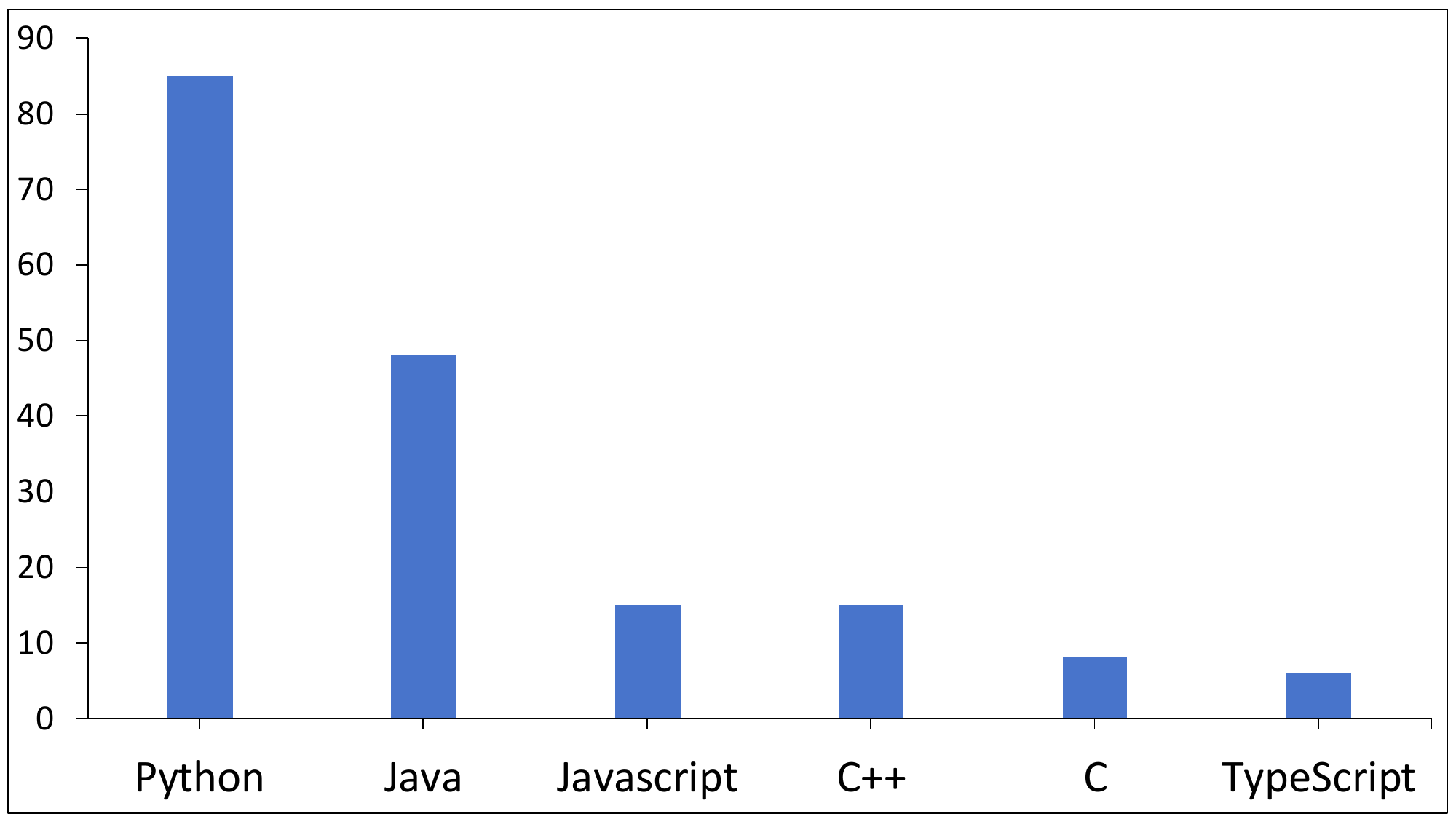}
        \label{fig:PL}
    }
    \hfill
    \subfigure[The frequency of datasets]{
        \includegraphics[width=0.45\textwidth]{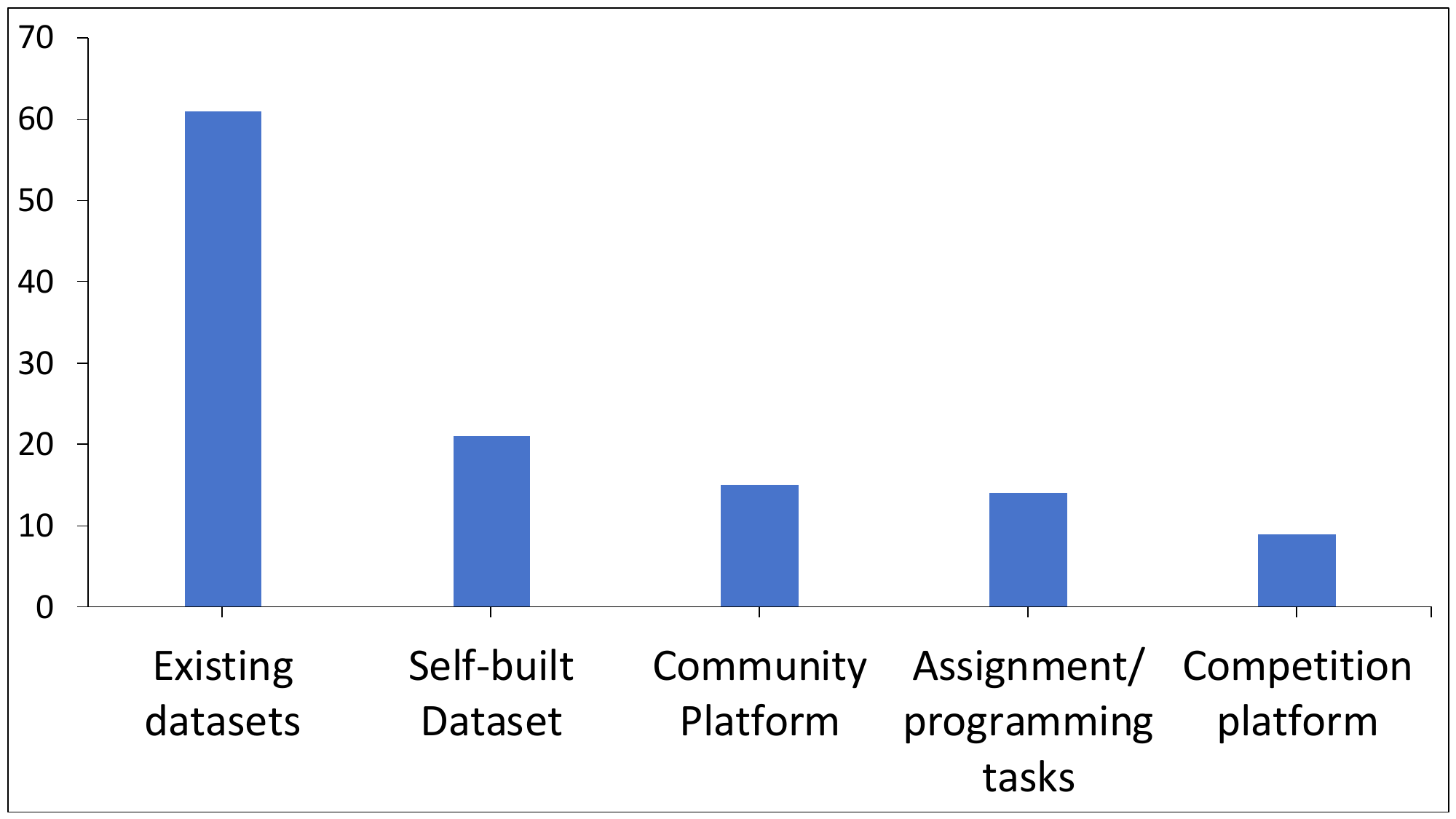}
        \label{fig:datasets}
    }

    \vspace{0.5em}

    \subfigure[The frequency of bug detection methods]{
        \includegraphics[width=0.45\textwidth]{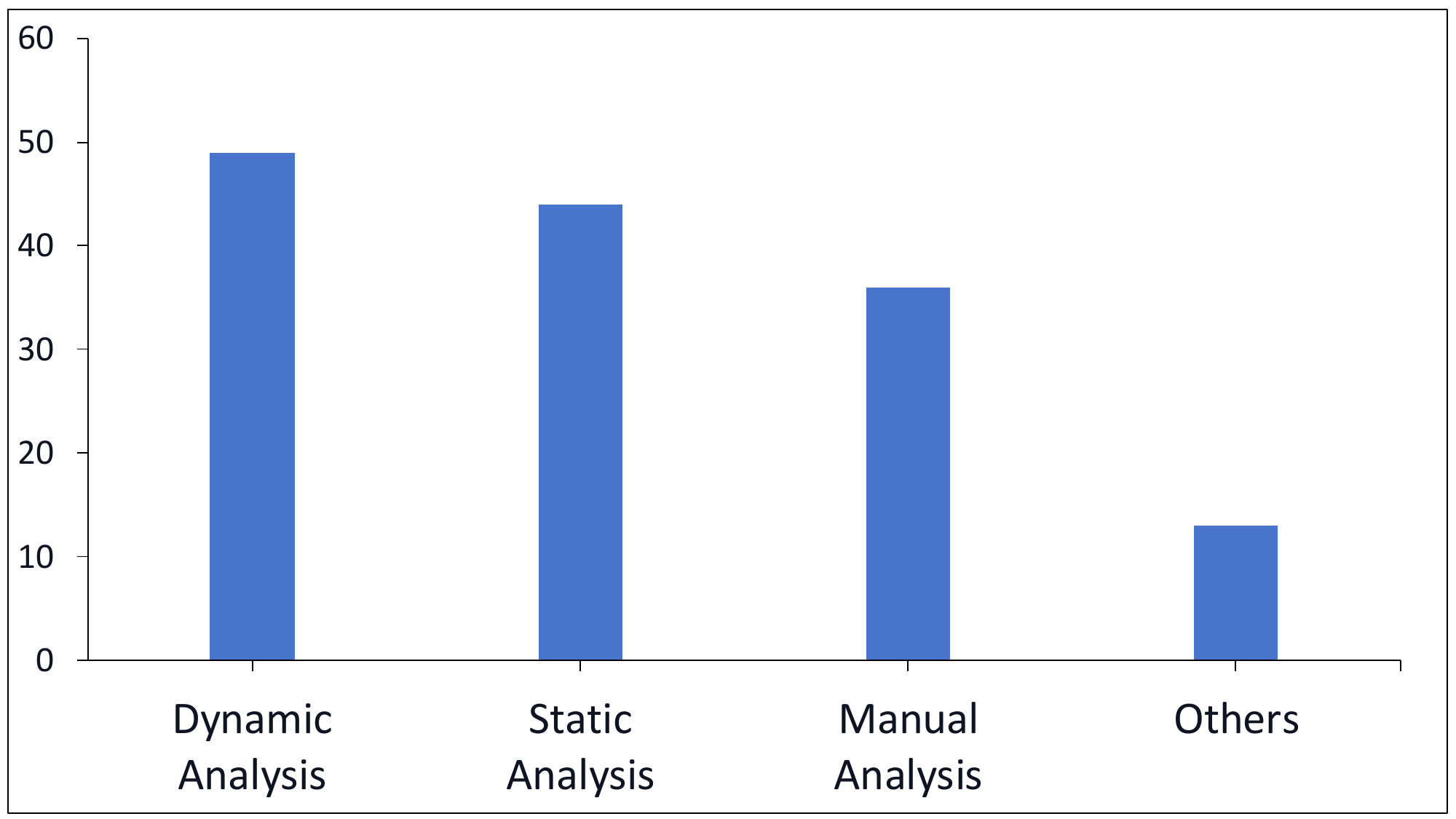}
        \label{fig:bugdetection}
    }

    \caption{The frequency of programming languages, datasets, and bug detection methods in the studies}
    \label{fig:PLanddatasetandbugdetection}
\end{figure}

Bug detection methods are fundamental to evaluating the quality and reliability of \aigc{} as they enable researchers to systematically identify bug types and patterns. Existing studies largely rely on traditional static and dynamic analysis rather than developing methods specifically tailored to \aigc{}. \update{As shown in Figure~\ref{fig:bugdetection}, the studies that provide this information employed three main types of bug detection approaches, with dynamic analysis (39\%) being the most commonly used (e.g.,~\cite{liu2023your, 10.1145/3756681.3756964}), followed by static analysis (e.g., \cite{moratis2024write, afrin2025quantization}) (35\%) and manual inspection (e.g., \cite{kou2024large, cassano2023multipl}) (28\%). This includes the use of popular static analysis tools such as PMD\footnote{\url{https://pmd.github.io/}}, Pylint\footnote{\url{https://pylint.pycqa.org/}}, and Bandit\footnote{\url{https://bandit.readthedocs.io/}}. Note that only about 3\% of the studies did not specify the bug detection method used. However, we confirmed that the bugs they reported cover all categories in our classification and are accompanied by adequately detailed descriptions to support accurate classification. Consequently, the lack of information about the bug detection methods does not affect the quality or granularity of our bug categorization, and its impact on the overall analysis remains negligible.}


\subsection{Categorization of Bugs in AI-Generated Code (RQ1)}
\begin{figure}[htbp]
\centering
\includegraphics[width=0.55\linewidth]{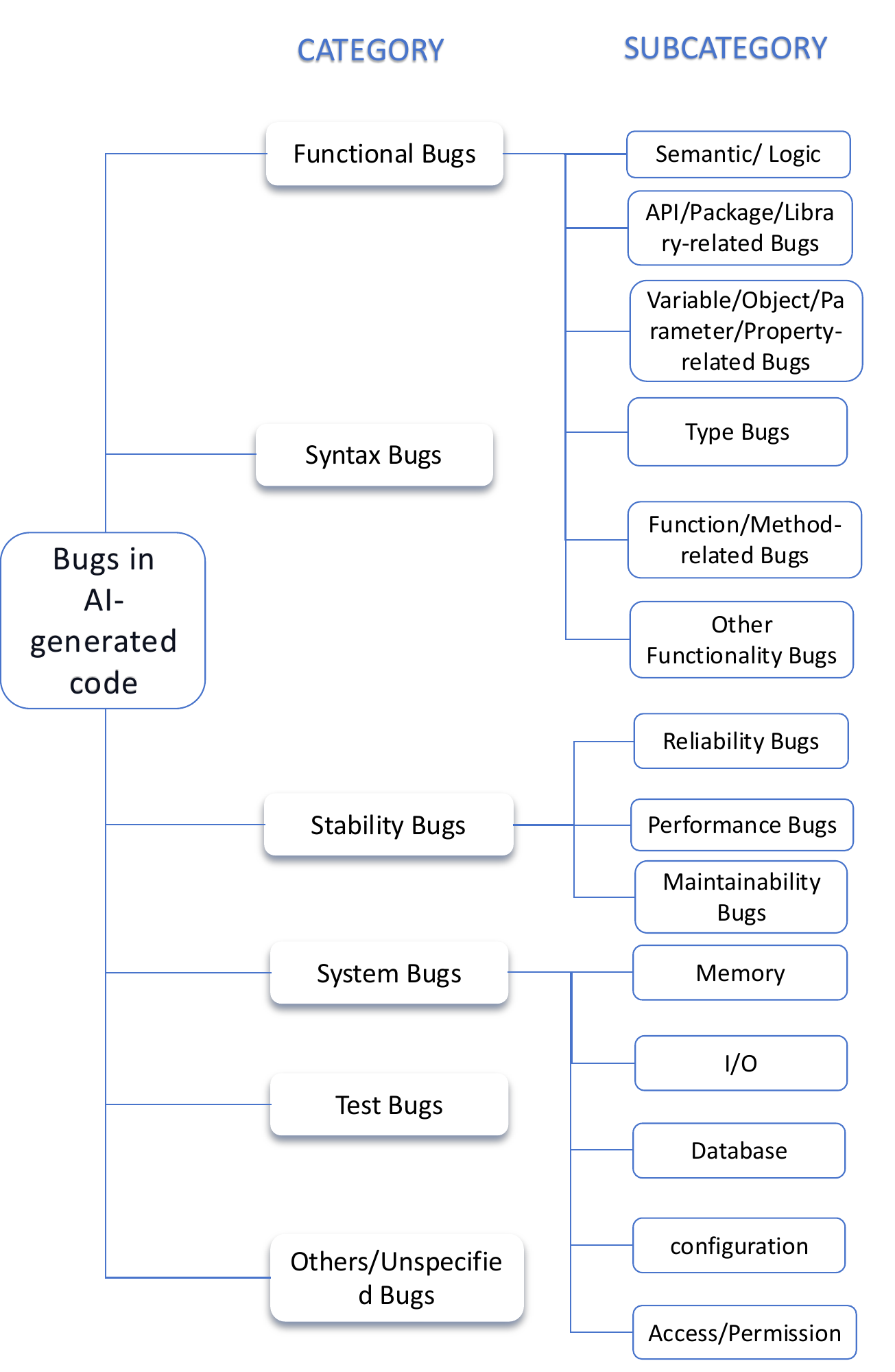}
\caption{Classification of bugs in AI-generated code}
\label{fig:taxonomy}
\end{figure}

\update{We present a comprehensive classification of the bugs identified in the studies, as shown in Figure~\ref{fig:taxonomy}. We further analyze the frequency and distribution of each bug category and discuss hallucination as a root cause of bugs in \aigc{}.}

\subsubsection{Classification of Bugs in AI-Generated Code}

\update{
As part of our analysis of bug types and patterns, we examined each reported bug based on the bug description and underlying cause. We categorized the identified bugs using established software engineering classifications (e.g., \cite{CATOLINO2019165, gyimesi2021bugsjs, tan2014bug}). We also supported our classification with insights from Tambon et al. \cite{tambon2025bugs}, which specifically examines bugs in \aigc{}. In particular, we drew on its empirical findings regarding common bug manifestations to validate and refine our classification. We map the bug patterns observed in \aigc{} to existing categories and use the Defect Type dimension of Orthogonal Defect Classification (ODC) \cite{chillarege1992orthogonal} as the basis for our classification, while extending it where necessary to capture AI-specific phenomena. Based on our analysis, we identified six major bug categories: \textit{Functional Bugs}, \textit{Stability Bugs}, \textit{Syntax Bugs}, \textit{System Bugs}, \textit{Test Bugs}, and \textit{Unspecified Bugs}. We map our bug classification to ODC defect types (Table~\ref{tab:ODCmapping}), aligning it with a well-established defect classification framework and thereby improving the interpretability and standardization of our classification. Although separating AI-specific root causes from their technical manifestations could provide a more comprehensive, multi-dimensional taxonomy, the studies included in our review rarely report AI-specific root causes explicitly. Consequently, evidence remains insufficient to reliably classify individual bugs by AI-specific root cause. Therefore, we focus on technical manifestations, which we can consistently identify and extract from the existing literature. As the body of research on \aigc{} matures and more studies explicitly report AI-specific root causes, future work could establish a dedicated Root Cause taxonomy to complement the current classification and provide a more comprehensive view of AI-generated bugs.
}
\begin{table}[h]
    \centering
    \footnotesize 
    \setlength{\tabcolsep}{4pt} 
    \renewcommand{\arraystretch}{1.05} 
    \caption{Mapping between our bug classification and ODC bug type}
    \begin{tabularx}{\textwidth}{XXX}
        \toprule
        \textbf{Bug Category} & \textbf{Bug Subcategory} & \textbf{ODC Bug Type} \\
        \midrule
        Functionality & Semantic/Logical Bugs & Algorithm\\
         & API/Package/Library-related Bugs & Interface \\
         & Variable/Object/Parameter/Property-related Bugs & Assignment \\
         & Type Bugs & Assignment \\
         & Function/Method-related Bugs & Function \\
         & Other Functionality Bugs & Function\\
         \cmidrule(lr){1-3}
         Stability & Reliability Bugs & Checking\\
          & Performance Bugs & Algorithm\\
        & Maintainability Bugs & (extension beyond ODC)\\
        \cmidrule(lr){1-3}
        Syntax Bugs &  & Build/Package\\
        \cmidrule(lr){1-3}
        System Bugs & Memory Bugs & Assignment\\
         & I/O Bugs & Interface\\
         & Database Bugs & Relationship\\
         & Hardware Bugs & Interface\\
         & Configuration Bugs & Build/Package\\
         & Access/Permission Bugs & Checking\\
         \cmidrule(lr){1-3}
         Test Bugs &  & (extension beyond ODC)\\
         \cmidrule(lr){1-3}
         Unspecified Bugs &  & (extension beyond ODC)\\
        \bottomrule
    \end{tabularx}
\label{tab:ODCmapping}
\end{table}

\textit{Functional Bugs} concern a function producing incorrect output or implementing the wrong feature or functionality, while \textit{Stability Bugs} refer to issues that affect a system's reliability, performance, and maintainability. Functional and Stability bugs (explained in detail later in this section) are the two primary bug categories. We classified \textit{Semantic and Logical bugs} as subcategories of functional bugs, since functional bugs often result from semantic or logical bugs. These bugs do not result in compilation errors but lead to incorrect functionality. The program may run as normal, but its behavior deviates from the intended behavior specified by the requirements. \update{To align our classification with the ISO/IEC 25010 quality model \cite{25010:2011} and to clearly define the scope of non-functional bugs, we group non-functional issues under \textit{Stability Bugs}. Within this category, we further introduce three subcategories: \textit{Reliability bugs}, \textit{Performance bugs}, and \textit{Maintainability bugs}. This hierarchical structure allows us to systematically organize non-functional concerns while maintaining a clear, mutually exclusive distinction from functionality-related bugs.} When the code violates the syntax rules of the programming language, the program will be interpreted, resulting in compilation errors. \textit{Syntax Bugs} always lead to failure at the compilation and interpretation stage. Given their prevalence and generality, we list them as an independent bug type. 
\textit{System bugs} affect the interaction between code and system resources, hardware, the operating system, or the network. Among these bug types, memory bugs are prevalent. We categorize \textit{Memory-related bugs} as a subcategory under \textit{System Bugs}. We also considered another category: \textit{Test Bugs}. This bug type encompasses all bugs related to the test code, regardless of the cause. In some cases, bug descriptions in the studies are not specific enough to categorize, or they are too broad, with multiple potential causes and lack clear clarification. We categorize such cases under \textit{Others/Unspecified Bugs}.
Below, we provide a detailed description of the six bug types. \\

\noindent \textbf{I. Functional Bugs:} \textit{Functional bugs} generally relate to program functionality and cause the program to fail to run as expected by the user's functional requirements or design specifications. They may result from design problems, unclear requirements, or incorrect code implementation. This type of bug indicates a problem with the program's core business logic or implementation, leading to unexpected results. 
We have further divided functional bugs into six subcategories to distinguish their specific manifestations or causes, as follows:

\begin{itemize}
    \item \textit{Semantic/Logic bugs}: 
    \textit{Semantic bugs} occur when the code is syntactically correct but does not accurately express its meaning or intent, so the program does not work as the developer expected. Listing \ref{lst:semantic} shows an example of \aigc{} containing semantic bugs. \textit{Logical bugs} refer to bugs in a program's algorithm or business logic that produce incorrect output or unexpected behavior for a given input. Even if the program executes as expected in some cases, the results still fail to meet requirements because of incorrect algorithms, conditional judgments, or data-processing bugs. Fan et al. \cite{fan2023automated} defined the problem of generating solutions that do not meet the user's intention as the misalignment problem. In our classification, we classify these as \textit{logic bugs}. Listing~\ref{lst:logic} provides an example. Many studies on code generation have reported semantic and logic bugs \cite{inproceedingsINTERVENOR, yuen2025prompting, wang2025semguard, calo2026measuring, bai2025collaboration}. 
    
    \item \textit{Function/Method-related Bugs}: Functions (or methods) are blocks of code that encapsulate specific functions. Function components typically include the function name, parameters, function body, return value, and so on. Therefore, we classify bugs involving related components as such bugs. Jesse et al. \cite{jesse2023large} investigated the presence of simple/single statement bugs (SStuBs) \cite{karampatsis2020often} in LLM-generated code. SStuBs are bugs that have single-statement fixes that match a small set of bug templates. For example, SWAP\_ARGUMENTS, DIFFERENT\_METHOD\_SAME\_ARGS, and OVERLOAD\_METHOD\_MORE\_ARGS are classified as \textit{function/method related bugs}. Song et al. \cite{song2023empirical} conducted an in-depth analysis of code-generation bugs in three representative LLMs on the HumanEval dataset. They found that a major bug type was \textit{method call bugs}. 
    
    \item \textit{API/Package/Library-related Bugs}: We group API, Library, and Package-related bugs together as all three are different manifestations of software interfaces and componentization, abstract mechanisms for organizing, encapsulating, and providing functionality. Bugs in this category relate to external libraries used in software development. Kabir et al. \cite{kabir2024stack} conducted an in-depth analysis of ChatGPT answers to programming questions on Stack Overflow and verified their correctness, revealing conceptual, factual, coding, and terminological bugs. Specifically, coding bugs include logical bugs, incorrect API/library/function usage, incomplete code, and syntax bugs. 
    
    \item \textit{Variable/Objects/Parameters/Properties-related Bugs}: We grouped bugs related to variables, objects, parameters, and properties, as they all play a fundamental role in storing and representing data in programming. Variables are the most general form of data storage, while objects encapsulate both data and behavior, providing a more comprehensive approach to data management. Properties represent internal states within objects, and parameters are a special kind of variable used to pass input data to functions. Cassano et al. \cite{cassano2023multipl} created the first large-scale multilingual code generation benchmark and evaluated the multilingual performance of three state-of-the-art code generation models. They conducted an evaluation study of bugs, presenting complete code examples generated by Codex containing various bugs, including some variable/object/parameter/properties-related bugs, such as \texttt{KeyNotFoundException}, which is typically thrown when attempting to access a non-existent key in a dictionary or collection, and \texttt{InvalidAssignment}, which is typically thrown when attempting to assign a value to a variable or object that is not allowed.
    
    \item \textit{Type Errors}: We found that many of the bugs we identified relate to types, regardless of the roles they play in the code, such as type mismatches, numeric type bugs, and function-call type bugs. These bugs occur when the program uses incompatible or incorrect data types, resulting in execution failures or unexpected results. Inala et al. \cite{inala2022fault} aim to improve LLMs' ability to generate code for various programming tasks. By parsing Python compiler error messages, they identified the 10 most common execution bug categories, including \texttt{TypeError}, where an operation or function is applied to an object of an inappropriate type.
    
    \item \textit{Other Functionality Bugs}: Some bugs we identified cannot be classified into any of the above subcategories. This is because these bugs are described in terms of requirements rather than specific details. Barbàra et al. \cite{barbara2024automatic} discussed various minor bugs in generated code in the context of smart contracts, including bugs resulting from a lack of correlation between prompts and results, incomplete functions and variable assignments, limited variations in function representation, and failures in currency understanding. Most are connected to program functionality. For example, the failure in currency understanding - where all values in the lease agreement were expressed in USD, while the smart contract represented all variables in Ether - is a clear functionality bug, as it directly affects the correctness of the contract's execution. 
\end{itemize}


In the example shown in Listing \ref{lst:semantic}, 
the task is expected to return \texttt{True} if all keys are lowercase strings or all keys are uppercase strings for a given dictionary; otherwise, it returns \texttt {False}. If the given dictionary is empty, the function should also return \texttt {False}. Line 6 will cause an \texttt{AttributeError}. The bug occurs because the \texttt{islower()} method is called on an integer, which results in the error. 

\begin{lstlisting}[ language=Python, caption={Semantic bug case in Python generated by GPT-3.5 \cite{inproceedingsINTERVENOR}}, label={lst:semantic}]
def check_dict_case(dict):
    if len(dict) == 0:
        return False
    else:
        keys = list(dict.keys())
        if keys[0].islower():
            for key in keys:
                if not key.islower():
                    return False
            return True
    else:
        for key in keys:
            if not key.isupper():
                return False
        return True
\end{lstlisting}

Listing~\ref{lst:logic} shows a logic bug example. In the example, the initialization of \texttt{dp[1]} in Line 4 is incorrect, as it does not consider whether \texttt{nums[0]} and \texttt{nums[1]} are equal. The core logic of the algorithm (lines 7 and 8) attempts to use dynamic programming to find the minimum number of operations required to transform an array (nums) into an alternating array, but its approach does not align with the problem. The program runs correctly, but the output is incorrect. Under our classification, this is classified as a logic bug.

\begin{lstlisting}[ language=Python, caption={Logic bug case in Python generated by Codex \cite{fan2023automated}}, label={lst:logic}]
public static int minimumOperations(int[] nums) {
    int n = nums.length; 
    int[] dp = new int[n]; 
    dp[0] = 0; dp[1] = 1; 
    for (int i = 2; i < n; i++){ 
        dp[i] = dp[i- 1] + 1; 
        if(nums[i] == nums[i- 2]) 
            dp[i] = Math.min(dp[i- 2] + 1, dp[i]); 
    } 
    return dp[n- 1]; 
}
\end{lstlisting}


\noindent \textbf{II. Stability Bugs:} \update{Stability bugs refer to issues that affect a software system’s ability to operate reliably and consistently over time, particularly under varying environments, workloads, or prolonged execution. These bugs may cause crashes, deadlocks, or other unpredictable behavior, ultimately degrading system availability and user experience. To better characterize their impact, stability bugs can be further divided into three subcategories: \textit{Reliability bugs} capture failures, crashes, and error-handling issues that cause software to behave incorrectly or unexpectedly under various environments and situations. \textit{Performance bugs} refer to software that fails to meet expected performance standards during execution, such as prolonged response times, excessive resource consumption, or insufficient system processing power; these are common in large systems requiring high concurrency or computationally intensive tasks. \textit{Maintainability bugs} arise when generated programs meet functional and non-functional goals but suffer from poor code readability, code smells, style inconsistency, or lack of adherence to best practices, complicating future development and collaboration.}

Zhou et al. \cite{zhou2025exploring} discussed problems practitioners encounter when using GitHub Copilot and proposed a new taxonomy of these problems. They identified a category called Suggestion Content Issues. One subcategory within this is less efficient suggestions - code that satisfies functional requirements and produces correct outputs but exhibits performance bugs due to unnecessarily complex implementations or a lack of optimization. The author provided an example where Copilot returns functionally correct code when prompted to identify ``the cup with the most water'' or ``any cup with the most water'' but these solutions do not necessarily reflect the most efficient approach to the problem. Ren et al. \cite{ren2023misuse} presented an example coding task: write a Java method to swap two elements in a vector (Listing~\ref{lst:exception}). The code involves incomplete exception-handling bugs for the two APIs, \texttt{java.util.Vector.get(int index)} and \texttt{java.util.Vector.set(int index, E element)}. When the passed index parameter is outside the valid range, an \texttt{ArrayIndexOutOfBoundsException} should be thrown; however, the generated code snippet does not perform any exception handling. Improper exception handling can lead to serious consequences, including crashes, data corruption, and security vulnerabilities.


\begin{lstlisting}[language=Java, caption={A performance-realted bug in Java generated by ChatGPT \cite{ren2023misuse}},label={lst:exception}]
public static void swap(Vector<Integer> v, int i, int j) { 
    int temp = v.get(i); 
    v.set(i, v.get(j)); 
    v.set(j, temp); 
}
\end{lstlisting}

Moratis et al. \cite{moratis2024write} studied the code quality bugs of the \aigc{} and classified those quality bugs into three categories of violations: (a) violations relevant to standard practice, such as avoiding the use of globals (Best Practices), (b) violations relevant to code readability, such as not using braces for if statements (Code Style), and (c) violations that can lead to bugs, such as using a trailing comma when declaring an array (Error Prone). Patel et al. \cite{patel2024comparative} specified some bugs including \texttt{NM\_CLASS\_NAMING\_CONVENTION}, \texttt{ClassNamingConventions}, \texttt{UnnecessaryImport}, and \texttt{ShortClassName}. In our classification, we classified all of them under the Code Style and Standards Issues category. Liu et al. \cite{liu2024refining} introduced a similar bug category called ``Code Style and Maintainability''. For example, the bug illustrated in Listing~\ref{lst:quality} (Line 1) shows that the parameter \texttt{maxSize} in function \texttt{maxFreq} is declared but never used.


\begin{lstlisting}[breaklines=true, language=Python, caption={Code quality and standard issue case in Python generated by ChatGPT \cite{liu2024refining}},label={lst:quality}]
def maxFreq(self, s: str, maxLetters: int, minSize: int, maxSize: int)->int: 
    count = defaultdict(int)
    res = 0  
    for i in range(len(s) - minSize + 1): 
        substring = s[i: i + minSize] 
        if len(set(substring)) <= maxLetters: 
            count[substring] += 1
            res = max(res, count[substring])  
    return res
\end{lstlisting}

\noindent \textbf{III. Syntax Bugs:} \textit{Syntax bugs} describe code that does not conform to the grammatical or writing rules of the programming language, making it impossible for the compiler or interpreter to understand or translate the code. Most syntax bugs are detected early in compilation (for statically typed languages like Java, C++, and Go) or interpretation (for dynamically typed languages like Python and JavaScript). The compiler or interpreter will immediately report an error, preventing the program from running. Syntax bugs usually stem from carelessness or unfamiliarity with language rules, such as spelling errors, incorrect operator use, and missing or extra punctuation.

Tian et al. \cite{tian-etal-2024-debugbench} classified Syntax Bugs into four types: 1) Syntax error, 2) Reference Error, 3) Logic Error, and 4) Multiple Error. We treated three of the minor bugs under the Reference Error category as Semantic/Logic bugs, while the remaining minor bug was classified as a Syntax Bug. We did not classify bugs under ``Multiple Error'', as this is not a bug category itself but simply a grouping of multiple bugs reported in a single snippet. We only found one case related to syntax bugs, as shown in Listing~\ref{lst:syntax}, Line 2: \texttt{``SyntaxError: '(' was never closed''}.


\begin{lstlisting}[language=Python, caption={Syntax bug case in Python generated by GPT-4 \cite{tian-etal-2024-debugbench}},label={lst:syntax}]
class Solution: 
    def dp(self,i,s,prev,k,ct,n,dct: 
        ...   
        exc=self.dp(i+1,s,prev,k-1,ct,n,dct)
        dct[(i,prev,ct,k)]=min(inc,exc) 
        return min(inc,exc)
\end{lstlisting}

\noindent \textbf{IV. System Bugs:} \textit{System bugs} involve system-level problems, including bugs related to hardware, the operating system, or the network. This includes incompatibility with the operating system interface, hardware driver problems, network delays, resource management problems, or improper configuration bugs. These problems often require cross-platform or system-level debugging and may affect the normal operation of multiple components or systems. We classified system bugs into the following subcategories:

\begin{itemize}
    \item \textit{Memory Bugs}: Memory management is a core aspect of program execution because it affects both functionality and performance. When the program accesses and manages the computer's memory abruptly, it can cause unexpected behavior, resulting in performance overhead, data corruption, and even system crashes. Wang et al. \cite{wang2025towards} discussed memory-related bugs in an in-depth analysis of code-generation bugs across six representative LLMs. Tu et al. \cite{tu2024isolating} utilized LLMs to generate effective test programs for compiler bug isolation. But they found some invalid test programs that contain invalid memory access, such as out-of-bounds reads or out-of-bounds writes. 
    
    \item \textit{I/O Bugs}: I/O-related bugs occur when a program performs input/output operations and involve problems that arise during interaction between the program and external resources (including files, networks, devices, etc.). Pan et al. \cite{PAN2025107602} found that current large-scale language models still struggle to handle all input/output (I/O) scenarios perfectly in efficient code-generation tasks. The main bug types in code-generation tasks include \texttt{EOFError}, which is generally considered an I/O-related bug. This bug occurs with files, data streams, or other I/O operations and usually means the program attempts to read data but reaches the end of the file or data stream.
    
    \item \textit{Database Bugs}: This group includes bugs that arise during the interaction between the program and the database, involving problems with database connection, query, or transaction processing. Du et al. \cite{du2024evaluating} automatically parsed the error logs generated during the interpretation and execution of incorrectly generated classes and analyzed the bug distribution across models. They found that a small number of models would encounter \texttt{sqlite4.OperationalError}, one of the exceptions that can be thrown by SQLite databases when performing operations, indicating that some operational bugs occurred during database operations.
    
    \item \textit{Hardware Bugs}: Hardware-related bugs usually involve physical component failures or design flaws. Chen and Huang \cite{chen2023forgetful} reported a preliminary exploration that empirically characterizes common bugs produced by LLMs in robot programming. They identified a bug in this category, known as Physical Error, which specifically causes the robot to perform impractical physical actions for the given task.
    
    \item \textit{Configuration Bugs}: Configuration-related bugs refer to problems caused by system configuration errors, improper settings, or configuration conflicts. Nettur et al. \cite{nettur2024cypress} identified a new bug - ``POM class file path not found'' - which occurs when the POM class needs to be referenced and the specified file path is incorrect or cannot be found.
    
    \item \textit{Access/Permission Bugs}: These refer to problems caused by a violation of access permissions or access control rules when the program attempts to access resources. Du et al. \cite{du2024evaluating} also found that incorrectly generated classes can encounter \texttt{PermissionError}, following unauthorized operation attempts.
\end{itemize}
 
 Listing~\ref{lst:memory} shows an example of a memory bug. The memory bug was caused by an infinite loop, meaning the function's recursive call lacked a clear termination condition. This can lead to infinite recursion and a stack overflow. 

\begin{lstlisting}[language=Python, caption={Memory bug case in Python generated by CodeGen-16b \cite{wang2025towards}},label={lst:memory}] 
def make_a_pile(n): 
    if n % 2 == 0: 
        return [n] + make_a_pile(n+2) 
    else:  
        return [n] + make_a_pile(n+1)

\end{lstlisting}



In the example shown in Listing~\ref{lst:memory2}, \texttt{memem\_access} represent undefined behavior. 
In line 10, the address of the 2-byte variable \texttt{b} (\texttt{short int b = 1;} in line 2) is incorrectly cast to an \texttt{int*} via \texttt{int *c = \&b;}. This causes the program to treat a short as a 4-byte integer, leading to strict aliasing violations, potential misaligned memory access, and out-of-bounds writes when \texttt{*c} is assigned on line 12. Together, these operations cause severe memory corruption.

\begin{lstlisting}[language=C, caption={Memory bug case in C generated by GPT-3.5 \cite{tu2024isolating}},label={lst:memory2}] 
int a; 
short int b=1; //int b=1; 
int main() { 
    int i; 
    for (i =0; i <56; i++){ 
        for (; a; a--){ 
            ; 
        } 
    } 
    int *c=&b; 
    if(*c){ 
        *c=1%(unsigned int)*c|5; 
    } 
    printf ("%d\n",b); 
    return 0; 
}
\end{lstlisting}

\noindent \textbf{V. Test Bugs:} are bugs or problems found in the test code rather than in the production code (i.e., the code under test). This class of bugs typically occurs in test scripts or test files, hindering testing effectiveness and leading to false positives and possible flakiness \cite{TAHIR2023111837}. It may cause test cases to fail to execute correctly, prevent accurate verification of software functionality, or even produce incorrect test results. In our classification, we categorized all bugs found in test code as test bugs, with assertion-related bugs as a typical example.

Wang et al. \cite{inproceedingsINTERVENOR} discussed the occurrence of \texttt{AssertionError}, which usually stems from semantic or logic bugs. Assertion-related faults are the dominant root cause of silent horror bugs, which occur when tests pass while the production code is incorrect or may be incorrect \cite{vahabzadeh2015empirical}. 
Siddiq et al. \cite{siddiq2024using} explored test smells in AI-generated unit tests, which means that although this test is correct, there is no explanation for the expected outputs passed to the assertions, which is a case of the Magic Number Test smell defined in this study. This can confuse other developers reading the test and does not help them understand the meaning of these inputs or the expected results. The study demonstrated the following Java example (Listing~\ref{lst:test}), which includes a unit test for a method in the LargestDivisor class. It checks whether the method under test returns the largest divisor of a number. Although this test is correct, it does not explain the expected output passed to the assertion. 

\begin{lstlisting}[language=Java, caption={Test bug case in Java  \cite{siddiq2024using}},label={lst:test}]
public class LargestDivisorTest { 
    @Test 
    void testLargestDivisor() {
        assertEquals(5, LargestDivisor.largestDivisor(15)); 
        assertEquals(1, LargestDivisor.largestDivisor(3)); 
    } 
}
\end{lstlisting}




Yuan et al. \cite{yuan2024evaluating} conducted an empirical study evaluating ChatGPT's unit test generation capabilities. They found that ChatGPT-generated tests still exhibited correctness issues, including compilation errors and execution failures. For example, Listing~\ref{lst:test2} shows a ChatGPT-generated test failure. The test threw a \texttt{NullpointerException} when executing Line 3. This bug occurred because the created object ``url" accessed a non-existent external resource \texttt{/test.jar} (in Line 2). This effectively exposed an inherent flaw in ChatGPT: its inability to perceive external resources during test generation.
\UseRawInputEncoding
\begin{lstlisting}[language=Java, caption={Test bug case in Java  generated by ChatGPT\cite{yuan2024evaluating}},label={lst:test2}]
public void testGetManifestFromJarURLConnection() throws IOException {
    URL url = getClass().getResource("/test.jar");
    //java.lang.NullpointerExeception
    JarURLConnection connection = (JarURLConnection)url.openConnection();
}
\end{lstlisting}


\noindent \textbf{VI. Others/Unspecified Bugs:} We classified all other bug types that we could not group in a single category from the previous list as \textit{Others/Unspecified Bugs} - i.e., the bug category is provided, but it cannot be mapped to our classification and lacks detailed descriptions.
For example, Cassano et al. \cite{cassano2023multipl} identified a bug referred to as \texttt{errorByGeneratedProgram} in the Racket programming language. However, due to the lack of explanation regarding the nature of the bug in the generated program, we could not assign it to a specific category and therefore classified it as an unspecified bug. Zhou et al. \cite{zhou2025exploring} categorized bugs related to the content of generated code into seven situations. One of these, \texttt{SUGGESTION WITH BUGS}, is described as cases where GitHub Copilot generates contextually relevant code that nonetheless contains bugs. However, because this category provides no specific bug types, we classify them as unspecified bugs.

\subsubsection{In-depth Analysis of Bug Types}
\label{sec:bugtypeanalysis}

\begin{figure}[htbp]
\centering
\includegraphics[width=0.7\linewidth]{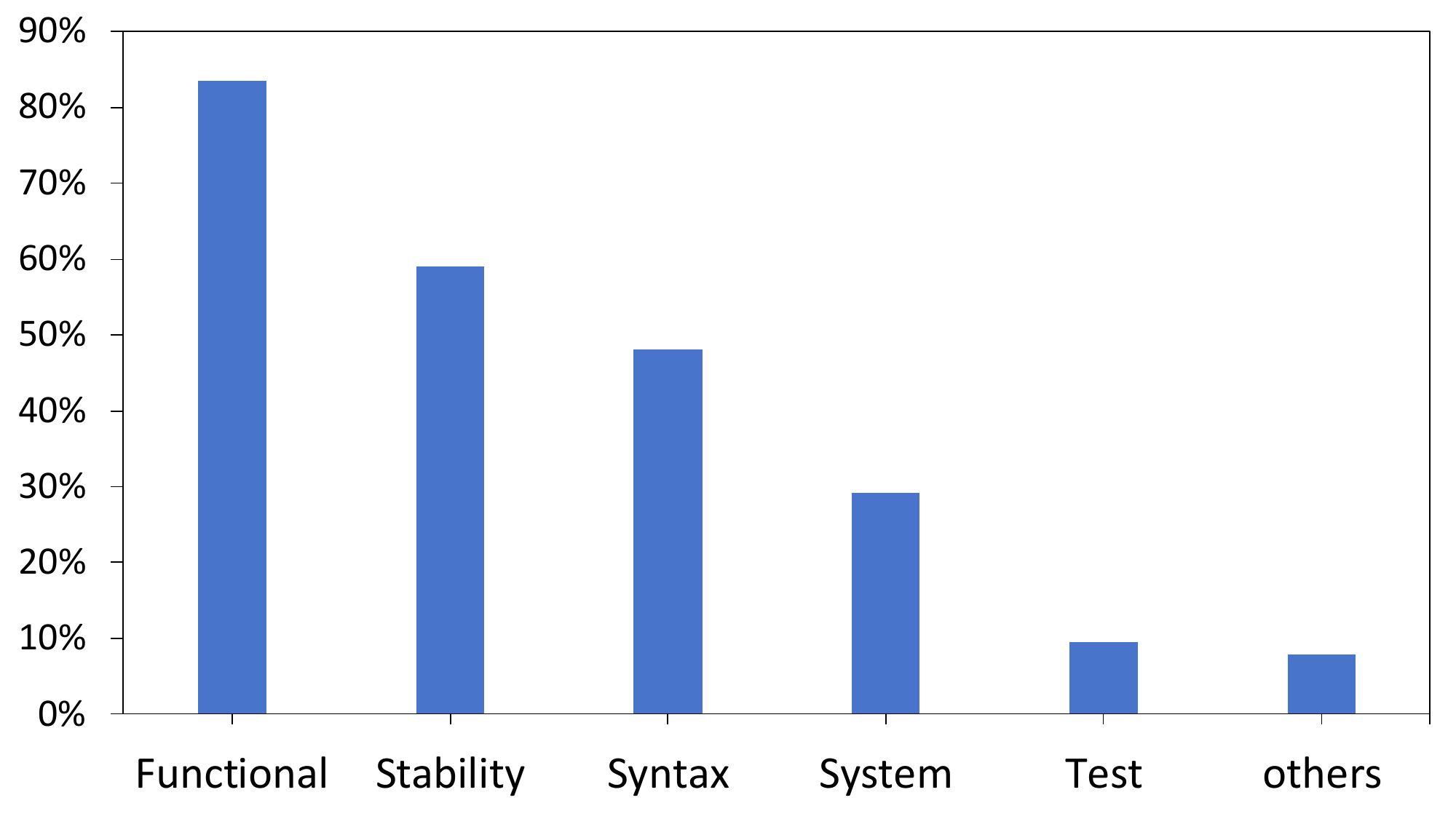}
\caption{The frequency of bug categories across studies}
\label{fig:frequencybugtype}
\end{figure}

\update{Figure~\ref{fig:frequencybugtype} shows the frequency of bug types reported in the analyzed studies. As shown in Table~\ref{tab:bugsinstudies}, Functional Bugs are the most commonly investigated bug category, with 83\% of the included studies reporting this bug type. Stability Bugs appear in about 59\% of studies. Although 48\% of studies report Syntax Bugs, most identify only the presence of syntax-related issues without describing specific bug types in detail. System Bugs appear in fewer than 30\% of studies, while Test Bugs appear in only 9\%. The Others category includes bugs reported in a few studies (7\%) that lack sufficient details or descriptions about bugs in \aigc{}.}

\begin{table}[h]
    \centering
    \scriptsize 
    \caption{Distribution of bug types across the selected studies}
    \begin{tabularx}{\textwidth}{p{2.5cm}p{1.2cm}X} 
        \toprule
        \textbf{Category} & \textbf{Studies} & \textbf{References} \\
        \midrule
        Functional Bugs & 106 & 
        \cite{tian-etal-2024-debugbench, barbara2024automatic, inproceedingsINTERVENOR, zhong2024can, zhou2025exploring, baralla2024assessing, fan2023automated, PAN2025107602, siddiq2024quality, khan2024assessing, jesse2023large, haindl2024does, song2023empirical, corso2024generating, tang2024developer, spinellis2024pair, wang2025towards, karanjai2023smarter, yu2024fight, tambon2025bugs, patel2024comparative, liu2024no, liu2024refining, rabbi2024ai, pan2024lost, dong2024self, nguyen2022empirical, yetistiren2022assessing, sagodi2024methodology, zheng2023codegeex, mastropaolo2023robustness, billah2024large, cipriano2024llms, ramos2024batfix, siddiq2022empirical, bai2025collaboration, inala2022fault, cassano2023multipl, zhang-etal-2023-self, mathews2024test, ouyang2025empirical, chen2023forgetful, ho2024predicting, weisz2022better, kou2024large, kabir2024stack, fan2024oracle, kouemo2024chain, almanasra2025analysis, omidvar2024evaluating, wu2023rustgen, du2024evaluating, tang2024chatgpt, feng2023investigating, jain2022jigsaw, zhang2024pair,liu2023your, 10.1109/ASE63991.2025.00195, 10.1145/3759425.3763383, yu2025realisticcodebench, midolo2025refactoring, ye2025solcontracteval, zhu2026assessing,liu2026beyond,wang2026specifications,huang2026hallucination,ma2026large,zhuo2026identifying,wang2026making,fayed2026prompt,shehab2026scalable,swaraj2026trust,eagal2025analyzing,liang2024can,bayram2025comparative,liu2025evaluating,ron2026galapagos,zhao2025hfuzzer,nekrasov2026iac,feischl2026large,bui2025correctness,pendyala2025performance,bui2025correctness,yang2025stop,nogueira2025beyond,zhang2026mpc,afrin2025quantization,song2025code,honarvar2025question,olivieri2025code,wang2025context,nouri2025large,almanasra2026comparative,jahan2025does,khojah2025impact,spracklen2025we,cotroneo2025human,abbassi2025taxonomy,yuen2025prompting,wang2025semguard,kabir2025well,dolcetti2026helping,dou2026wrong,calo2026measuring,bai2025collaboration,sun2026quality} \\
        \cmidrule(lr){1-3}
        Stability Bugs & 75  & \cite{ren2023misuse,zhou2025exploring,fan2023automated,PAN2025107602,siddiq2024quality,khan2024assessing,haindl2024does,corso2024generating,tang2024developer,spinellis2024pair,karanjai2023smarter,patel2024comparative,liu2024refining,rabbi2024ai,dong2024self,nguyen2022empirical,sagodi2024methodology,billah2024large,cipriano2024llms,siddiq2022empirical,zhang2024copilot,siddiq2024franc,inala2022fault,weisz2022better,wu2023rustgen,feng2023investigating,cassano2023multipl,zhang-etal-2023-self,mathews2024test,ho2024predicting,almanasra2025analysis,du2024evaluating,tang2024chatgpt,zhang2024pair,depalma2024exploring,10.1109/ASE63991.2025.00195,10.1145/3759425.3763383,midolo2025refactoring,he2026speed,liu2026beyond,wang2026specifications,huang2026hallucination,ma2026large,zhuo2026identifying,wang2026making,fayed2026prompt,shehab2026scalable,swaraj2026trust,eagal2025analyzing,liang2024can,bayram2025comparative,ron2026galapagos,feischl2026large,bui2025correctness,11025871,pendyala2025performance,bui2025correctness,zhang2026mpc,borg2026echoes,afrin2025quantization,song2025code,honarvar2025question,olivieri2025code,wang2025context,nouri2025large,almanasra2026comparative,li2026performance,cotroneo2025human,abbassi2025taxonomy,yuen2025prompting,kabir2025well,dolcetti2026helping,dou2026wrong,bai2025collaboration,sun2026quality}  \\ \cmidrule(lr){1-3}
        Syntax Bugs & 61  & \cite{tian-etal-2024-debugbench,inproceedingsINTERVENOR,zhou2025exploring,fan2023automated,PAN2025107602,song2023empirical,corso2024generating,spinellis2024pair,wang2025towards,tambon2025bugs,liu2024no,liu2024refining,rabbi2024ai,hendrycks2021measuring,pan2024lost,dong2024self,sagodi2024methodology,zheng2023codegeex,mastropaolo2023robustness,cipriano2024llms,ramos2024batfix,siddiq2022empirical,fakih2024llm4plc,siddiq2024franc,inala2022fault,cassano2023multipl,zhang-etal-2023-self,chen2023forgetful,weisz2022better,kabir2024stack,feng2023investigating,zhang2024pair,10.1145/3759425.3763383,midolo2025refactoring,ye2025solcontracteval,wang2026specifications,ma2026large,wang2026making,fayed2026prompt,shehab2026scalable,swaraj2026trust,liang2024can,bayram2025comparative,liu2025evaluating,ron2026galapagos,nekrasov2026iac,feischl2026large,11025871,pendyala2025performance,nogueira2025beyond,afrin2025quantization,song2025code,honarvar2025question,wang2025context,nouri2025large,almanasra2026comparative,abbassi2025taxonomy,wang2025semguard,dolcetti2026helping,dou2026wrong,bai2025collaboration}  \\ \cmidrule(lr){1-3}
        System Bugs & 37  & \cite{inproceedingsINTERVENOR,fan2023automated,PAN2025107602,song2023empirical,wang2025towards,liu2024no,liu2024refining,yetistiren2022assessing,zheng2023codegeex,billah2024large,inala2022fault,cassano2023multipl,zhang-etal-2023-self,mathews2024test,nettur2024cypress,ho2024predicting,almanasra2025analysis,du2024evaluating,zhang2024pair, 10.1109/ASE63991.2025.00195,wang2026specifications,nekrasov2026iac,pendyala2025performance,yang2025stop,nogueira2025beyond,afrin2025quantization,honarvar2025question,olivieri2025code,nouri2025large,almanasra2026comparative,cotroneo2025human,abbassi2025taxonomy,yuen2025prompting,kabir2025well,dou2026wrong}  \\ \cmidrule(lr){1-3}
        Test Bugs & 12  & \cite{inproceedingsINTERVENOR,tu2024isolating,karanjai2023smarter,dakhel2024effective,zilberman2024no,nettur2024cypress,siddiq2024using,yuan2024evaluating,yang2024evaluation,10.1109/ASE63991.2025.00223,10.1145/3756681.3756964,zhang2026automated}  \\ \cmidrule(lr){1-3}
        Others/Unspecified Bugs & 10 &
        \cite{zhou2025exploring,tang2024developer,tambon2025bugs,pan2024lost,inala2022fault,cassano2023multipl,weisz2022better,liu2026beyond,swaraj2026trust,nouri2025large} \\
        \bottomrule
    \end{tabularx}
\label{tab:bugsinstudies}
\end{table}

\begin{figure}[h]
    \centering

    \subfigure[Functional bug type frequency]{
        \includegraphics[
            width=0.7\textwidth,
            height=4cm,
            keepaspectratio
        ]{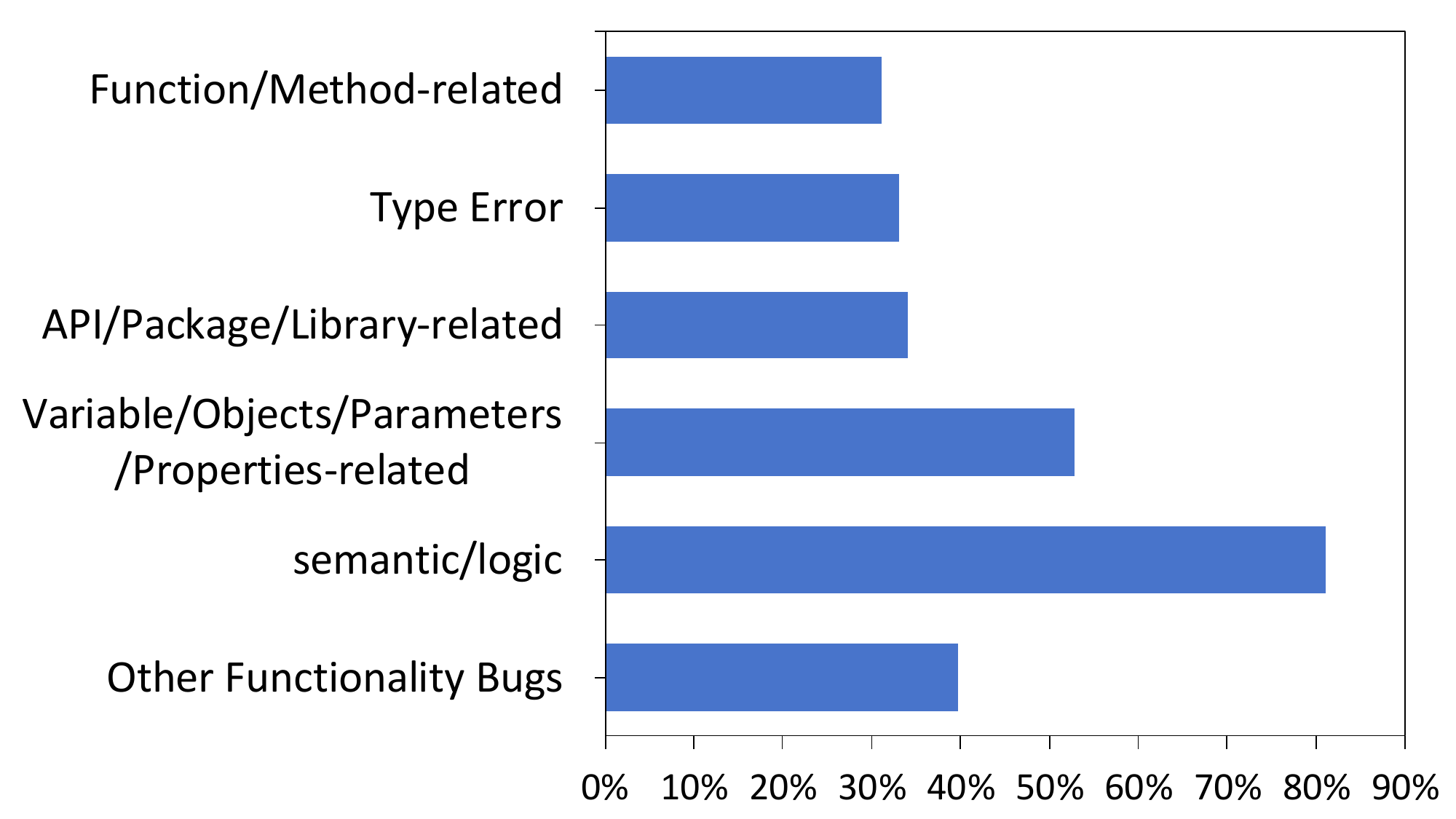}
        \label{fig:functionalbugtype}
    }

    \vspace{0.5em}

    \subfigure[Stability bug type frequency]{
        \includegraphics[
            width=0.45\textwidth,
            height=5cm,
            keepaspectratio
        ]{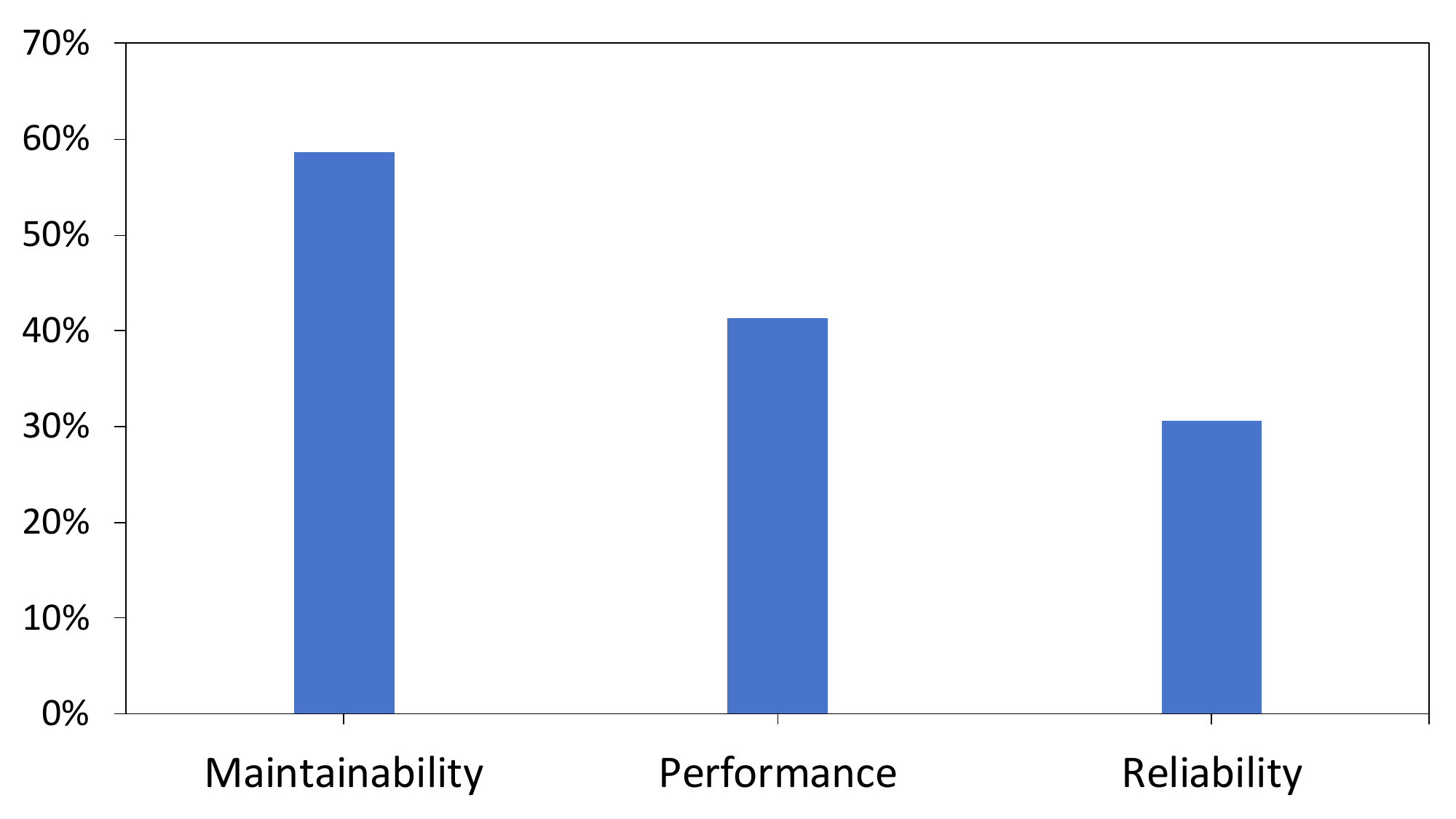}
        \label{fig:stabilitybugtype}
    }
    \hfill
    \subfigure[System bug type frequency]{
        \includegraphics[
            width=0.45\textwidth,
            height=5cm,
            keepaspectratio
        ]{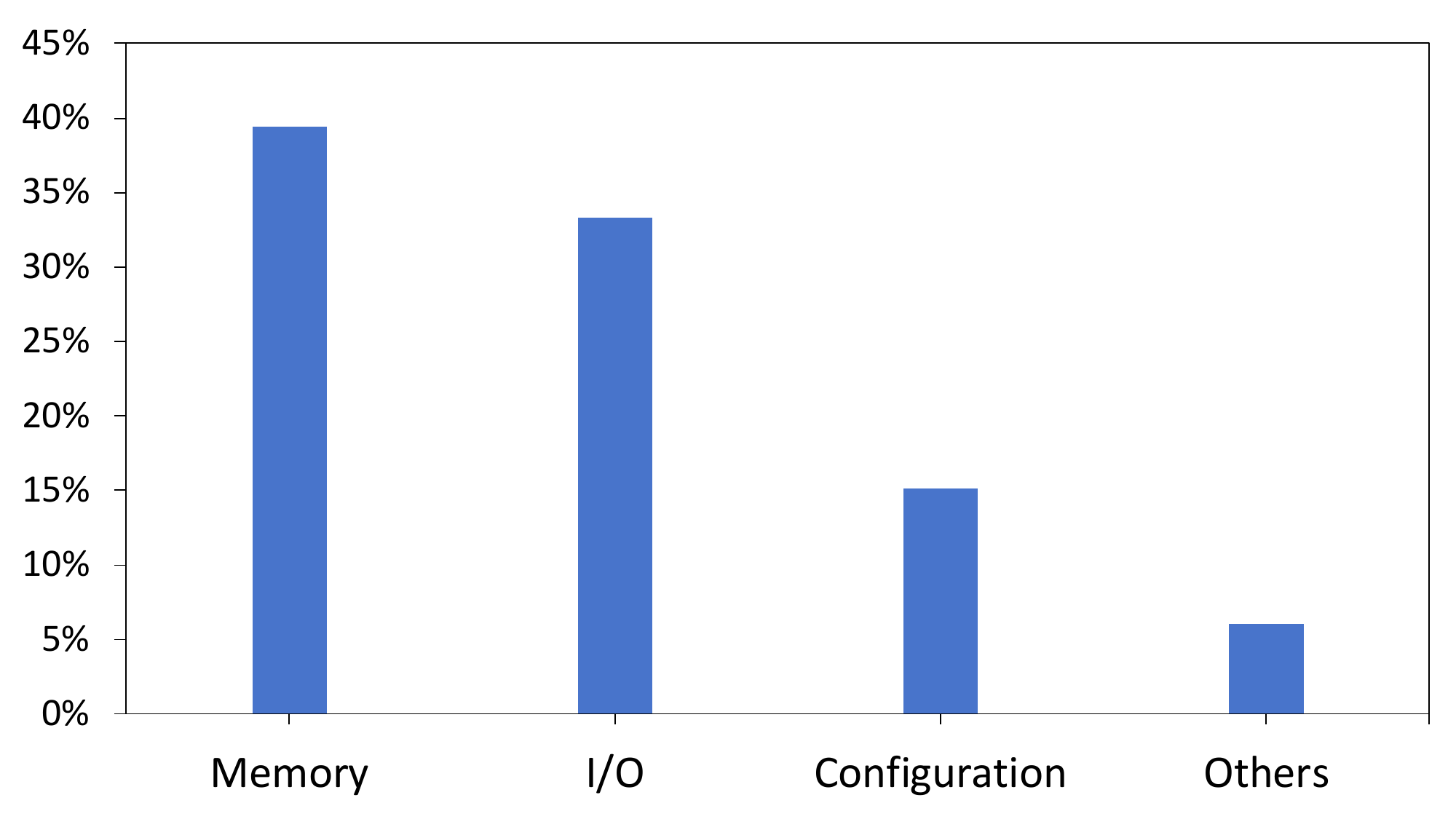}
        \label{fig:systembugtype}
    }

    \caption{Frequency of bug subcategories}
    \label{fig:functionalandsystem}
\end{figure}

\update{Functional, Stability, and System Bugs each contain multiple subcategories, whose distributions within their respective main categories were further analyzed. Figure~\ref{fig:functionalbugtype} shows the distribution of the six Functional Bug subcategories. Regarding the Functional Bug subcategories, Semantic/Logical Bugs (81\%) and Variable/Object/Parameter/Property-related Bugs (53\%) are the two most frequently reported subcategories. Figure~\ref{fig:stabilitybugtype} presents the distribution of the three Stability Bug subcategories. Among the Stability Bug subcategories, Maintainability Bugs are the most commonly reported type, as they include issues related to code quality and coding standards. Figure~\ref{fig:systembugtype} shows the distribution of four System Bug subcategories, with Memory Bugs being the most frequently reported. Other System Bug subcategories, including Access/Permission, Database, and Hardware Bugs, were reported in only one or two studies and therefore represent relatively minor components of the overall distribution. These findings suggest that the primary limitations of models used in the surveyed studies lie in understanding programming task requirements and reasoning about program logic, while code quality and runtime resource management represent additional areas of concern.}


\update{Our analysis shows that most bug categories in \aigc{} can also occur in manually written code. However, \textbf{hallucination} represents a phenomenon that is particularly relevant to \aigc{}. Hallucination is defined as generated content that contradicts or cannot be verified against its source \cite{filippova2020controlled}. More broadly, in AI and NLP research, hallucination refers to generating inaccurate, fabricated, or unverifiable information that may still appear plausible \cite{ji2023survey}. In code generation, hallucinations may manifest as outputs that are syntactically valid and stylistically appropriate but are inconsistent with real-world APIs, libraries, program context, or intended semantics \cite{zhang2025llm, liu2026beyond}. For example, a model may ``invent'' non-existent functions, objects, libraries, or behaviors, which can cause the generated code to fail. 

Because hallucination is broadly defined and our study focuses specifically on \aigc{}, its manifestations may overlap with other bug categories. For example, several studies have reported package hallucinations and references to undefined variables, objects, or methods \cite{3766078.3766268, siddiq2022empirical, tian-etal-2024-debugbench, cassano2023multipl, dou2025what}. Such bugs would traditionally be classified as functional bugs, but in the context of \aigc{}, they may also reflect hallucinated entities or behaviors. Hallucination may be relevant to a broad range of bug categories, although the nature and extent of this relationship vary across categories. Importantly, only a small proportion of the surveyed studies explicitly identified hallucination as the cause of the reported bugs. This suggests that the role of hallucination in introducing bugs into \aigc{} remains largely underexplored in the studies included in our review. Further research is therefore needed to systematically investigate when and how hallucinations contribute to different types of bugs in \aigc{}.}


\begin{tcolorbox}[title=RQ1 Key Findings]
\update{We classified all the bugs that occurred in \aigc{} based on prior bug classification studies in software engineering, comprising six major categories and fourteen subcategories. \textit{Functional}, \textit{Stability}, and \textit{Syntax} Bugs are the most frequently reported categories in the reviewed studies, whereas research on \textit{System} or \textit{Test}-related Bugs remains limited. Some existing studies treat \textit{Hallucination} as a distinct bug type. In this study, we treat it as a root cause that may lead to other bug types.}
\end{tcolorbox}

\subsection{Bug-Proneness of AI Coding Systems (RQ2)}
\update{In this section, we discuss the relationship between the \cgm{} and the generated bugs. Our goal is to investigate how the LLMs, coding tools, or coding agents used in the surveyed studies contribute to the \aigc{} and what common characteristics they share.}

\subsubsection{Overview of the Prominent LLM Families}
\begin{figure}[htbp]
\centering
\includegraphics[width=0.7\linewidth]{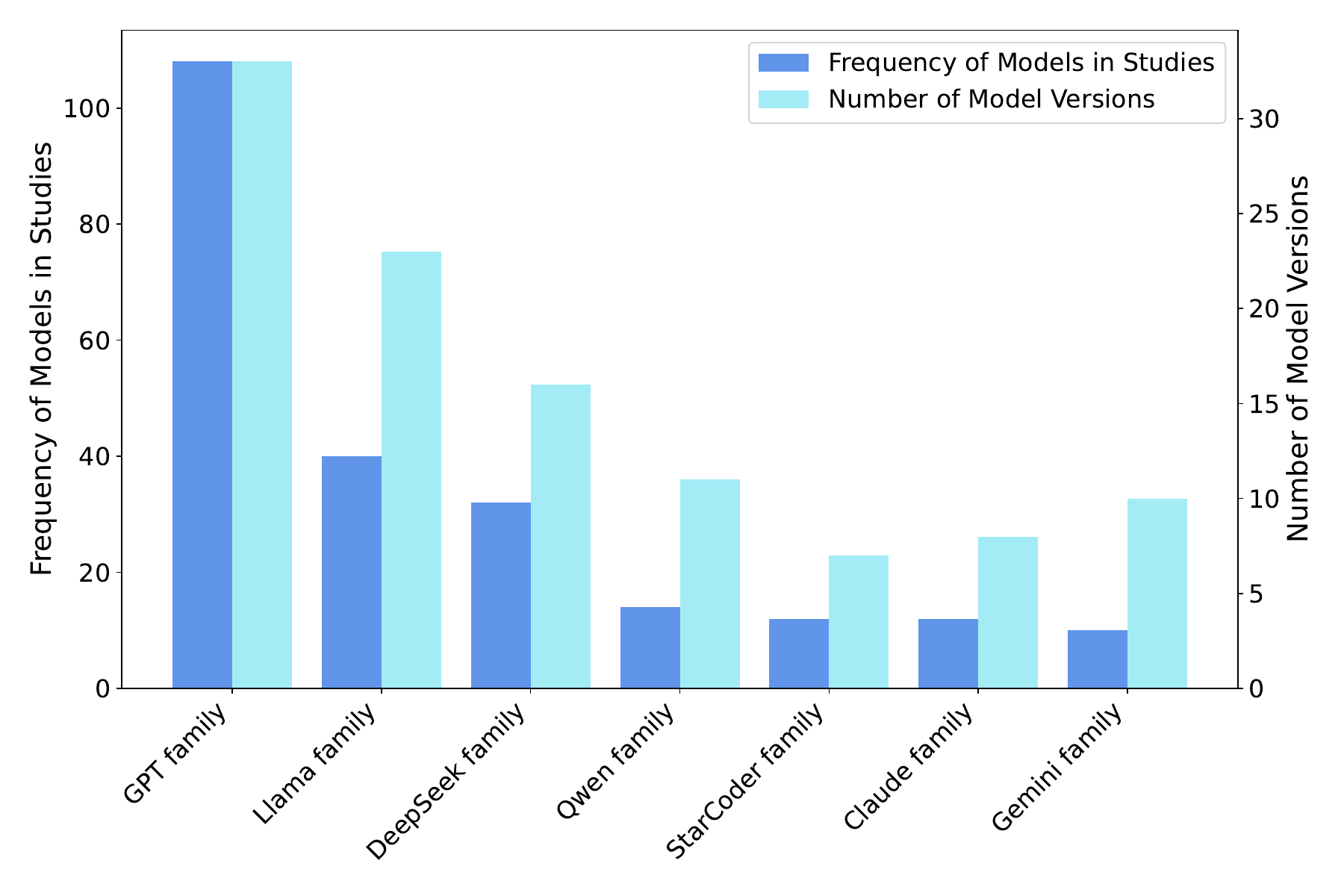}
\caption{Frequency and versions of models for code generation}
\label{fig:models}
\end{figure}

\update{In our analysis, we examine the prominent LLM families represented in the surveyed studies in the code generation context. Most models are based on decoder-only architectures. Consequently, grouping models by architectural type would add little differentiation to our dataset, since most models fall into the same architectural category. We therefore categorized models by name, developing organization, and identified series. In total, we identified 43 distinct model families, including the well-known GPT \cite{ouyang2022training} and Llama \cite{rozière2024codellamaopenfoundation} families, as well as several AI coding tools/assistants such as GitHub Copilot and Cursor. Figure~\ref{fig:models} shows the frequency with which different model families appear across the selected studies, along with the number of distinct versions each family includes (only families appearing in more than 10 studies are shown).
Our results show that the GPT series was examined in 108 studies (85\%), covering 33 versions and reflecting its prominence in AI code generation research. The Llama family appeared in 40 studies (31\%), covering 23 versions; in addition to the general-purpose Llama series, the fine-tuned variant Code Llama is also widely used for programming tasks. The DeepSeek family \cite{guo2024deepseekcoderlargelanguagemodel} was reported in 32 studies (25\%), covering 16 versions, while the Qwen family \cite{bai2023qwen} appeared in 14 studies, covering 11 versions. Fewer than 10\% of studies investigated the StarCoder \cite{li2023starcoder}, Claude \cite{anthropic2024claude}, and Gemini \cite{team2023gemini} families, comprising 7, 8, and 10 versions, respectively. Several other models were mentioned less frequently but remain noteworthy, including InCoder \cite{fried2022incoder}, Mistral \cite{jiang2023mistral}, PolyCoder \cite{xu2022systematic}, CodeGen \cite{xia2024aicodereval}, TransCoder \cite{lachaux2020unsupervised}, GLM \cite{du2021glm}, Vicuna \cite{chiang2023vicuna}, SantaCoder \cite{allal2023santacoder}, PanGu-Coder \cite{christopoulou2022pangu}, and CodeGeeX \cite{zheng2023codegeex}.}

\update{The \textit{GPT-family} comprises a series of models developed by OpenAI, ranging from earlier versions such as GPT-2 and GPT-3 to more advanced models including GPT-3.5, GPT-3.5 Turbo, and the GPT-4 series (e.g., GPT-4, GPT-4 Turbo and GPT-4o-mini) and o-series / reasoning family  (e.g., o3-mini, o4-mini, and GPT-oss-120B). In addition, several open-source models developed by EleutherAI, such as GPT-J and GPT-Neo (125M and 1.3B), along with their fine-tuned variants, are designed to approximate GPT -3's capabilities, although they are not part of the official GPT family. In addition, Codex extends the GPT architecture to coding, including models such as code-davinci-002 and Codex 5.1. 

The \textit{Llama-family} of models, developed by Meta, is a series of open-source transformer models designed to provide an efficient and accessible alternative to proprietary large-scale language models. 
Llama 3, Llama 3-8B, 70B, Llama 3.1-8B, 70B, 405B, and Llama 3.3-70B show incremental enhancements in performance, scalability, and multilingual capabilities. Building upon this foundation, the Code Llama family, designed specifically for coding tasks, includes Code Llama-7B, CodeLlama-13B, and CodeLlama-Python. These models are fine-tuned on code datasets to optimize code comprehension and generation. Furthermore, Phind-CodeLlama-34B-v2 (PD-34B) is a further fine-tuned version of this family designed to improve real-world code generation performance.

The \textit{DeepSeek-family} of models, developed by DeepSeek, encompasses a diverse range of models, including general-purpose models, such as DeepSeek-V2-Lite, DeepSeek-V2.5, and DeepSeek-V3, designed for broad natural language understanding and generation tasks. Reasoning-oriented models, such as DeepSeek-R1 and DeepSeek-Reasoner, focus on complex, multi-step reasoning and problem-solving tasks. Code-specialized models, such as DeepSeek-Coder-1.3B and DeepSeek-Coder-V2, are specifically designed or adapted for programming-related tasks, including code generation, understanding, and transformation. 

The \textit{Qwen-family} of models, developed by Alibaba, includes models such as CodeQwen 1.5 and QwenCoder-3B, with relatively small parameter sizes suited for lightweight deployment and resource-constrained code generation scenarios. 

The \textit{StarCoder-family} was developed by the BigCode Project, an open-source community initiative jointly launched by Hugging Face and ServiceNow. It is a family of open-source LLMs designed for code generation tasks. 
The first-generation models include StarCoder and the smaller-scale StarCoder-7B. This was followed by the second-generation StarCoder2 models, which improved training data quality, context length, and code comprehension. 

The \textit{Claude-family} of models, developed by Anthropic, includes earlier versions such as the Claude 3 series, Claude 3.5 Sonnet, Claude 3.5 Haiku, and Claude 3.7 Sonnet. 
Claude models adopt a consistent naming scheme. Haiku emphasizes a lightweight design and fast responses, making it suitable for low-latency and high-concurrency scenarios. Sonnet strikes a balance between performance and cost, serving as the primary model for high-frequency use. Opus represents the flagship tier with the most comprehensive capabilities, particularly excelling at complex reasoning and long-horizon task execution.

The \textit{Gemini-family} of models, developed by Google, include the early releases models such as Gemini 1.0 Pro and its later iterations, Gemini-Pro and Gemini-Flash, which established the family's dual focus on general-purpose reasoning (Pro) and low-latency, cost-efficient inference (Flash). 

All these model families are built upon the transformer architecture, which remains the core foundation of modern LLMs. In terms of scale evolution, early models were characterized by relatively few parameters, with training objectives narrowly focused on specific programming tasks. Later code generation models gradually evolved into large-scale general-purpose models and large-scale code-specific models. The surveyed studies cover a broad spectrum of model sizes, ranging from lightweight models with fewer than one billion parameters to very large models with hundreds of billions of parameters. Most studies focus on mainstream model families such as GPT, Llama, DeepSeek, and Qwen, which in turn span a correspondingly wide range of parameter scales. Beyond scale, each model family also reflects a distinct application focus. The Claude and Gemini families focus primarily on general-purpose tasks, whereas the GPT, Llama, DeepSeek, and Qwen families support both general and programming tasks, with Codex, CodeLlama, DeepSeek-Coder, and Qwen-Coder specifically optimized for code generation. StarCoder, by contrast, is the only model family exclusively focused on research into code generation.
In terms of accessibility, these families also diverge in their release strategies. GPT, Claude, and Gemini are primarily proprietary, API-accessible model families, whereas Llama and StarCoder provide open weights for local deployment. DeepSeek and Qwen occupy an intermediate position, offering both openly available model weights and API-based access, thereby combining the flexibility of open-source deployment with the convenience of hosted services.}

\subsubsection{Bug Distribution across Application-focused Categories.}  

\update{To better understand how the characteristics of AI coding systems influence the types of bugs they generate, we categorize the systems examined in the reviewed studies into three groups: general-purpose LLMs, code-specific LLMs, and AI coding tools and agents. For the two LLM categories, we further distinguish between proprietary and open-weight models.}\\



\begin{figure}[htbp]
    \centering

    \subfigure[General-Purpose LLMs]{
        \includegraphics[width=0.6\textwidth]{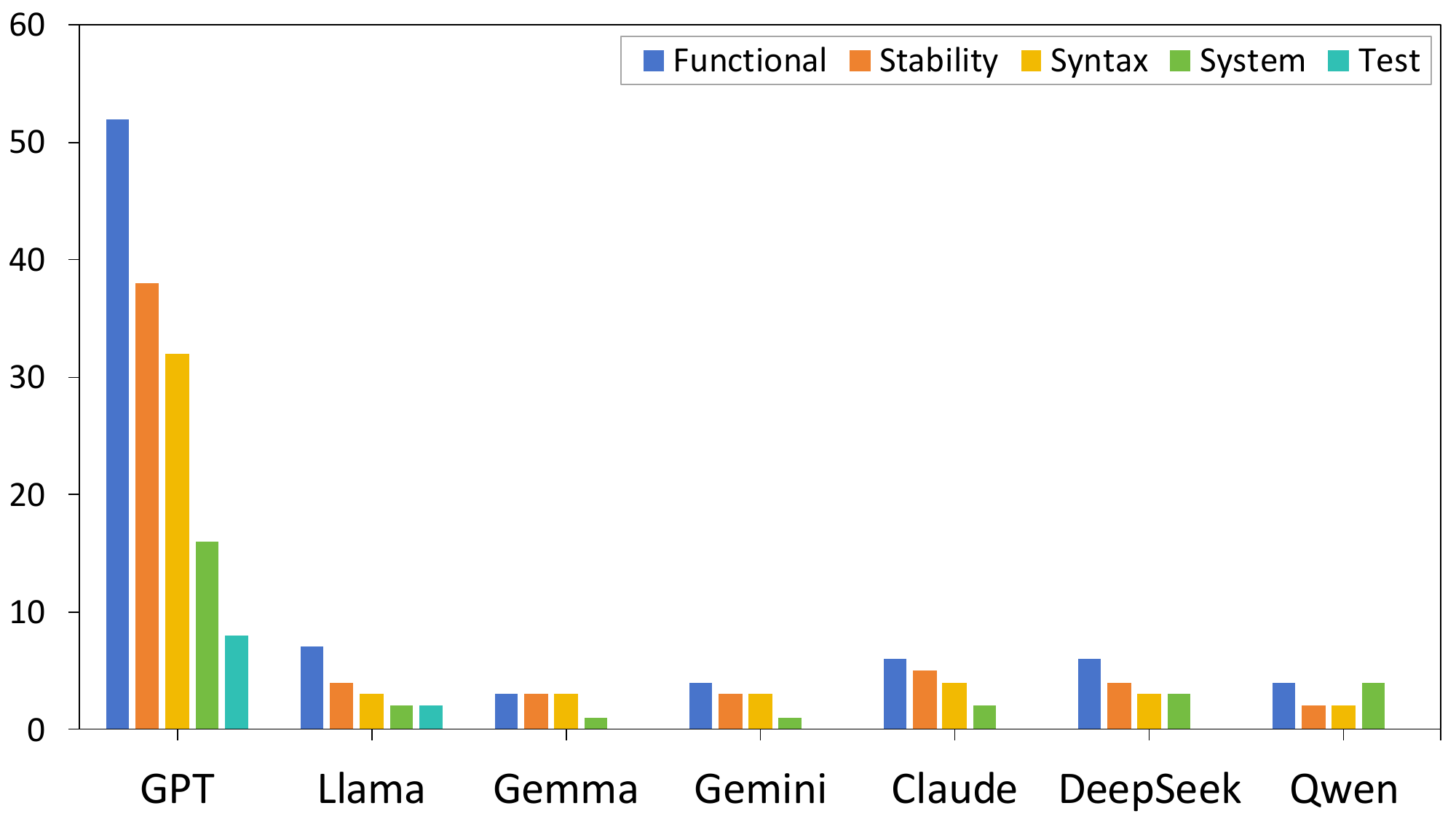}
        \label{fig:generalllm}
    }


    \subfigure[Code-Specific LLMs]{
        \includegraphics[width=0.6\textwidth]{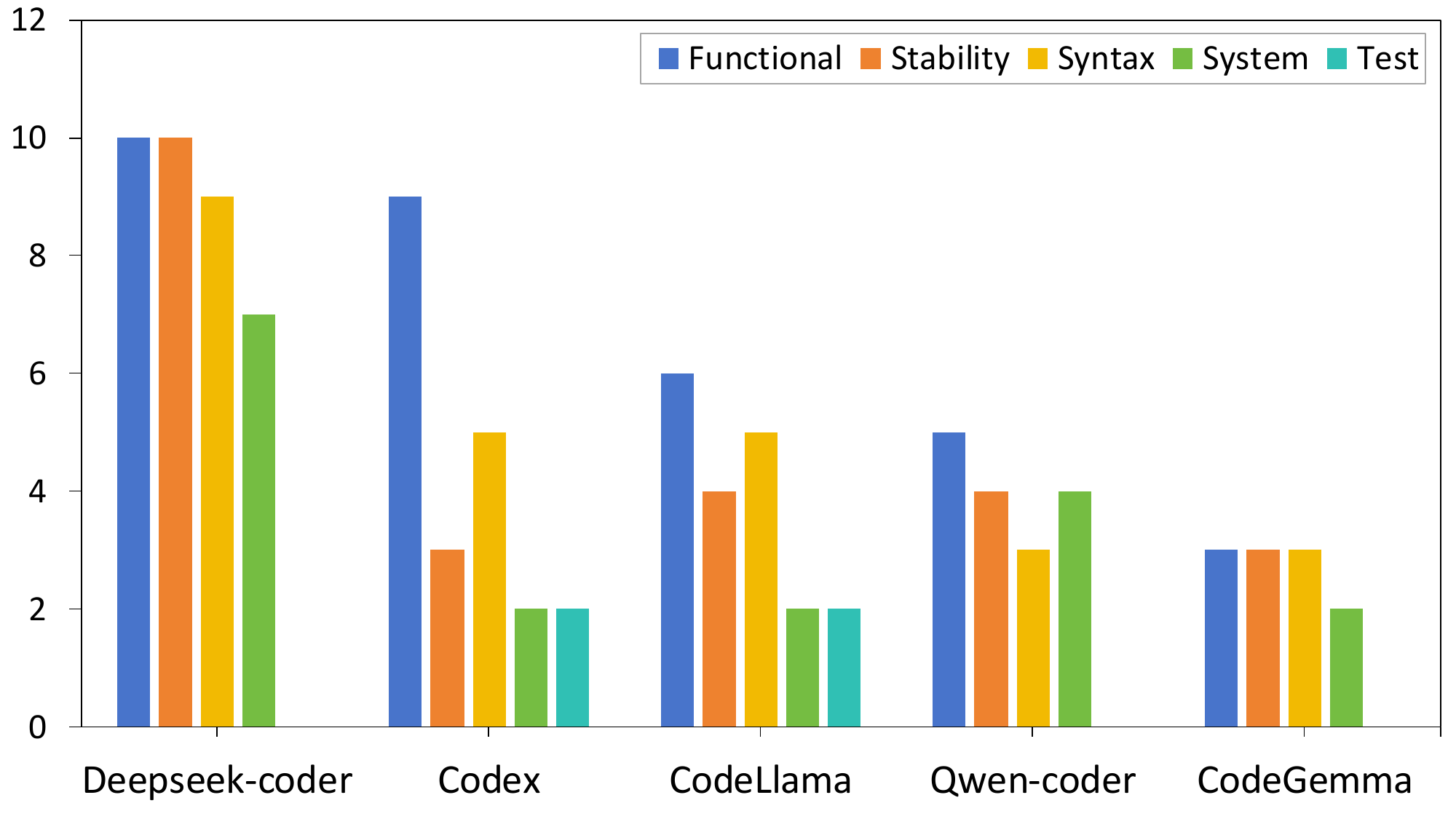}
        \label{fig:codespecificllm}
    }


    \subfigure[AI Coding Tools and Agents]{
        \includegraphics[width=0.6\textwidth]{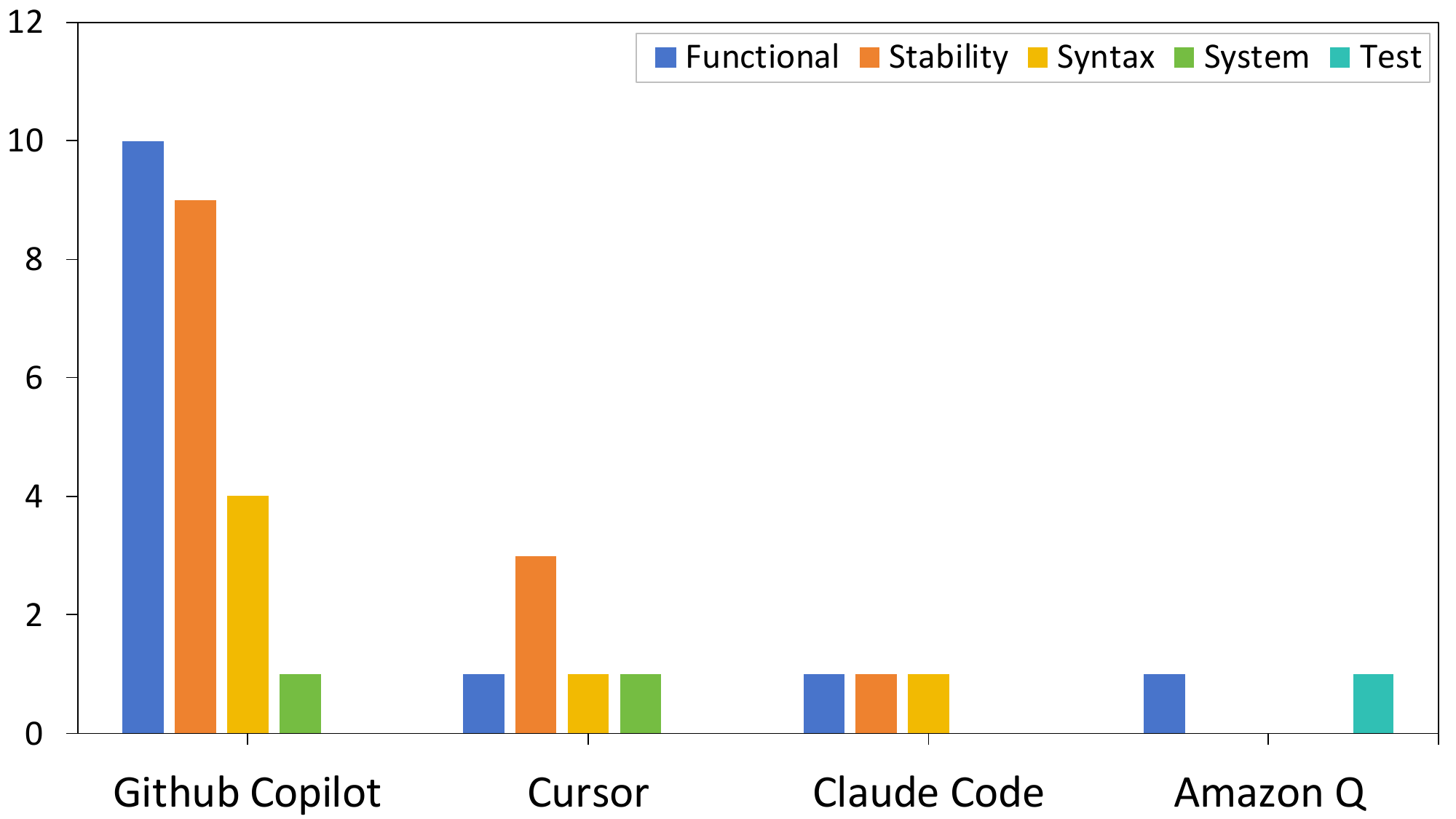}
        \label{fig:AItools}
    }

    \caption{The frequency of bug categories across LLMs and tools.}
    \label{fig:modelsandbugs}
\end{figure}

\update{\noindent \textbf{General-Purpose LLMs:} As shown in Figure~\ref{fig:generalllm}, among the studies included in our review, GPT models were the most extensively studied general-purpose LLMs, appearing in 67 studies after excluding code-specific variants. Within these studies, Functional Bugs remain the dominant bug type, reported in 52 studies (78\%), with semantic or logical bugs, variable/objects/parameters/properties-related bugs being the most frequently mentioned. Stability Bugs are the second most common, appearing in 38 studies (57\%), where they primarily involve failure to adhere to coding conventions and improper exception handling. Syntax Bugs appear in 32 (48\%) studies, though most studies note their presence without further elaboration of the bug type or the nature of their occurrence. System Bugs and Test-related Bugs are the least reported, appearing in 16 (24\%) and 8 (12\%) studies, respectively. Among the studies examining Llama models, Functional Bugs are the most frequently reported (7 studies, 78\%), followed by Stability Bugs (4 studies, 44\%). Syntax, System, and Test-related Bugs are reported in 3, 2, and 2 studies, respectively. The remaining model families are represented in fewer than 10 studies each. Gemma, Gemini, Claude, and DeepSeek exhibit a similar bug distribution, with the reported frequency decreasing in the following order: Functional, Stability, Syntax, System, Test-related Bugs. Qwen shows a slightly different pattern, where System Bugs are reported more frequently than Stability Bugs among the studies, resulting in the order Functional, System, Stability, Syntax, Test-related Bugs. Notably, except for GPT and Llama, none of the other five model families are associated with Test-related Bugs, largely because relatively few studies included in our review investigated AI-based test generation. Therefore, the limited sample size prevents definitive conclusions regarding differences in Test-related Bugs across model families.

We also compared model families by accessibility. Among the API-accessible families, GPT is by far the most extensively studied (67 studies), whereas Claude and Gemini appear in only 6 and 4 studies, respectively. Among open-weight families, Llama was investigated in 9 studies, substantially more than Gemma (3 studies). For the open-weight + API category, DeepSeek was examined in 6 studies, compared with 4 studies for Qwen. Notably, among the studies involving Llama, one identified a factual error bug type, which the authors interpreted as a manifestation of model hallucination. 

\noindent \textbf{Code-Specific LLMs:} As shown in Figure~\ref{fig:codespecificllm}, 12 studies examined DeepSeek-Coder, the code-specific variant of DeepSeek. Functional Bugs and Stability Bugs are the most frequently reported (10 studies, 83\%), followed by Syntax Bugs (9 studies, 75\%) and System Bugs (7 studies, 58\%). No study explicitly reported Test-related Bugs. Nine studies examined Code Llama, the code-specific variant of Llama. Functional Bugs were the most frequently reported (6 studies, 67\%), followed by Syntax Bugs (5 studies, 56\%) and Stability Bugs (4 studies, 44\%), while System Bugs and Test-related Bugs each appear in two studies (22\%). Codex, OpenAI's code-specific model family (distinct from the Codex coding agent), was examined in 11 studies. Functional Bugs are reported in 9 studies (82\%), followed by Syntax Bugs in 5 studies (45\%) and Stability Bugs in 3 studies (27\%). System Bugs and Test Bugs are each reported in 2 studies (18\%). CodeGemma, the code-specific variant of Gemma, was examined in 3 studies, with Functional, Stability, and Syntax Bugs reported in all three studies. Qwen-Coder, the code-specific variant of Qwen, was examined in 5 studies, with Functional Bugs reported in all five studies (100\%), while Stability and System Bugs were also frequently reported (4 studies each). CodeGen and InCoder were examined in 8 and 7 studies, respectively, although only 3 studies for each model provided model-specific bug distributions. Across these studies, Functional Bugs and Syntax Bugs were reported for both models. Notably, all three studies that used InCoder and two of the three used CodeGen also reported System Bugs.

Overall, the code generated by these code-specific models remains susceptible to both high-level functional or logical failures and lower-level syntax, stability, and system-related problems despite their specialization for coding tasks. Except for the DeepSeek family, general-purpose models and their corresponding code-specific variants in the Llama, Gemma, and Qwen families were examined in broadly similar numbers of studies. In contrast, DeepSeek-Coder appear in about twice as many studies as general-purpose DeepSeek models. This higher prevalence may partly reflect its earlier availability and explicit specialization for code-related tasks, making it a natural candidate for studies using code-generation benchmarks. In terms of accessibility, most of the code-specific LLMs discussed above provide publicly available model weights, whereas Codex is a notable exception as a proprietary code-specific model developed by OpenAI. Neither Claude nor Gemini has a separately designated code-specific model family. Anthropic, however, provides Claude Code, an agentic coding system built around the Claude family. We discuss such AI coding systems separately in the subsequent analysis of AI coding tools and agents.\\

\noindent \textbf{AI Coding Tools and Agents:} 
AI coding tools primarily assist developers by generating code snippets or suggesting edits, whereas agentic coding systems can operate across multiple files and handle complex development tasks with varying degrees of human supervision \cite{sapkota2025vibecodingvsagentic}. Tools such as GitHub Copilot have evolved to support agentic capabilities alongside traditional code-completion features. As shown in Figure~\ref{fig:AItools}, the surveyed studies identified several AI coding tools and agents, with GitHub Copilot being the most extensively studied. Originally introduced as a real-time code completion assistant, Copilot has gradually evolved toward agentic coding by incorporating an Agent mode that can autonomously performs multi-step development tasks. Across the 13 studies that examined Copilot, Functional Bugs are the most frequently reported, appearing in 10 studies (77\%). Stability Bugs appear in 9 studies (69\%), followed by Syntax Bugs in 4 studies (31\%), whereas System Bugs appear in only one study. This distribution of bug categories broadly matches that observed for GPT-based code generation models, consistent with Copilot's reliance on underlying GPT-family models. Cursor provides a similar combination of assistive and agentic capabilities. In addition to real-time code completion, Cursor provides a built-in Composer agent that can understand the codebase and perform project-level code editing. Cursor appear in 3 studies, all of which reported Stability Bugs (100\%), while Functional, Syntax, and System Bugs were each reported in one study. Claude Code is a command-line coding agent developed by Anthropic. It can read files, execute shell commands, and modify code across an entire codebase directly from the terminal. However, Claude Code appears in only one study, which reported Functional, Stability, and Syntax Bugs. Amazon Q Developer provides a broader software engineering workflow, having evolved from a code completion assistant into a development assistant with agentic capabilities spanning code generation, testing, code transformation, and software modernization. Its reported findings also show how the types of bugs investigated can vary by development task. One study \cite{10.1145/3756681.3756964} reported several Test-related Bugs, including Lack of Cohesion of Test Cases, Assertion Roulette, Duplicated Assert, and Conditional Test Logic. In addition, Omidvar et al. \cite{omidvar2024evaluating} reported that Amazon Q Code Transformation sometimes produced hallucinated or incorrect outputs, generated correct but incomplete responses, and produced outputs that appeared correct but resulted in unintended consequences. 

Although these agentic systems provide greater autonomy than traditional code-completion assistants, the limited number of studies makes it difficult to determine whether their bug profiles differ systematically from those of conventional AI coding tools.}



\subsubsection{Comparative Analysis of Bugs across LLMs.} \update{In this section, we provide a broader analysis of bugs in LLM-generated code, focusing on their evolution across models, empirical comparisons among code generation models, and the potential factors underlying their occurrence.}\\

\noindent\textbf{Evolution of Bugs Across Models.} 
To explore how bugs evolve with model development, we analyzed the evolution of bugs over time concerning specific models -the GPT family - from model GPT-2 (the oldest model in the reviewed studies) to GPT-4o (the latest available model in the reviewed studies). We observed that bug diversity increases as models evolve. For GPT-2 and GPT-3, the studies mainly observed \textit{Functional, Stability, Syntax, System,} and \textit{Test} Bugs. 
For Codex, bug types cover almost all categories. As the first generation of GPT models specifically designed for code, Codex has attracted extensive research on its weaknesses, including API misuse, semantic misunderstandings, and the first emergence of hallucination bugs. For GPT-3.5 and GPT-4, studies report indicate that, aside from unspecified bugs, most other bug types are present. GPT-3.5 is the most studied, producing a wide variety of bug types, with logical bugs remaining the most common. In GPT-4, aside from Functional and System Bugs, other bug types occur less frequently than in Codex and GPT-3.5, suggesting that logical reasoning failures remain a persistent challenge even as model capabilities improve. Research on GPT-4o is still limited, but early findings suggest its bugs mainly involve hallucinations and reasoning errors. 
This suggests that as models become stronger, low-level bugs decrease, while higher-level bugs, particularly logical and semantic, become more prominent and may occur in more complex scenarios. It is important to note that the frequency with which a model is studied directly affects the number and types of bugs observed. Early models such as GPT-2 and GPT-3 have been used in only a few studies, limiting the diversity of observed bugs. By contrast, Codex, GPT-3.5, and GPT-4 have been extensively investigated, leading to broader bug coverage. Therefore, bug-type coverage is closely tied to the number of studies and research focus, and caution is required when interpreting the evolution of model bugs.\\

\noindent\textbf{Empirical Comparison of Different Code Generation Models.}
\update{We analyzed studies that conducted empirical comparisons of code generation across different models, which discussed the presence of bugs in the generated code. The detailed results are included in \Cref{app:A} - \Cref{tab:bugtypeempirical,tab:contextempirical,tab:promptempirical,tab:testtaskempirical}. Several key findings emerge despite substantial differences in models, datasets, and evaluation setups. Evidence from the reviewed studies suggests that increasing model size generally improves correctness and reduces certain bug categories, but larger models still struggle with deeper semantic reasoning. Furthermore, different models - and even different sizes of the same model - tend to produce distinct bug types when performing the same task, indicating that bug patterns are highly dependent on model architecture and training \cite{wang2025towards, liu2026beyond}. Much of the inconsistency across studies can be attributed to variations in evaluation design, highlighting that conclusions about LLM reliability are highly sensitive to experimental setup. When faced with tasks and datasets of varying complexity, LLMs often struggle with more complex problem descriptions. Within the same model family, newer versions typically perform better and generate fewer bugs overall \cite{kabir2025well}. When generating unit tests, some LLMs tend to produce a wide variety of test smells and can generate syntactically invalid code \cite{zilberman2024no, siddiq2024using}. In test generation, providing context or additional supporting information usually reduces test smells. Finally, different prompting strategies yield distinct bug patterns \cite{tambon2025bugs, khojah2025impact, yuen2025prompting}. For example, manually written prompts often lead to minor mistakes, such as syntax bugs, whereas structured or task-oriented prompts are more likely to trigger semantic or logical bugs.} \\

\noindent\textbf{Reasons behind Bugs in the Generated Code.}
\update{To further understand why LLMs vary in their propensity to generate buggy code, we examine the underlying factors identified across the surveyed literature, as summarized with representative examples in \Cref{app:B} -- \Cref{tab:bugreasons}. Several studies (e.g., \cite{song2023empirical, yang2024evaluation}) suggest that the code-generation capabilities of different LLMs primarily depend on model size and their ability to understand instructions, and several studies (e.g., \cite{zheng2023codegeex, fan2023automated}) instead highlight the training data itself as the dominant factor. Others (e.g., \cite{du2024evaluating, bai2025collaboration}) suggest that model size, instruction dataset, and training data jointly determine the model's code generation capabilities. In summary, the code generation capabilities of different models are primarily influenced by model size, instruction-following ability, and the quality and distribution of their training data. Smaller models often struggle to understand complex task requirements and therefore tend to generate more erroneous code. A model's strengths across programming languages are closely tied to distributional biases in its training data. For example, some models excel in specific programming languages simply because those languages constitute a larger portion of their training corpus. Furthermore, because many models are trained on large datasets of human-written code that may contain mistakes, bug patterns in generated code often overlap with common human programming errors. Importantly, designing a model specifically for coding tasks does not guarantee a lower error rate. Overall performance remains strongly dependent on model size, as well as the quantity, quality, and diversity of the training data. These factors collectively play a critical role in reducing the bugs and defects present in generated code.}


\begin{tcolorbox}[title= RQ2 Key Findings]

All models tend to generate buggy code across different tasks and programming environments to varying extent. Model size, instruction-following capability, and the quality and diversity of training data are key factors that influence bugs in the generated code. 
As models evolve, the diversity of bugs in the generated code can change significantly. Bug patterns depend heavily on model architecture and training-data distribution. 

\end{tcolorbox}

\subsection{Bug Mitigation Approaches (RQ3)}

After analyzing discussions of bugs in \aigc{} in relevant studies, we further explore bug mitigation approaches specifically proposed for \aigc{}. As shown in Table~\ref{tab:mitigation}, these approaches can be broadly categorized into four groups: 1) prompt engineering, 2) code enhancement modules/frameworks, 3) autonomous coding agents, 4) program analysis and repair techniques, and 5) mitigation approaches for task-specific code generation. We describe each category in detail in the following sections. We classify each study by its primary mitigation mechanism. We also summarize and introduce studies on bug mitigation for specialized code generation tasks.

\begin{table}[h]
    \centering
    \footnotesize 
    \setlength{\tabcolsep}{4pt} 
    \renewcommand{\arraystretch}{1.05} 
    \caption{Mitigation approach distribution across studies}
    \begin{tabularx}{\textwidth}{p{3.3cm}p{1.2cm}>{\raggedright\arraybackslash}X}
        \toprule
        \textbf{Category} & \textbf{Studies} & \textbf{References} \\
        \midrule
        Prompt Engineering and Strategies & 45 & \cite{tian-etal-2024-debugbench, ren2023misuse, moratis2024write, zhong2024can, khan2024assessing, jesse2023large, wang2025towards, karanjai2023smarter, tambon2025bugs, liu2024no, liu2024refining, pan2024lost, billah2024large, zhang2024copilot, bai2025collaboration, chen2023forgetful, nettur2024cypress, yuan2024evaluating, yang2024evaluation, zhu2026assessing, liu2026beyond, wang2026specifications, fayed2026prompt, liang2024can, feischl2026large, pendyala2025performance, yang2025stop, nogueira2025beyond, zhang2026automated, borg2026echoes, song2025code, wang2025context, li2026performance, khojah2025impact, spracklen2025we, yuen2025prompting, dolcetti2026helping, dou2026wrong, sun2026quality, wang2026making, nekrasov2026iac} \\
        \cmidrule(lr){1-3}
        Code Enhancement Modules / Frameworks & 14 & 
        \cite{dakhel2024effective, ramos2024batfix, siddiq2024franc, inala2022fault, zhang-etal-2023-self, wu2023rustgen, siddiq2024using, jain2022jigsaw, yang2024evaluation, zhuo2026identifying, shehab2026scalable, liu2025evaluating, ron2026galapagos, bui2025correctness} \\
        \cmidrule(lr){1-3}
        Autonomous Coding Agents & 5 & 
        \cite{inproceedingsINTERVENOR, bai2025collaboration, zhang2026mpc, bai2025collaboration, wang2026making} \\
        \cmidrule(lr){1-3}
        Program-analysis-based Methods & 9 & 
        \cite{inproceedingsINTERVENOR, fan2023automated, liu2024refining, fakih2024llm4plc, mathews2024test, kouemo2024chain, yang2024evaluation, huang2026hallucination, nekrasov2026iac} \\
        \cmidrule(lr){1-3}
        Mitigation Approaches for Task-Specific Code Generation & 8 & 
        \cite{karanjai2023smarter,chen2023forgetful,yuan2024evaluating,dakhel2024effective,siddiq2024using,tang2024chatgpt, 10.1109/ASE63991.2025.00223, zhang2026automated} \\
        \bottomrule
    \end{tabularx}
\label{tab:mitigation}
\end{table}

\subsubsection{Prompt Engineering and Strategies}
\label{sec:promptstrategies}

\update{Prompt engineering techniques guide LLMs to generate outputs aligned with expectations through carefully designed input prompts. Researchers often enhance prompts by incorporating additional information and context such as API documentation, function signatures, or error messages. In other cases, prompts leverage in-context learning through few-shot examples and can be further augmented with Chain-of-Thought (CoT) reasoning to enhance the correctness and completeness of generated code. Due to its lightweight, cost-effective, and highly adaptable nature, prompt engineering has become a widely adopted strategy in LLM-based code generation and automated bug repair. We discuss the prompt engineering strategies identified in the reviewed studies in detail below, with the corresponding studies and illustrative examples summarized in \Cref{app:B} -- \Cref{tab:mitigationprompt}. 

\textbf{Few-Shot Examples} refer to providing a small number of examples in the prompt to help the model understand the task pattern and expected output \cite{wang2020generalizing}. \textbf{CoT} guides the model to reason step by step in prompts instead of answering directly \cite{wei2022chain}. Some studies use CoT to guide LLMs in step-by-step reasoning and progressively fix bugs in the code. The effectiveness of CoT prompting depends on both model architecture and task characteristics, and does not exhibit consistent improvements in code generation tasks. While CoT can improve performance for certain models or tasks, it may also degrade performance in others \cite{yang2024evaluation}. Moreover, by explicitly exposing intermediate reasoning steps, CoT improves interpretability but can also lead to output non-determinism \cite{ouyang2025empirical}. \textbf{Contextual Information} such as coding task description, relevant documentation, API descriptions, and function signatures is provided to the model, enabling it to understand the coding task accurately. Some studies focus on specific bug types and implement targeted repairs by embedding corresponding bug feature information or annotations in prompts \cite{jesse2023large}. Incorporating contextual information is more effective than simply scaling model size to improve code repair performance. Prompt granularity also matters: Fine-grained prompts, which provide specific and detailed instructions, reduce errors more effectively than coarse-grained prompts, which provide general, high-level instructions \cite{ren2023misuse}. Furthermore, LLM-based approaches and traditional APR techniques are complementary rather than interchangeable, with LLMs demonstrating greater flexibility in adapting to different repair requirements \cite{fan2023automated}. Several studies adopted \textit{Error Feedback}-based iterative repair mechanisms to continuously optimize code quality in the process of providing feedback to LLMs. Iterative feedback mechanisms can significantly improve code repair performance, but their benefits depend largely on feedback quality. No single feedback strategy is universally optimal, as more feedback does not necessarily yield better repairs; clear, relevant feedback matters more. Different bug types require tailored feedback approaches \cite{pan2024lost}. Despite improvements through multiple iterations, most functional and complex errors in \aigc{} remain difficult to resolve \cite{tian-etal-2024-debugbench}. Therefore, supplementing LLM-based bug repair with human knowledge or external tools, such as static analysis tools, representing an important direction for improving repair performance \cite{liu2024no}.} \\

\noindent \noindent\textbf{Summary of Prompt Engineering and Strategies:} When using prompt engineering for bug mitigation or repair, providing task-relevant, correct examples can substantially reduce model-generated bugs. Most studies incorporate error feedback within the CoT process to guide models toward alternative solutions. Combining CoT with RAG can further enrich prompts; however, CoT may increase output uncertainty \cite{ouyang2025empirical}, and RAG can degrade performance in certain scenarios \cite{yang2024evaluation}. The prompt's style also matters. Natural language prompts (instructional descriptions) and code-style prompts (example-based code templates) tend to work better for different model types \cite{yang2024evaluation}. In addition, explicitly highlighting bug characteristics can improve repair effectiveness, while knowledge-base-driven iterative checking and fixing can resolve most bugs. Compared to traditional APR tools, model-based repair approaches offer greater flexibility and efficiency. Meanwhile, static analysis remains effective as supplementary information during the fixing process to address maintainability bugs, but repairing logical bugs remains particularly challenging.

\subsubsection{Code Enhancement Modules/Framework} 
\label{sec:codeenhancement}

Those models and frameworks integrate multiple techniques, such as fault-aware neural code rankers \cite{inala2022fault}, syntax checking \cite{fakih2024llm4plc}, and program analysis \cite{siddiq2024franc}, forming a more advanced automated code generation workflow. Examples include RustGen \cite{wu2023rustgen}, Jigsaw \cite{jain2022jigsaw}, LLM4PLC \cite{fakih2024llm4plc}, and FRANC \cite{siddiq2024franc}. Specifically, Inala et al. \cite{inala2022fault} proposed CODERANKER, a fault-aware neural code ranker that predicts code correctness and reorders programs without execution. Trained on outputs from Codex, GPT-J, and GPT-Neo, it distinguishes intent bugs from execution bugs and can even locate faulty lines. Experiments show significant gains on datasets such as APPS, HumanEval, and MBPP, outperforming binary classifiers. Although CODERANKER identifies execution bugs more effectively than intent bugs, using cross-model data further improves performance. Building on error feedback, Wu et al. \cite{wu2023rustgen} introduced RustGen, which improves Rust code generation by treating the LLM as a black-box reasoning service. RustGen incorporates compilation errors back into the workflow through three mechanisms: history-based search to avoid repeated failures, prompt engineering to optimize generation, and grammar skeletonization to reduce prompt size. Compared to GPT-3.5 and GPT-4 baselines, RustGen improves compilation success, especially when combined with grammar skeletonization. However, logical bugs and missing functionality may still remain undetected without manual review. Similarly, Jain et al. \cite{jain2022jigsaw} proposed Jigsaw, which augments a pre-trained LM with pre- and post-processing modules and user feedback. Pre-processing refines natural language intent via context selection, while post-processing repairs code syntactically and semantically through checks, variable conversions, and AST rewriting. These mechanisms address reference and semantic bugs, improving accuracy by 15\% to 40\%. With continuous user feedback, GPT-3 accuracy improved by over 30\%, and Codex's solution rates rose substantially, particularly in TensorFlow tasks. 

Fakih et al. \cite{fakih2024llm4plc} presented LLM4PLC, a pipeline for verifiable PLC code. By combining prompt engineering, syntax checking, formal verification, and LoRA fine-tuning, LLM4PLC progressively corrects both syntactic and functional bugs under user guidance. Experiments demonstrate significant improvements in code success rate, compilation pass rate, and quality, highlighting the advantages of integrating verification tools with LLMs. Siddiq et al. \cite{siddiq2024franc} developed FRANC, a framework that filters, ranks, and repairs LLM-generated code. It applies static filtering to discard uncompilable code, quality-based ranking to prioritize stronger snippets, and LLM-driven repair to address bugs flagged by static analysis. FRANC is particularly effective for fixing vulnerabilities in Python (e.g., XML validation, subprocess API misuse, and Flask debug mode) and code smells in Java (e.g., array toString misuse and suspicious comparisons). However, it struggles with other complex bugs, such as infinite loops or out-of-bounds bugs. Zhang et al. \cite{zhang-etal-2023-self} proposed Self-Edit, a fault-aware code editor for code generation and a generation-editing framework that simulates human programming. In this approach, an LLM first generates code, then executes and tests it. The test results are converted into structured feedback comments, which a neural editor uses to revise the code. Self-Edit, along with its contextual learning variant, effectively corrects syntax, runtime, and semantic bugs, achieving up to 89\% performance improvement across nine models. Remarkably, smaller LLMs equipped with the neural editor can match or even outperform larger models, highlighting the importance of structured feedback and iterative editing. Zhuo et al. \cite{zhuo2026identifying} proposed Dr.Fix as a multi-stage LLM prompting approach that integrates detection and reasoning capabilities. It first detects potential API misuse, classifies it into predefined categories such as intent and hallucination misuse, validates the detected misuse through contextual reasoning, and finally generates syntactically correct and semantically meaningful repair suggestions aligned with developer intent. Bui et al. \cite{bui2025correctness} proposed Openia as a framework that leverages internal representations of code LLMs during code generation to assess the correctness of generated code, enabling earlier bug detection than post-hoc validation. It analyzes internal signals from DeepSeek-Coder, Code Llama, and Magicoder across standalone and repository-level code generation tasks. Experimental results show that Openia consistently outperforms classification-based and zero-shot approaches, improving accuracy and F1-score across diverse benchmarks, particularly in complex, context-dependent repository-level generation.\\

\noindent \textbf{Summary of the Code Enhancement Modules/Framework:} Structured error feedback is more effective than simple natural-language prompts for improving code generation quality. Automated, iterative pipelines are common in code-enhancement frameworks, and a staged approach involving generation, checking, and fixing is generally superior to single-step generation. Compared with preprocessing, post-processing mechanisms tend to deliver more direct and substantial benefits; in fact, a strong post-processing module can compensate for model-size limitations, making enhanced small models a cost-effective alternative to large models. 
Hybrid methods that combine static analysis with LLM-based repair can address most common bugs \cite{li2024enhancing}, but still struggle with complex logic bugs that may require in-depth dynamic analysis (which is known to be more expensive \cite{ernst2003static}). Ultimately, these techniques function more as patch-style enhancements to existing models rather than fundamental solutions to the underlying bugs in code generation.

\subsubsection{Autonomous Coding Agents} 
\label{sec:autonomousagent}
Coding agents are systems based on LLMs or other intelligent agents that can independently complete programming tasks and manage the task flow. From requirements understanding to design, coding, debugging, and testing, they function like ``self-driving programmers''. Previous work also implemented some autonomous program repair agents. For example,  Wang et al. \cite{inproceedingsINTERVENOR} proposed INTERVENOR, an interactive repair chain that iteratively leverages compiler feedback to guide repairs. It employs two LLM-based agents: a code \textit{learner} that generates initial code and a code \textit{teacher} that produces natural language repair instructions (CoRs) from compiler error reports. The learner applies these CoRs, and the process repeats until compilation succeeds or it reaches a round limit. Experiments show that INTERVENOR not only outperforms GPT-3.5 by approximately 18\% and 4.3\% on both generation and translation tasks, but also surpasses self-debugging and self-collaboration methods. It proves effective across languages and excels in \texttt{AssertionError} repair, highlighting the strength of precise, execution-based feedback in driving high-quality code fixes. Building on the idea of collaboration, Dong et al. \cite{dong2024self} introduced a self-collaborative framework in which LLMs adopt specialized ``expert'' roles-analyst, coder, and tester-mirroring human teamwork. Analysts decompose tasks and provide high-level guidance; coders implement and refine solutions; and testers evaluate functionality, readability, and maintainability, offering targeted feedback. This division of labor enables substantial improvements: GPT-3.5 with self-collaboration achieves state-of-the-art results on HumanEval and MBPP (including extended benchmarks) and improves ChatGPT's performance by $\sim$ 30\%. 
These results demonstrate that structured role assignment and intra-agent dialogue can significantly enhance LLM code generation on both standard and challenging tasks. 

Inspired by pair programming practices, Zhang et al. \cite{zhang2024pair} proposed PairCoder, where two LLM-driven agents collaborate as Navigator and Driver. The Navigator handles high-level planning, reflection, and exploration of multiple solution strategies, evaluating execution feedback to decide whether to retain or abandon candidate plans, and suggesting repair strategies. The Driver implements the Navigator's instructions, executes tests, and performs bug fixes as needed. Through iterative alternation, PairCoder integrates multi-plan exploration with feedback-driven refinement, substantially outperforming GPT-3.5-Turbo and DeepSeek-Coder across all benchmarks, with particularly strong improvements on challenging datasets such as CodeContest, which contains real competitive programming tasks that demand complex algorithmic reasoning and strict functional correctness. The success of this study lies in alleviating the difficulty of generating correct solutions on the first attempt by systematically combining exploration and iterative repair. In parallel, Mathews et al. \cite{mathews2024test} incorporated test-driven development into AI-assisted code generation through the TGen framework. Given a problem statement and test cases, an encoder agent first generates candidate code. A repair agent then proposes modifications based on verifier feedback, which the encoder applies, and PyTest runs the updated code. If the tests fail, the repair loop iterates until it succeeds. 
The repair loop brings further incremental gains, with GPT-4 improving by 5\% on private tests and Llama 3 by 9\%, and an overall +7.27\% on CodeChef. These results confirm that test cases help models better capture requirements, verify logic, and handle edge cases, while iterative repair provides steady refinement, particularly benefiting weaker models. Extending this collaborative paradigm, Bai et al. \cite{bai2025collaboration} proposed a multi-phase agent-LLM collaboration framework. Here, an intelligent agent assigns developers and reviewer roles to the LLM. User requirements are first refined into precise task descriptions, and then the developer generates solutions guided by CoT reasoning. The reviewer then checks syntax, semantics, and function calls. This approach nearly halves logical error rates on HumanEval and MBPP relative to GPT-3.5, with further reductions when integrated with GPT-4. Moreover, it significantly improves code quality scores, approaching full marks with GPT-4 and boosts security, as reflected in marked improvements in the Secure@1 metric. Zhang et al. \cite{zhang2026mpc} proposed MPC-Coder, a dual-knowledge-enhanced multi-agent framework for code generation in thermal processing control. It consists of five specialized agents for parsing, planning, coding, verification, and fixing, which operate in a closed-loop generation–verification–repair process. The verification agent checks generated code against logical specifications using formal verification tools, while the fixing agent iteratively addresses detected issues based on verification feedback, thereby improving code reliability. \\

\noindent \textbf{Summary of the Autonomous Coding Agents: }Overall, autonomous coding agents are evolving beyond simple code generation to increasingly handle the entire software development lifecycle, including requirement understanding, design, implementation, debugging, and testing \cite{dong2024self,bai2025collaboration}. 
This shift reflects a clear trend toward end-to-end automation. Within these systems, iterative feedback loops consistently outperform single-shot generation. Execution feedback, such as compiler errors and failing test cases, provides deterministic, verifiable signals that substantially improve the accuracy and reliability of code repair. Another key advancement lies in multi-agent collaboration with explicit role specialization. By assigning different roles to multiple LLMs, such as analyst, developer, and reviewer, and enabling structured communication among them, these systems establish more systematic reasoning and self-correction processes. Such collaborative frameworks significantly reduce logical bugs and yield greater improvements in code safety and robustness.

\subsubsection{Program Analysis and Repair Techniques}
\label{sec:programanalysisbased}
Program analysis-based repair methods use static, dynamic, or hybrid analysis techniques to identify and fix bugs in code. They typically use structured program information, such as compiler error messages, control flow, call graphs, or ASTs, to locate potential problems and generate repair strategies. For instance, BatFix \cite{ramos2024batfix} repairs LLM-generated code translations using program repair and synthesis. It takes the original program, the translated program, and sample inputs as input, then detects syntactic or semantic bugs via compiler errors and control-flow differences. It replaces problematic statements with placeholders and synthesizes correct replacements through statement mapping. Experiments show BatFix fixes 41\% of the Java-to-C++ programs from TransCoder, 50\% from Codex, and 22\% of Python-to-C++ programs from Codex. 
BatFix effectively repairs multi-line bugs such as variable initialization, type declarations, and API calls, but struggles when translations are of very low quality. Compared with GPT-3.5, it achieves similar overall fix rates but better addresses subtle bugs, showing complementarity with LLMs. In contrast, Ngassom et al. \cite{kouemo2024chain} proposed a test-free, fully automated method to improve the reliability of LLM-generated code. They identified error-prone nodes in the AST and generated targeted verification questions (Targeted VQs) to guide ChatGPT (gpt-3.5-turbo) in localized repairs. This approach reduced attribute errors by 66\%, name errors by 86\%, and increased runnable code by 13\% compared to no-VQ and general-VQ baselines. Although it introduced a few minor bugs, the false positive rate remained low, confirming its effectiveness in improving functional correctness.\\


\noindent \textbf{Summary of the Program Analysis and Repair Techniques: }Program-analysis-based repair methods rely on structured syntactic and semantic information to accurately locate bugs, offering far greater stability than the probabilistic nature of LLM-generated output. By leveraging compiler diagnostics, control-flow structures, and AST, these approaches excel at handling fine-grained bugs and multi-line semantic inconsistencies, making them a natural complement to LLM-based repair. Moreover, unlike agent-based systems that depend on execution feedback, program analysis can operate independently of any test environment or runnable code. This makes it particularly suitable for fix scenarios with no execution environment, cross-language code, or limited computational and memory resources.

\subsubsection{Mitigation Approaches for Task-Specific Code Generation}
\label{sec:taskspecificmitigation}
Several studies have combined multiple techniques to enhance LLM-based code generation for tasks such as smart contracts, robotics, and testing, aiming to improve code reliability and mitigate generation bugs. Karanjai et al. \cite{karanjai2023smarter} adopted a guided prompt approach for smart contract generation, parsing Solidity code to extract comments and file names and form structured prompts. They found that including meaningful function names, comments, and code templates improved code correctness, whereas placeholders reduced ChatGPT's performance, suggesting that detailed prompts positively influence the quality of generated code. Chen et al. \cite{chen2023forgetful} identified that LLMs in robot programming suffer from forgetfulness, where they forget prompt-provided information or misinterpret it as non-factual, leading to execution errors. By enhancing prompts with specialized functions, key constraints, and clearer instructions, completion rates and bug reductions improved significantly. 
However, simply reiterating prompt instructions did not yield significant improvement. 

Building on prompt-guided testing, Yuan et al. \cite{yuan2024evaluating} introduced ChatTester, which generates unit tests using intention and generation prompts. A verify-fix mechanism iteratively corrects compilation and assertion errors, improving compilation rate by 24\%, pass rate by 11\%, and increasing correctly asserted tests by 19\%. This approach showed similar benefits for CodeLlama-Instruct and CodeFuse. In parallel, Dakhel et al. \cite{dakhel2024effective} proposed MuTAP, which combines LLMs with mutation testing to improve test effectiveness. Syntax Bugs are corrected via re-prompting or line omission, and failing assertions are replaced using expected outputs. The process iterates until all mutants are killed or no unused mutants remain. MuTAP reduced unintended output behaviors by 44-52\% across Codex and Llama-2-Chat, demonstrating the benefit of iterative LLM-guided mutation testing. Similarly, Siddiq et al. \cite{siddiq2024using} improved JUnit test generation through two strategies: (1) varying context styles (e.g., JavaDoc inclusion, method signatures) and (2) applying heuristic fixes to address compilation, structural, and test smell bugs. For models like Codex, GPT-3.5-Turbo, and StarCoder, these methods significantly improved compilation rates-for instance, Codex increased from 5\% to 99\% on SF110, although LLMs still underperformed traditional tools like EvoSuite in test correctness and coverage. Tang et al. \cite{tang2024chatgpt} used manual repair, where undergraduate students attempt to fix bugs in generated unit test cases with the help of IntelliJ. Most bugs involved accessing private/protected fields, undefined methods, interface instantiation, or incorrect parameter types. This highlights that certain domain-specific or structural bugs still require human intervention.\\

\noindent \textbf{Summary of the Mitigation Approaches for Task-Specific Code Generation: }Overall, in task-specific code generation, research on bug mitigation has largely focused on improving prompts, as prompts are not merely auxiliary hints but one of the decisive factors shaping the quality of generated code. For smart contract generation, structured, information-rich prompts significantly improve correctness, with meaningful function names, comments, and templates providing clear benefits. Adding essential constraints and clarifying the task structure greatly improves task completion rates and reduces execution bugs. For test generation, simple context-enhanced prompts combined with lightweight program analysis can dramatically boost compilation rates; however, semantic correctness and coverage still lag behind traditional test generators. LLMs tend to produce tests that are syntactically valid but semantically insufficient. Moreover, certain domain-specific bugs still require human intervention, indicating that current LLMs struggle to fully address these bugs.

\subsubsection{Relationship between Mitigation Approaches and Bug Types} \update{
We identified 67 studies that explicitly report bug mitigation strategies and their effectiveness in fixing bugs in \aigc{}.

\textit{Prompt-based Strategies} (Section \ref{sec:promptstrategies}) are the most frequently adopted approaches in the surveyed studies, with 45 studies falling into this category. Among these studies, Functional Bugs are the most commonly addressed category (35 studies, 78\%), followed by Stability Bugs (24 studies, 53\%) and Syntax Bugs (19 studies, 42\%). In comparison, System Bugs (9 studies) and Test Bugs (4 studies) are discussed much less frequently. This suggests that prompt-based approaches can, to some extent, address high-level semantic issues as well as structural errors in generated code, likely because carefully designed instructions can guide model behavior. \textit{Code Enhancement Modules/Frameworks} (Section \ref{sec:codeenhancement}) are reported in 14 studies. Like prompt engineering, these studies mainly focus on Functional (9 studies), Stability (7 studies), and Syntax Bugs (7 studies), with relatively few addressing System or Test Bugs. This suggests that existing framework-based approaches are more commonly applied to improving the correctness and structural quality of generated code, although studies evaluate their effectiveness in different ways. \textit{Program Analysis and Repair Techniques} (Section \ref{sec:programanalysisbased}) appear in 9 studies. Functional Bugs are the most frequently investigated category (7 studies), followed by Stability (4 studies) and Syntax Bugs (4 studies), whereas only two studies consider System Bugs and one study addresses Test Bugs. The emphasis on functional and execution-related issues likely reflects the use of static or dynamic program analysis to identify defects in program behavior, rather than indicating superior repair capability. Finally, \textit{Autonomous Coding Agents } (Section \ref{sec:autonomousagent}) are the least represented approach, with only 5 studies. Four studies apply agent-based methods to repair Functional Bugs, while two studies each report applications to Stability and Syntax Bugs. Their limited representation primarily reflects the recent emergence of agentic coding systems, which have received limited attention in the literature due to their recency. Therefore, the available evidence is insufficient to draw conclusions about their comparative mitigation effectiveness across different bug categories.

Overall, Functional, Stability, and Syntax Bugs emerge as the most commonly addressed and mitigated bug types across existing bug mitigation strategies for AI-generated code in our reviewed studies. This distribution aligns with the prevalence of these bug types in our collected dataset. We note that many studies did not discuss the mitigation approaches they applied to fix or remove bugs, since the studies we analyzed primarily focus on bugs in \aigc{}, where bug mitigation strategies are applied after bugs are identified.
 The results reflect the distribution of research attention in this survey rather than the frequency or effectiveness of bug mitigation. Therefore, when interpreting these findings, one should exercise caution: the observed distribution does not necessarily reflect the full landscape of existing bug mitigation strategies, but rather the characteristics of the subset of studies included in our analysis. The actual mitigation performance varies across studies and cannot be inferred from these counts alone.}

\begin{figure}[htbp]
\centering
\includegraphics[width=0.9\linewidth]{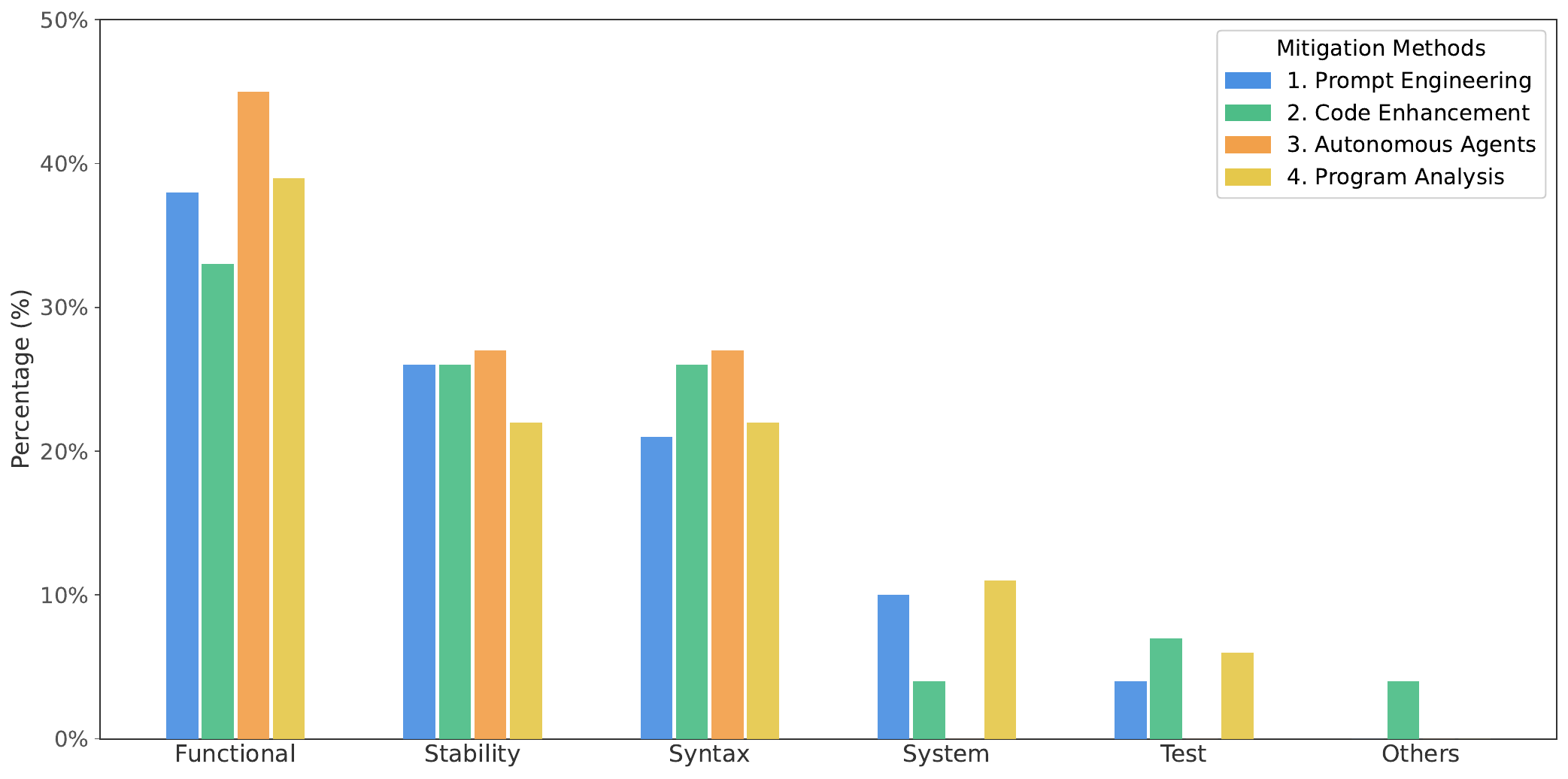}
\caption{Distribution of Bug Mitigation Methods Across Bug Types}
\label{fig:mitigationmapping}
\end{figure}

\begin{tcolorbox}[title=RQ3 Key Findings]
Prompt engineering techniques are widely employed to prevent bugs in \aigc{}. These methods are easy to apply but are sensitive to model instability and offer limited effectiveness for more complex bugs. 
Traditional program analysis techniques have been used as a complementary approach, offering more deterministic and interpretable signals for bug detection and repair. Certain domain-specific or semantically complex bugs will still require human intervention.

\end{tcolorbox}
\section{Discussion}
\label{sec:discussion}
\subsection{Interpretation of Findings}
We observed from the surveyed studies that, while \cgm{} are being increasingly used and adopted in practice, several challenges remain regarding the quality of the generated code. Bugs have been widely reported in \aigc{}, regardless of the models used. However, those \textbf{bugs vary in nature, frequency, and severity. Because the training corpora of almost all models contain substantial amounts of code from open-source projects, not all of this code adheres to rigorous logical standards, and it is known to contain many bugs} \cite{verdi2020empirical,rokon2020sourcefinder}.

In terms of bug types, \textit{functional bugs} consistently emerged as the predominant bug, encompassing seven subcategories that capture the majority of function-related bugs. Our analysis reveals that \textit{logic} and \textit{semantic bugs} were among the most frequently reported categories, both in their distribution within relevant studies and in their proportion among all specific bugs. This predominance indicates that LLM models continue to struggle with deep program semantics and logical reasoning. \textit{Logic bugs} can result from misunderstanding the code's underlying intent and require deep semantic understanding. However, \textbf{LLMs typically lack immediate execution or unit-testing feedback during the code generation process.} Consequently, even when the generated code is syntactically correct, it cannot guarantee the correctness of algorithmic or business logic. In contrast, \textit{syntax bugs} are generally easier for models to learn, as they can capture rule-based patterns in large-scale code corpora more effectively.

For RQ1, we focused in particular on \textit{hallucination bugs} in \aigc{}, as they pose one of the most significant threats to the credibility and practical usability of LLMs in software development.
We analyzed bugs that could potentially be attributed to model hallucinations, based on a broad definition of hallucination bugs.  
However, under a narrow definition, only bugs explicitly attributed to hallucination by the original studies would be classified as hallucination bugs. This narrower scope would limit the number of cases compared to the broader definition. Several studies provide clear examples under this narrow definition. For instance, Chen et al. \cite{chen2023forgetful} identified \texttt{``Factual Error''} as a bug type explicitly caused by model hallucination, in which the model fabricated numerical values rather than using the specific numbers provided by users in the prompts. 
Under this narrow definition, only bugs with explicit hallucination attribution from the original studies would be classified as hallucination-related, significantly reducing the number of such cases compared to our broader categorization.

\textbf{The GPT family of models has been the most extensively studied} in the selected studies, with their generated bugs covering all bug types in our taxonomy, mainly due to their early availability and widespread adoption, making them more integrated into tools such as GitHub Copilot and also easily accessible for researchers to experiment with. Moreover, many \AItools{}, including Copilot and Cursor, are directly built on GPT models or their fine-tuned variants, further increasing the frequency of studies focusing on these models.

As discussed above, \textbf{the predominance of \textit{functional bugs} poses a major challenge, as these deep logical bugs cannot be fully addressed by relying solely on static analysis or syntax checking.} With the known limitation of static analysis regarding soundness and precision \cite{ernst2003static}, we note that \textbf{more application of dynamic (or hybrid) analysis techniques is needed to verify the correctness and quality of the program, especially with the presence of dynamic program features that cannot be efficiently modeled through pure static analysis} \cite{sui2020recall}. 
Consequently, relatively few studies attempt to use program analysis-based methods in isolation to mitigate bugs in LLM-generated code. Instead, the research community has increasingly adopted hybrid approaches that combine multiple mitigation strategies. Prompt engineering and coding agents have emerged as mainstream methods. Prompt engineering techniques guide LLMs toward more logically coherent outputs by providing clearer contextual information, while coding agents can autonomously generate, modify, test, and even debug code. \textbf{Given the complexity of the bugs, multi-layered intervention strategies are necessary to operate effectively at different stages of the code generation pipeline.}
\subsection{Limitations}

\noindent \textbf{Taxonomy Limitation}. While analyzing bug data in \aigc{}, we observed that the boundaries between certain bug types are inherently ambiguous, leading \aigc{} to exhibit cross-categorical or hybrid bugs that do not neatly fit into a single classification category. For example, a \textit{memory bug} may also affect performance and reliability, potentially causing performance degradation or even crashes. \textit{Hallucinations} may also be related to functional bugs, as they sometimes result in missing features or deviation from the task objective. In the case of \textit{Test Bug}, many of those bugs can be related to any of the other bug types (i.e., a test bug can also be related to code style and standard issues). For frequently occurring bug types such as \textit{functional bugs}, we provide fine-grained subcategories because many studies offer detailed descriptions and examples of these bugs. In contrast, \textit{syntax bugs} are classified at a coarser level despite their high occurrence frequency. This is because most studies simply report the presence of syntax bugs in generated code without providing detailed bug descriptions or specific bug subtypes. However, this does not affect the validity of our taxonomy.\\

\noindent \textbf{Comparability Limitation}. When analyzing models' tendency to generate buggy code, some studies focus on a single model, even when different studies examine the same model; direct comparison remains challenging because experimental conditions vary, including datasets, model configurations, and prompt designs. Each of these factors can significantly influence the types and frequencies of bugs produced, making cross-study comparisons of single-model results inherently difficult and potentially misleading. \\

\noindent \textbf{Temporal Limitation}. Given the rapid evolution of LLMs and their training data, models are frequently updated or replaced by newer versions, while research publication cycles introduce an inevitable lag. Consequently, by the time a study is published, the model under investigation may already be outdated or replaced by successor models with different capabilities and bug characteristics. This temporal mismatch limits our ability to conclude whether current or emerging models tend to produce specific bug types. Although longitudinal studies examining how bug patterns evolve over time are lacking, we synthesize findings from relevant studies and analyze the types of bugs representative model families tend to produce across different model releases.\\

\noindent \textbf{Coverage Limitation}. In examining bug mitigation approaches, we specifically focused on studies that implemented an integrated framework encompassing AI-based code generation, bug detection, and targeted repair. Research adopting such end-to-end workflows ensures analytical consistency and facilitates the analysis of the interplay between generation and mitigation, revealing how different generation strategies influence the types of bugs produced and the effectiveness of subsequent mitigation efforts. However, this focus also means we did not include studies that focus solely on using LLMs for program repair.
We considered this to be outside the scope of this review, given the extensive body of work already studying the use of LLMs in program repair (e.g., \cite{zhang2024systematic}).

\subsection{Future Research}
For researchers, a promising direction for future research is to build a unified, scalable benchmark dataset for evaluating bugs in \aigc{}. This dataset should evolve with new LLM releases and tool updates, enabling comparisons of models over time. This would establish a more consistent and reproducible basis for evaluating whether model improvements affect bug patterns and the effectiveness of bug mitigation.

Moreover, most existing studies focus on relatively simple code-generation tasks, with limited attention to more complex, context-dependent development scenarios. Future work could explore how to integrate AI more deeply into the entire software engineering process. For example, by examining its performance and limitations in tasks such as configuration file generation, build script creation, and system integration.

Building on these findings, future research could focus on targeted model training and bug-fix strategies for the most frequent bug types, aiming to enhance models' understanding of specific bug patterns, particularly those that are difficult to fix, such as logic bugs. In addition, hybrid approaches and autonomous coding agent systems represent promising directions for enhancing the reliability of AI-generated code and more effectively mitigating such bugs.

For practitioners, it is important to exercise caution regarding the reliability and correctness of the generated code. Particular attention should be paid to logical bugs and hallucination phenomena, which can easily compromise code quality. At this stage, developers cannot fully rely on AI-generated outputs. Instead, they should critically review the generated code using their domain knowledge, continuously monitor the behavior of AI-assisted development tools, and use AI to boost productivity.
\section{Threats to Validity}
\label{sec:threats}
\noindent\textbf{Internal Validity:} The study selection process may affect the reliability of our findings and threaten internal validity, as our study may not cover all relevant literature. From the initial search strategy implementation to the subsequent screening phase, there is an inherent risk of missing relevant studies. We acknowledge that the subjective nature of our inclusion/exclusion criteria, as well as the reliance on researchers' knowledge during data extraction, may influence the robustness of our results. Moreover, some individual studies provide insufficient information regarding datasets, experimental setups, or evaluation metrics, which may affect the accuracy of our data extraction and bug classification. To mitigate these threats, we implemented several strategies. We rigorously followed the SLR guidelines of Kitchenham and Charters~\cite{KitchenhamGuidelines} and employed snowballing techniques~\cite{wohlin2014guidelines} to expand our sample and ensure comprehensive coverage of relevant studies. The research team explicitly defined and agreed on all selection criteria before beginning screening. The authors resolved disagreements through discussion among at least three authors until they reached a consensus. These measures help enhance the reliability and reproducibility of our sample selection, though we recognize that some degree of selection bias remains unavoidable in any systematic review. \update{Additionally, the reviewed studies rely on a diverse set of datasets, including HumanEval \cite{chen2021evaluatinglargelanguagemodels}, APPS \cite{coooper2021code}, DevGPT \cite{xiao2024devgpt}, MBPP \cite{austin2021program} and several self-constructed datasets. These datasets differ in task complexity, evaluation protocols, and test coverage, which may affect the direct comparability of bug frequencies across studies. However, our analysis does not aim to provide precise quantitative comparisons between models or datasets. Instead, it aims to synthesize and summarize studies and reports indicative trends to provide a comprehensive overview of existing research findings regarding bugs in \aigc{}.}\\

\noindent\textbf{External Validity:} 
\update{We applied Q1/Q2 journal rankings and CORE A*/A conference rankings as a venue-quality screening criterion. This may introduce selection bias, as venue ranking does not necessarily reflect the methodological quality of individual studies. Consequently, some relevant or high-quality studies published in other venues may have been excluded at this stage. Beyond this potential selection bias, the representativeness of our sample may also be affected by imbalances in the distribution of research domains, programming languages, and models. In particular, research is unevenly distributed across domains, with general code generation tasks studied far more than specialized areas such as test generation or smart contracts. Similarly, programming languages are unevenly represented. Our dataset is dominated by widely used languages such as Python and Java, while other languages are relatively underrepresented. Although many bug categories (e.g., logic, semantic, and performance bugs) are largely language-agnostic, others (e.g., syntax, type-related, and API-related bugs) depend on language-specific features. As a result, bug patterns in underrepresented languages may not be fully captured. A similar imbalance exists in model coverage, where popular LLMs are overrepresented, potentially biasing the observed distribution of bug types. However, such imbalances are largely unavoidable in systematic reviews and reflect broader publication trends in the literature \cite{kitchenham2004procedures}. To mitigate these limitations, we defined the bug classification at a high level of abstraction, focusing on bug characteristics rather than language- or model-specific constructs. This design allows us to provide a comprehensive view of bugs in AI-generated code, rather than being restricted to a specific domain, language, or model.\\}

\noindent\textbf{Construct Validity:} Our classification covers the majority of bug types found in \aigc{}. However, it cannot perfectly align with each original authors' understanding of bug classification across all selected studies, which could pose a threat to construct validity. Different researchers adopt varying definitions and conceptualizations of what constitutes a ``bug'', leading to inconsistencies in how studies categorize and describe errors. We mitigate this threat by explicitly defining the relevant concepts in our classification before classifying the studies, thereby reducing the potential impact of inconsistent bug definitions across studies. \update{Additionally, some studies included in the review (3\%) did not explicitly report their bug detection methods. As a result, the process by which these studies identified bugs remains unclear, which may introduce inconsistencies in the extracted bug types. To mitigate this issue, we examined each study's context and included only bugs clearly described in the reported results. We observed that these studies still provide detailed descriptions of the reported bugs, enabling reliable mapping to our classification. Therefore, while this remains a potential threat, we believe that its effect on the construct validity of our results is minimal.}

\section{Related Work}
\label{sec:relatedwork}
An increasing number of studies have explored the use of AI models for code generation tasks. In this section, we discuss prior literature surveys that focused specifically on \aigc{}.

Dehaerne et al. \cite{dehaerne2022code} explored the use of machine learning in automatic code generation and identified three code-generation paradigms: description-to-code, code-to-description, and code-to-code. The study found that the most commonly used machine learning models included Recurrent Neural Networks (RNNs), Transformers, and Convolutional Neural Networks (CNNs). To evaluate the quality of the generated code, the review compared the synthesized code with actual code statements from real-world applications.

Previous studies also explored in more detail how LLMs are transforming source code generation from natural language. Jiang et al. \cite{jiang2024survey} introduced a taxonomy to categorize and discuss recent developments in using LLMs for code generation, covering areas such as data management, recent advances, performance evaluation, ethical considerations, environmental impact, and practical applications.
Additionally, the survey compared various models on popular coding benchmarks to demonstrate trends in performance improvement.

AI-driven code-generation models are increasingly used in programming education. Educators mainly used LLM-based models to generate and evaluate assignments; students used them as virtual assistants to generate exercises, sample solutions, and alternative solutions. However, these models had limitations in accuracy, and the generated content often contained minor errors, manageable for teachers, but potentially risky for beginners. Academic misconduct and over-reliance on AI-generated content were also identified as key concerns when integrating these models into education \cite{cambaz2024use}.

The performance of AI for code generation tasks was not always reliable. Chen et al. \cite{chen2024survey} summarized the methods and metrics used to evaluate LLMs' performance in code-generation tasks. The article provided a detailed discussion of similarity-based evaluations (e.g., BLEU, ROUGE, METEOR, CodeBLEU), execution-based evaluations (e.g., compilation/interpreter success rate, unit test pass rate, and performance efficiency), and feedback-based evaluations (e.g., blind reviews, usability assessments, and readability evaluations). It also reviewed the main benchmarks and datasets used for code generation evaluation, including HumanEval, MBPP, and CodeXGLUE. It provided an in-depth analysis of the metrics covered in these benchmark suites.

Many literature reviews also focused on the security issues of AI-generated code. Ramirez et al. \cite{ramirez2024state} comprehensively analyzed the security issues in code generated by LLMs and identified common vulnerabilities, including simple bugs (SStuBs), code smells, security flaws (e.g., CWEs), and improper exception handling. Although there was a lack of research directly mapping these vulnerabilities to specific attacks, various mitigation strategies were proposed, including prompt engineering, iterative repair, and static analysis tools.

Similarly, Negri-Ribalta et al. \cite{negri2024systematic} examined the impact of AI models on code-generation security. The study found that current mainstream models (e.g., Codex, ChatGPT, CodeGen) generally produced insecure code with common MITRE CWE vulnerabilities and explored both the practical consequences and potential mitigations. It showed that prompt design had a significant impact on the security of the generated results. Developers, especially those without expertise, were often found to overtrust model outputs. While tools such as static analyzers, prompt optimization, and middleware protections were available, the overall security assurance remained insufficient.

Notably, some studies have focused on exploring data smells for coding tasks. Vitale et al. \cite{vitale2025catalog} identified and classified common data smells in programming tasks, including syntax errors, logical errors, runtime errors, semantic issues, compilation errors, performance inefficiencies, and maintainability concerns. Using specific examples, they proposed the corresponding mitigation strategies. The findings emphasized the importance of improving training datasets, optimizing training processes, and developing best practice guidelines. It also suggested that software developers should maintain a critical perspective toward AI-assisted code generation and adapt code review methods to detect specific types of vulnerabilities.

Compared with our survey, existing studies either comprehensively explore the application of AI in coding tasks or focus on the security issues associated with AI-generated code. In contrast, our work focuses on analyzing and identifying common bugs in AI-generated code and summarizing the methods used to mitigate or fix these bugs. To the best of our knowledge, this is the first survey that focuses explicitly on bugs in AI-generated code.
\section{Conclusions}
\label{sec:conclusions}
This study presents the results of a systematic literature review on bugs in \aigc{}. We surveyed studies that utilized \AItools{} and models and reported bugs and other issues in the generated code. We summarized a classification of bugs in \aigc{}, providing a unified analytical framework for future research. Furthermore, the review analyzed how prone various models and tools are to generating buggy code and summarized corresponding mitigation strategies. Our findings underscore that LLMs, while improving at generating working code, still have inherent limitations in reasoning about complex code, leading to bugs and defects in the generated code. To address this fundamental challenge, code-generation research has shifted away from isolated reliance on static analysis toward multi-layered intervention strategies that combine program analysis, prompt engineering, and coding agents.

This survey lays a foundation for understanding bugs in \aigc{} and offers practical insights for researchers and practitioners dedicated to improving the reliability of current and future code-generation methods. Future work should focus on building more robust datasets and expanding the scope to broader software engineering tasks, such as configuration file generation and project building, to fully explore AI's potential role in the end-to-end development workflow. Despite the challenges facing AI code generation, through systematic bug analysis and the continuous evolution of mitigation strategies, we are accelerating toward a new, highly reliable, AI-driven paradigm for software development.                                                                                                                                                                                                                                            

\section*{Data Availability}
We created a dedicated public repository (\url{https://zenodo.org/records/19650238}) to provide all relevant materials for this survey that are required for future replications or comparisons.

\begin{acks}
We thank Kai Tian for his help with data collection and processing in the early stages of this work.
\end{acks}

\bibliographystyle{ACM-Reference-Format}
\bibliography{ref}

\clearpage
\appendix
\section{Empirical Comparison of Different Code Generation Models - Additional Results}
\label{app:A}
\setcounter{table}{0}

\renewcommand{\thetable}{\thesection.\arabic{table}}
\makeatletter
\@addtoreset{table}{section}
\makeatother

\begin{xltabular}{\textwidth}{
    >{\raggedright\arraybackslash}p{1.3cm}
    >{\raggedright\arraybackslash}p{1.8cm}
    >{\raggedright\arraybackslash}p{2.3cm}
    >{\raggedright\arraybackslash}X
}
\caption{Bug Type-based Empirical Comparison}
\label{tab:bugtypeempirical} \\

\toprule
\textbf{Studies} &
\textbf{Bug type} &
\textbf{Models/Tools} &
\textbf{Findings} \\
\midrule
\endfirsthead

\toprule
\textbf{Studies} &
\textbf{Dataset} &
\textbf{Models/Tools} &
\textbf{Findings} \\
\midrule
\endhead

\midrule
\multicolumn{4}{r}{\textit{Continued on next page}} \\
\endfoot

\bottomrule
\endlastfoot
        Song et al. \cite{song2023empirical} & Semantic and Syntactic Bugs & CodeGen-16B, InCoder-1.3B, and ChatGPT-3.5 & The study found similar overall distributions of bug types across different models, with ``garbage code'' as the most common semantic bug, ``code block error'' as the most common syntactic bug. Notably, different LLMs sometimes produced entirely different semantic bugs for the same task.  \\
        \cmidrule(lr){1-4}
        Wang et al. \cite{wang2025towards}  & Semantic and Syntactic Bugs & CodeGen-16B, InCoder-1.3B, ChatGPT-3.5, GPT-4, SantaCoder, and StarCoder & They observed that all models struggle with complex logical conditions, regardless of size, leading to bugs such as incorrect conditions and logical direction bugs. Smaller models (InCoder, CodeGen) generated incorrect or incomplete code more often, while larger models (GPT-3.5, GPT-4) exhibited more consistent values and fewer arithmetic bugs. Importantly, smaller models exhibited a wider variety of bug types. Syntactic bugs followed a similar pattern, with all models struggled to produce complete code blocks. Specific bugs included GPT-3.5's hallucinated method calls, incorrect function names in CodeGen and InCoder, and incorrect parameters in GPT-3.5, SantaCoder, and StarCoder. \\
        \cmidrule(lr){1-4}
        Corso et al. \cite{corso2024generating} & Semantic and Syntactic Bugs & Copilot (Codex), Tabnine, ChatGPT (GPT-3), Bard (PaLM 2)  & They found Copilot produced the most correct methods, followed by ChatGPT, Bard, and Tabnine. While all tools performed similarly on self-contained or class-related methods, performance dropped sharply on external dependency methods, especially for ChatGPT and Bard, which cannot access external contexts. Notably, each tool generated unique methods that others could not replicate. \\
        \cmidrule(lr){1-4}
        Zhong et al. \cite{zhong2024can} & API misuse &  GPT-3.5, GPT-4, LLaMA-2, and Vicuna-1.5. GPT-4 & GPT-4 had the highest misuse rate, while LLaMA-2 had the lowest, with a poor compilation rate. The study also found that changing the temperature had a negligible effect on model improvement.\\
        \cmidrule(lr){1-4}
        Liu et al.\cite{liu2026beyond} & Hallucinations & DeepSeek-R1, DeepSeek-Coder-7B, DeepSeek-Coder-1.3B, CodeLlama-7B & The frequency of code hallucinations may be negatively correlated with the model's parameter size. The number of hallucinations identified from the code generated by DeepSeek-Coder-1.3B, DeepSeek-Coder7B, and DeepSeek-R1-671B decreases as the parameter size increases. DeepSeek-R1, despite exhibiting the fewest hallucinations overall, has the highest proportion of hallucinations in Library/Project. \\
        \cmidrule(lr){1-4}
        Ma et al. \cite{ma2026large} & Semantic and Syntactic Bugs & GPT-4o, GPT-4o mini, DeepSeek-V2-Lite-Chat, Gemma-2-9B-Instruct, Llama-3.1-8B-Instruct, DeepSeek-Coder-V2-Lite-Instruct, CodeLlama-7B-Instruct, CodeGemma-7B-Instruct & GPT-4o-mini and GPT-4o perform best in terms of code syntax, with GPT-4o-mini achieving the highest mean score and GPT-4o exhibiting the fewest Warning and Error issues. API import issues and variable management issues are the most frequently observed problems. In terms of semantic accuracy, GPT-4o consistently outperforms the other evaluated models across different prompting methods. In contrast, CodeLlama-7B demonstrates the weakest performance. These results indicate substantial differences among models in both syntactic and sematic quality. \\
        \cmidrule(lr){1-4}
        Zhuo et al. \cite{zhuo2026identifying} & API misuse & StarCoder-7B, Qwen2.5-Coder-7B, and GitHub Copilot & They identify four recurring misuse patterns in LLM-generated API code: Intent misuse, Hallucination misuse, Missing item misuse and Redundancy misuse. While Copilot and Qwen2.5-Coder generally outperform StarCoder, their shared error patterns indicate common challenges in API usage across models.  \\
        
        \cmidrule(lr){1-4}
        Abbassi et al.\cite{abbassi2025taxonomy} & Inefficiency & CodeLlama, DeepSeek-Coder, CodeGemma & 33.54\% of the studied sample exhibited multiple inefficiencies, indicating that inefficiencies in LLM-generated code are diverse and interconnected. General Logic and Performance inefficiencies are the most frequent and often co-occurring with Maintainability, and Readability inefficiencies. This co-occurrence suggests that limitations in LLMs' ability to generate correct and efficient code contribute to broader inefficiencies affecting code quality. The surveyed participants largely validated the relevance and prevalence of identified inefficiencies, confirming that it is aligned with realworld inefficiencies in LLM-generated code regardless of the model used. Additionally, while participants acknowledged these inefficiencies, they emphasized that in practice, they prioritize correctness over other code aspects. \\

        \cmidrule(lr){1-4}
        Kabir et al. \cite{kabir2025well} & Maintenance issues & ChatGPT-3.5, ChatGPT-4o & ChatGPT-3.5 successfully revised most Java files but failed on some tasks because the files were too long for the prompts to be processed properly. Among the revised files, some were semantically equivalent to the ground truth, while the remaining failures were mainly caused by unnecessary code truncation, missing project-specific logic, and insufficient handling of corner cases. ChatGPT-4o showed a similar overall ability to revise the Java files, although its outcomes differed from ChatGPT-3.5 on several tasks. Overall, ChatGPT-4o achieved slightly better semantic equivalence with the ground truth than ChatGPT-3.5. \\
        
        \bottomrule

\end{xltabular}
\begin{xltabular}{\textwidth}{
    >{\raggedright\arraybackslash}p{1.6cm}
    >{\raggedright\arraybackslash}p{1.9cm}
    >{\raggedright\arraybackslash}p{2.2cm}
    >{\raggedright\arraybackslash}X
}
\caption{Context-based Empirical Comparison}
\label{tab:contextempirical} \\

\toprule
\textbf{Studies} &
\textbf{Dataset} &
\textbf{Models/Tools} &
\textbf{Findings} \\
\midrule
\endfirsthead

\toprule
\textbf{Studies} &
\textbf{Dataset} &
\textbf{Models/Tools} &
\textbf{Findings} \\
\midrule
\endhead

\midrule
\multicolumn{4}{r}{\textit{Continued on next page}} \\
\endfoot

\bottomrule
\endlastfoot

Bai et al. \cite{bai2025collaboration} &
HumanEval and  MBPP dataset &
CodeGeeX, CodeLlama, DeepSeek-Coder-V2, GPT-3.5 and GPT-4 &
The study comprehensively evaluated the quality of the generated code, focusing on code readability, efficiency, and robustness. The study used the logical bug rate to measure the proportion of logical bugs in the generated code among failing test cases, directly reflecting the model's understanding of the task.  \\
\midrule

Zheng et al. \cite{zheng2023codegeex}  &
HumanEval-X benchmark &
CodeGeeX, GPT-J-6B, GPT-NeoX-20B, InCoder-6.7B, and CodeGen-Multi-6B/16B  &
CodeGeeX outperformed mixed-corpus models like GPT-J-6B and GPT-NeoX-20B, surpassed smaller code-specific models such as InCoder-6.7B and CodeGen-Multi-6B, and achieved performance comparable to the larger CodeGen-Multi-16B. In terms of bug types, the most common bug across InCoder, CodeGen, and CodeGeeX was Wrong Answer, with the proportion of such bugs exceeding 50\% across 200 sampled outputs. CodeGeeX had the highest rate of Wrong Answer, yet its generated code contained fewer runtime or syntax/semantic bugs. In contrast, InCoder was more prone to producing code with syntax or semantic bugs, while CodeGen consistently ranked in the middle among the three LLMs in terms of overall bug ratio. \\

\midrule

Billah et al. \cite{billah2024large} &
Programming platforms: Codeforces and LeetCode &
Gemini, Meta AI and ChatGPT &
They found that LLMs may struggle to understand larger problem descriptions, which can lead to poor performance on Codeforces. As for bug distribution, Gemini and Meta AI had the most difficulty generating compilable solutions to the problem. In contrast, ChatGPT struggled to generate accurate solutions within the time limits specified by the programming platform.\\
\midrule

Yu et al. \cite{yu2025realisticcodebench}  &
HumanEval, RealisticCodeBench, which was constructed from 2100+ real-world ChatGPT/Copilot-generated code samples from GitHub. & GPT-4.1, GPT-4o-mini, DeepSeek-V3, Llama 3.1, Phi-3, Mistral, ChatGLM, CodeGeeX4, DeepSeek-Coder, StarCoder2, CodeGen2.5, CodeLlama.
&
While models like CodeGeeX4-9B reach impressive pass@1 scores on HumanEval and DeepSeek-Coder-6.7B, their performance drops substantially on RealisticCodeBench's Python subset. This suggests that current LLMs may either be overly specialized for HumanEval-style problems or lack optimization for practical coding tasks.\\
\midrule

Ye et al. \cite{ye2025solcontracteval} &
SolContractEval, the first contract-level code generation benchmark & Claude-3.7-Sonnet, o4-mini, GPT-4o, Gemini-2.0-Flash-Exp, DeepSeek-R1, Qwen2.5-Coder
&
Claude-3.7-Sonnet [17] and o4-mini [18] demonstrate the strongest performance in contract-level code generation, followed by GPT-4o [19], DeepSeek-R1 [20], and Qwen2.5-Coder-32b [21]. Second, current LLM shows weaker performance on Solidity contract-level generation than on class-level generation in general-purpose languages. Third, models perform relatively well on the most common ERC20 contract type, but show significant shortcomings in generating more complex contracts like ERC721 and DeFi. \\
\midrule

Bayram et al. \cite{bayram2025comparative} &
HumanEval & GPT-3.5 Turbo, GPT-4 Omni, Claude 3.5 Sonnet, Claude 3.7 Sonnet, Claude Sonnet 4, and Claude Opus 4
&
The results indicate that Claude Sonnet 4 achieved the highest performance , followed closely by Claude Opus 4. Across all metrics, models developed by Anthropic Claude consistently outperformed those developed by OpenAI GPT. Anthropic Claude models were observed to generate more sophisticated and maintainable solutions with superior syntactic accuracy. In contrast, OpenAI GPT models tended to adopt simpler strategies but exhibited notable limitations in terms of reliability. \\
\midrule

Li et al. \cite{li2026performance} &
HumanEval, AixBench, MBPP, and the performance-oriented benchmark EvalPerf. &  GitHub Copilot, Copilot Chat, CodeLlama, and DeepSeek-Coder 
&
They identify several root causes that contribute to these regressions, including inefficient function calls, inefficient looping, inefficient algorithm, and inefficient use of language features. They conducted this root cause analysis for all models across all datasets, highlighting how different models exhibit distinct patterns of performance regressions depending on the dataset.  \\

\midrule

Dou et al. \cite{dou2026wrong} & HumanEval+, MBPP+, and APPS+ & GPT-4, GPT-3.5, Claude-3, DeepSeek-V3, DeepSeek-R1, Llama-3-Instruct2, Phi-3-Instruct, StarCoder-2, DeepSeekCoder & Experimental results indicate that DeepSeek-R1 and DeepSeek-V3 achieve the strongest code generation performance, while Phi-3 performs the worst. Closed-source models generally outperform open-source models, particularly on complex problems and real-world projects, with functional bugs being the most prevalent and syntax bugs the least frequent. LLM-generated code tends to be shorter but more complex than canonical solutions, while complex tasks often lead to algorithmic inefficiencies and timeout errors. Furthermore, bug distributions differ between standard benchmarks and real-world projects, highlighting the limitations of existing benchmarks in reflecting real-world code generation quality.  \\

\end{xltabular}
\begin{table}[H]
    \centering
    \footnotesize 
    \setlength{\tabcolsep}{4pt} 
    \renewcommand{\arraystretch}{1.05} 
    \caption{Test Generation Tasks-based Empirical Comparison}
    \begin{tabularx}{\textwidth}{p{2.0cm}p{1.8cm}p{2.2cm}>{\raggedright\arraybackslash}X}
        \toprule
        \textbf{Studies} & \textbf{Datasets} & \textbf{Models/Tools} & \textbf{Findings} \\
        \midrule
        Zilberman et al. \cite{zilberman2024no} & HumanEval, APPS, and ReCA &  GPT-3, GPT-3.5, and GPT-4 & GPT-3 produced the highest proportion of functional programs (likely due to the simplicity of HumanEval), while GPT-4 improved over GPT-3.5. Under black-box settings, all models generated longer test suites with irrelevant statements, whereas white-box prompts improved coverage by providing more context. The authors also analyzed test smells, finding that Magic Number (Hard-coding specific values in test code without using constant names or explanations) and Assertion Roulette (multiple assertions in the same test with no description, which makes it difficult to determine which assertion statement is the cause of the test failure) are the most common. GPT-3 produced the fewest smells under white-box conditions, while GPT-3.5 and GPT-4 performed worse on APPS and ReCA. \\
        \cmidrule(lr){1-4}
        Siddiq et al. \cite{siddiq2024using}   & HumanEval & Codex (2K/4K), GPT-3.5-Turbo, and StarCoder &  Less than half of their HumanEval unit tests compiled, though StarCoder reached 70\%. On SF110, only $2.7 - 12.7$\% compiled. The most frequent bugs were unknown symbols, incompatible conversions, and abstract instantiation, with unknown symbols causing over 62\% of the failures. For HumanEval, StarCoder produced the most correct tests, while GPT-3.5-Turbo generated the most tests with at least one passing case. Codex (4K) achieved the highest line and branch coverage, though still below manual tests. Familiar test smells included Magic Number and Assertion Roulette, AR, Lazy Tests, and Eager Tests; GPT-3.5-Turbo generated the fewest, while StarCoder's tests almost always contained at least one smell.  \\
        \cmidrule(lr){1-4}
        Du et al. \cite{du2024evaluating}  & ClassEval: class-level code generation benchmark  & Llama 3, Gemma, CodeLlama, CodeGemma, DeepSeek-V3, GPT-4o and Claude-3.5-sonnet  & The authors' analysis found that most incorrect code encountered \texttt{AttributeError} and \texttt{TypeError}, indicating that the model understood the error message and satisfied the syntactic or semantic constraints of the code context. Furthermore, in a few cases, \texttt{KeyError} was raised due to incorrect operations on dictionary variables. \\
        \cmidrule(lr){1-4}
        Yang et al. \cite{yang2024evaluation} & 17 Java projects &  CodeLlama, ChatGPT and DeepSeekCoder & CodeLlama-7B-Instruct generated many duplicate tests, while CodeLlama-13B-Instruct produced fewer but not better results. Many generated unit tests were syntactically invalid, mainly due to unresolved symbols, parameter mismatches, and abstract instantiations - all linked to hallucinations. CodeLlama-7B/13B-Instruct mostly produced bugs from runtime exceptions, reflecting a weaker ability in generating correct assertions.\\
        \cmidrule(lr){1-4}
        Alves et al. \cite{10.1145/3756681.3756964} & HumanEval &  GPT-4o, Amazon Q, and LLama 3.3 & GPT-4o exhibited the fewest errors, Amazon Q Developer the most, and LLaMA an intermediate level. Some errors were model-specific; for example, IndentationError and NameError occurred only in tests generated by Amazon Q Developer, whereas AssertionError was observed across all three LLMs. Amazon Q Developer exhibited the highest number and diversity of test smells, with some issues occurring exclusively in its generated tests. LLaMA exhibited only three types of test smells, while GPT-4o exhibited four, with a notable concentration of Suboptimal Assertion.\\
        \bottomrule
    \end{tabularx}
\label{tab:testtaskempirical}
\end{table}

\begin{table}[H]
    \centering
    \footnotesize 
    \setlength{\tabcolsep}{4pt} 
    \renewcommand{\arraystretch}{1.05} 
    \caption{Prompting Strategy-based Empirical Comparison}
    \begin{tabularx}{\textwidth}{p{2.0cm}p{1.8cm}p{2.2cm}>{\raggedright\arraybackslash}X}
        \toprule
        \textbf{Studies} & \textbf{Prompting strategy} & \textbf{Models/Tools} & \textbf{Findings} \\
        \midrule
        Tambon et al. \cite{tambon2025bugs} & Docstring-based prompts and human-labeled prompts &  CodeGen, PanGu-Coder, and Codex & For robust models like Codex, the most common bug pattern is Missing Corner Case. In contrast, for relatively weak and open-source models like PanGu-Coder or CodeGen, the most common category for both prompt types is Misinterpretation. Among all models, CodeGen has the highest number of bug samples in Silly Mistake and Syntax bug categories, which may be due to CodeGen's tendency to regenerate code (including signatures) from scratch. Codex has the highest number of samples in the Non-Prompted Consideration category across all prompt types. For both prompt types, the second and third-most-common bug patterns in PanGu-Coder-generated code are Missing Corner Case and Hallucinated Object. For the code generated by PanGu-Coder, changing the prompt type reduces the occurrence of Hallucinated Object but increases the number of Prompt-biased code. Considering all three studied models together, Misinterpretation is the most common bug pattern, followed by Missing Corner Cases, Hallucinated Object, Incomplete Generation, and Wrong Attribute. The incorrect code generated with human-labeled prompts appears to contain fewer ``trivial'' bugs, such as Syntax Bugs and Silly Mistakes, than the code generated with docstring-based prompts. \\
        \midrule
        Khojah et al. \cite{khojah2025impact} & Few-shot, persona, chain-of-thought, function signature, list of packages & GPT-4o, Llama3, Mistral & Including a function signature or few-shot examples significantly improves code correctness and reduces common errors, particularly TypeError, AttributeError, and ImportError. Providing package information further decreases import-related errors, although it may introduce hallucinated imports. Combining multiple prompt techniques offers little additional benefit. Llama 3 tends to generate simpler solutions, while signatures and few-shot examples often produce more complex code with additional code smells. Adding a relevant persona slightly improves code quality but comes with a small reduction in correctness. \\
        \midrule
        Yuen et al. \cite{yuen2025prompting} & Zero-Shot, Few-Shot, Chain-of-Thought, and Chain-of-Thought-Few-Shot & GPT-4o, Llama3-70B & CoT and CoT Few-Shot consistently achieved higher correctness for both Llama 3 and GPT-4o than Zero-Shot prompting. They also generated code that was more similar to human-written code while maintaining comparable complexity and structure. In contrast, Zero-Shot prompting produced substantially more errors across both models, with Logic and Control Flow errors being the most prevalent error type across prompting strategies. \\
        \bottomrule
    \end{tabularx}
\label{tab:promptempirical}
\end{table}

\begin{xltabular}{\textwidth}{
    >{\raggedright\arraybackslash}p{1.8cm}
    >{\raggedright\arraybackslash}X
    >{\raggedright\arraybackslash}X
}
\caption{Reported Reasons for Bugs in LLM-Generated Code}
\label{tab:bugreasons} \\

\toprule
\textbf{Studies} &
\textbf{Findings} &
\textbf{Reported Reasons} \\
\midrule
\endfirsthead

\toprule
\textbf{Studies} &
\textbf{Findings} &
\textbf{Reported Reasons} \\
\midrule
\endhead

\midrule
\multicolumn{3}{r}{\textit{Continued on next page}} \\
\endfoot

\bottomrule
\endlastfoot

\multirow{2}{=}{Song et al. \cite{song2023empirical}}
& InCoder-1.3B generated more logical bugs than CodeGen-16B and ChatGPT-3.5.
& The authors attributed this difference to the substantially smaller model size of InCoder-1.3B, which may limit its ability to understand complex task requirements. \\
\cmidrule(lr){2-3}
& Both open-source LLMs (CodeGen-16B and InCoder-1.3B) exhibited a high rate of ``missing multiple sentences'' bugs, whereas such bugs were relatively rare in ChatGPT-3.5.
& The authors proposed that ChatGPT-3.5 may be better at understanding natural-language input and task requirements, which could reduce this type of error. \\
\midrule

Yang et al. \cite{yang2024evaluation}
& The two largest open-source LLMs (Phind-CodeLlama-34Bv2 and DeepSeekCoder-33B-Instruct) achieved higher code coverage than the remaining smaller-scale LLMs, while GPT-4 generally outperformed the studied open-source models.
& The authors suggested that larger-scale LLMs tend to have stronger instruction-understanding capabilities, which may contribute to better code-generation performance. \\
\midrule

Zheng et al. \cite{zheng2023codegeex}
& CodeGeeX performed best on Python, whereas CodeGen-Multi-16B showed its strongest performance on Java. Models exhibited higher syntax error rates when generating Go code because Go's strict syntactic constraints caused many logically correct programs to fail compilation.
& The authors suggested that model performance on specific programming languages is closely related to the distribution of programming languages in the models' training data. \\
\midrule

\multirow{2}{=}{Fan et al. \cite{fan2023automated}}
& The bug classifications of Codex-generated code overlapped with those in Codeflaws, suggesting that Codex tends to make programming mistakes similar to human programmers.
& The authors attributed this similarity to Codex being trained on large amounts of human-written code, which may itself contain buggy programs. \\
\cmidrule(lr){2-3}
& Codex-generated solutions frequently exhibited syntax and algorithm-related bugs, including incomplete code, undefined variables, unmatched brackets, and incorrect algorithm selection caused by misleading identifier associations.
& The findings suggest that patterns in the training corpus may influence generated code, causing the model to reproduce misleading associations between identifiers, algorithms, and task contexts. \\
\midrule

\multirow{2}{=}{Du et al. \cite{du2024evaluating}}
& GPT-4 and GPT-3.5 outperformed other LLMs in class-level code generation and produced fewer buggy implementations, whereas smaller models such as PolyCoder and ChatGLM generated substantially more buggy code.
& The authors associated the higher bug rates of smaller and general-purpose models with their limited model capacity and weaker generalization capabilities. \\
\cmidrule(lr){2-3}
& All evaluated models achieved lower method-level accuracy on ClassEval than on HumanEval.
& The authors suggested that context-dependent class-level generation introduces additional opportunities for logical and structural bugs compared with standalone method generation. \\
\midrule

\multirow{2}{=}{Bai et al. \cite{bai2025collaboration}}
& DeepSeek-Coder-V2 achieved better code quality and task understanding and generated fewer bugs overall.
& The authors primarily attributed this improvement to its larger parameter size and extensive pre-training data, which may enhance its ability to avoid logical and structural bugs. \\
\cmidrule(lr){2-3}
& Although CodeGeeX was specifically designed for coding tasks, it still generated more buggy code than GPT-3.5, GPT-4, and CodeLlama.
& The findings suggest that being designed specifically for coding does not necessarily guarantee lower bug rates; instead, parameter size and the quantity, quality, and diversity of training data play important roles in reducing bugs. \\
\midrule

Kou et al. \cite{kou2024large}
& Only three GPT-4 bugs could be explained using attention-misalignment patterns, suggesting that weaker models such as CodeGen-2.7B are more susceptible to attention misalignment and that their generation bugs are therefore more explainable.
& The authors suggested that as language models become larger, they may develop different mechanisms for interpreting prompts and generating content, making their bugs less readily explained through attention-misalignment patterns alone. \\

\end{xltabular}
\clearpage

\section{Bug Mitigation Approaches - Additional Results}
\label{app:B}
\setcounter{table}{0}

\begin{xltabular}{\textwidth}{
    >{\raggedright\arraybackslash}p{1.8cm}
    >{\raggedright\arraybackslash}p{1.8cm}
    >{\raggedright\arraybackslash}X
}
\caption{Prompt Engineering and Strategies for Bug Mitigation}
\label{tab:mitigationprompt} \\

\toprule
\textbf{Prompting technique} &
\textbf{Studies} &
\textbf{Details} \\
\midrule
\endfirsthead

\toprule
\textbf{Prompting technique} &
\textbf{Studies} &
\textbf{Details} \\
\midrule
\endhead

\midrule
\multicolumn{3}{r}{\textit{Continued on next page}} \\
\endfoot

\bottomrule
\endlastfoot

Few-shot Examples
& Zhong et al. \cite{zhong2024can}
& They evaluated API misuse and compilation rates under different prompting strategies: zero-shot (no examples), one-shot unrelated (example from an unrelated task), and one-shot related (example using the same API correctly). In the zero-shot setting, Llama showed the lowest misuse rate, largely because many responses contained no code. For the DeepSeek-Coder series, larger specialized models produced more compilable code but did not reduce API misuse, indicating that better code generation does not guarantee higher reliability. The one-shot unrelated strategy was largely ineffective and even increased misuse, while the one-shot related strategy significantly reduced misuse for GPT-3.5, GPT-4, and Vicuna, but had little impact on Llama. Since LLMs are trained on codebases that may include API violations, they can easily reproduce such misuse, explaining the higher error rates in zero-shot and one-shot unrelated settings. \\

\cmidrule(lr){2-3}
& Li et al. \cite{li2026performance} & 
Few-shot prompting that targets the root causes of performance regressions helps identify potential performance bottlenecks and guides LLMs toward more efficient code. Prompt effectiveness varies by metric: Chain-of-Thought (CoT) tends to improve static regression analysis on benchmarks such as HumanEval and MBPP, whereas few-shot exemplars more consistently improve dynamic metrics such as execution time and CPU utilization. However, improvements in memory usage remain limited, indicating the need for further investigation.\\

\cmidrule(lr){2-3}
& Khojah et al. \cite{khojah2025impact} & 
Providing a function signature or few-shot examples has the clearest positive impact on correctness, while other prompt techniques show no statistically significant effect and may sometimes hurt performance. These techniques also reduce common errors, particularly TypeError, AttributeError, and ImportError; package information further reduces ImportErrors but may introduce hallucinated imports. Although CoT, persona, and package information can slightly reduce correctness, they tend to improve code quality in terms of code smells. Llama 3 consistently generates simpler code, with LLM-generated solutions generally no more complex than human-written code.\\

\midrule

Chain of Thought (CoT)
& Billah et al. \cite{billah2024large}
& They adopted a chained thinking prompt strategy, where it checks if the LLM's initial submission is rejected, it provides the model with error feedback from the online evaluation system (such as wrong answer, time limit exceeded, memory limit exceeded, compilation error, or runtime error). The model then generates alternative solutions based on this feedback, usually making up to 5 attempts. Through the feedback mechanism, ChatGPT and Gemini successfully solved over 12\% of the 98 LeetCode problems. ChatGPT and Meta AI successfully solved $\sim$5\% of the 126 Codeforces problems. In addition, LLMs showed considerable success in solving memory-limited problems, but ChatGPT showed challenges in time efficiency, while Gemini and Meta AI had difficulties in generating compilable solutions.  \\

\cmidrule(lr){2-3}

& Yang et al. \cite{yang2024evaluation} & They primarily improved code generation quality by optimizing prompt design and evaluating in-context learning (ICL) methods \cite{dong2024survey}. Prompt design focused on both description style (natural language (NL) and code language (CL)) and code features (such as method bodies, parameters, constructors, fields, and related methods). ICL methods include CoT, which guides the model to understand the target method before generating unit tests, and Retrieval-Augmented Generation (RAG), which enriches prompts by retrieving similar methods and their tests. The study evaluated the performance of five open-source LLMs (based on CodeLlama and DeepSeek-Coder, ranging from 7B to 34B parameters), GPT-4, and the traditional tool EvoSuite in generating unit tests. In prompt design, the NL style was more suitable for some models (such as CL-7B and CL-13B), while the inclusion of target class methods in code features improved grammatical validity but limited coverage. Regarding ICL methods, CoT improved DeepSeekCoder but weakened the CodeLlama series, while RAG generally led to performance degradation. \\

\cmidrule(lr){2-3}

& Ouyang et al. \cite{ouyang2025empirical} & They explored the impact of different prompt engineering strategies on the non-determinism of code generation. They found that while the CoT strategy may help reasoning in some tasks, in code generation tasks, CoT may actually increase output nondeterminism, particularly in low-temperature settings. This is counterintuitive, as it is generally believed that asking a model to ``think'' leads to more consistent output. Although output nondeterminism is not a bug itself, it can increase the likelihood of generating buggy code due to inconsistent reasoning paths across generations. \\

\cmidrule(lr){2-3}
& Vieira et al. \cite{10.1109/ASE63991.2025.00195}  & Standard zero-shot (S-ZS) and curated zero-shot (C-ZS) outperform Chain of Thought (CoT) across compilation, runtime, and test success. General-purpose models such as Llama 3.1 70B are particularly sensitive to verbose prompting, with performance declining substantially under CoT in both Java and Python. These results suggest that CoT may introduce unnecessary verbosity without the structured inductive bias required for program synthesis, highlighting prompt engineering as an important tuning layer alongside model architecture and training. \\

\midrule

Contextual Information & Jesse et al. \cite{jesse2023large} & They studied the impact of natural language annotations on suppressing simple/single statement bug generation in Codex. Using CodeTrans, they automatically generated annotations highlighting incorrect or fixed lines. Results showed annotations substantially improved Codex's matching and patch generation, while smaller LLMs saw little benefit. In PolyCoder and CodeGen, bugs decreased, and patch rates increased. In Codex models (Cushman, Davinci), although bugs rose, patch rates increased even more, nearly halving the error-to-patch ratio. This suggests annotations are more effective than simply scaling model size. The most common fixed bugs involved CHANGE\_IDENTIFIER and DIFFERENT\_METHOD\_SAME\_ARGS. Even with flawed annotations, Codex produced better code, whereas smaller models were easily misled.  \\

\cmidrule(lr){2-3}

& Wang \cite{wang2025towards} & They compared GPT-4's repair performance of its incorrect code solutions with and without labeled semantic and syntactic bug characteristics. On HumanEval, GPT-4 corrected 7 out of 18 buggy solutions with labels versus 4 out of 18 without; on BigCodeBench-Hard, it corrected 12 out of 18 with labels versus only 3 without. These results demonstrate the value of bug taxonomies in improving repair.  \\

\cmidrule(lr){2-3}

& Ren et al. \cite{ren2023misuse} & They examined the granularity of prompts for exception handling. General prompts reduced some incompleteness but often introduced misuse. Coarse-grained prompts (partial API exception specs) reduced bugs further but still left gaps. Fine-grained prompts (complete exception types and triggering conditions) nearly eliminated misuse, with only minor incompleteness. Building on this, they proposed the Knowledge-driven Prompt Chain (KPC), which iteratively checks and fixes exceptions using an API knowledge base. KPC typically resolves most bugs within ten iterations and outperforms ChatGPT in exception handling.  \\

\cmidrule(lr){2-3}

& Fan et al. \cite{fan2023automated}  & They compared traditional APR tools with prompt engineering. Using TBar and Recoder on Codex-generated code, Recoder produced fewer plausible but more correct patches, repairing more tasks overall. However, both struggled with multi-line edits due to reliance on single-line fault localization. To address this, the authors introduced Codex-e, with three strategies: Codex-ebug (error notification), Codex-eline (fixing a line), and Codex-estm (fixing suspicious statements). Codex-estm achieved the best performance by leveraging program text to locate bugs. Overall, Codex-e generated more correct patches than APR tools, offering more flexible and efficient repair capabilities.  \\

\cmidrule(lr){2-3}

& Zhu et al. \cite{zhu2026assessing}  & Overall, surrounding context provides limited overall improvement, but benefits Complex and context-dependent cases, particularly for GPT-o3-mini and Qwen2.5-32B. Qwen2.5-7B also benefits notably in cases where no replacement API is explicitly specified, while stronger models show fewer gains. Deprecated API documentation mainly benefits cases where no replacement API is explicitly specified, whereas replacement API documentation provides more substantial improvements, especially for Complex cases. Providing both types of documentation generally yields slightly greater gains, although the benefits are less consistent for Qwen2.5-7B. \\
\midrule

Iterative Prompting/Error Feedback & Liu et al. \cite{liu2024refining} & They investigated iterative repair with two types of feedback: simple feedback and feedback enhanced with static analysis and found that ChatGPT could resolve most style and maintainability bugs in Python code when guided by static analysis and runtime errors, since tools like Pylint provide precise locations and even suggested fixes. In contrast, most performance and efficiency bugs in Java were better solved with simple feedback. Interestingly, for bugs tied to incorrect output or efficiency, simpler prompts often outperformed detailed ones, likely because compiler feedback tends to be vague or unhelpful. \\

\cmidrule(lr){2-3}
& Pan et al. \cite{pan2024lost} & They introduced an iterative prompting approach that reuses contextual information from failed translations, including the original source, erroneous output, bug details, expected behavior, and even model-specific keywords. This strategy reduced compilation errors for GPT-4 after the first iteration, but showed weaker effects on runtime and functional bugs. Moreover, some translations deteriorated in later iterations, with functional bugs being converted into runtime or compilation errors, indicating that repeated repairs can introduce trade-offs and even new bugs. \\

\cmidrule(lr){2-3}
& Tian et al. \cite{tian-etal-2024-debugbench}  & They explored LLMs' own debugging capabilities by prompting them to generate multiple responses for a single debugging query. They found that increasing the number of samples significantly improved debugging performance, especially for condition, variable, and operation errors. Providing runtime information, such as program outputs and tracebacks, further enhanced repair: tracebacks worked well for syntax and reference errors by pinpointing error locations, but offered limited help for logic bugs, which remained difficult to debug. Overall, syntax and reference errors were much easier to fix, while logic or multi-error cases remained nearly as challenging as full code generation. \\

\cmidrule(lr){2-3}
& Liu et al. \cite{liu2024no}  & They manually examined ChatGPT's iterative debugging performance across different bug types, including incorrect answers, compilation errors, runtime errors, and timeout bugs. They observed that while compilation and runtime errors could often be fixed after multiple rounds, most incorrect answers and difficult tasks remained unsolved because ChatGPT lacked logical reasoning. They also pointed out stability problems, such as avoiding empty code bodies, and consistency bugs with human intent, like missing method signatures. The authors suggested that supplementing LLMs with human knowledge, for example, bug locations or repair hints, could help overcome these limitations.  \\

\cmidrule(lr){2-3}
& Feischl et al. \cite{feischl2026large}  & The iterative repair procedure generally decreases the pass rate across the evaluated LLMs. While consecutive repairs may further reduce the pass rate for CodeLlama-70B, they can improve performance over previous attempts for GPT-3.5; however, after three iterations, the pass rate remains below the initial level for all models. Selecting the best translation from the initial output and repair iterations can improve the overall pass rate. The procedure also slightly reduces code smells, particularly after the first iteration, although the extent varies across LLMs. \\


\end{xltabular}

\end{document}